\documentclass[trackchanges,twocolumn,resetfootnote]{aastex7}

\usepackage{wasysym}
\usepackage{comment}
\usepackage{siunitx}
\usepackage{soul}
\shorttitle{NIRCam GTO 2780 -- Debris Disks}
\shortauthors{G\'asp\'ar et al.}

\begin{document}

\title{The JWST/NIRCam Scattered Light Disks GTO 2780 program: panchromatic coronagraphic imaging of the
HD~10647, HD~32297, HD~61005, HD~107146, and HD~181327 debris disk systems}

\author[orcid=0000-0001-8612-3236,sname='G\'asp\'ar',gname='Andr\'as ']{Andr\'as G\'asp\'ar}
\affiliation{Steward Observatory and the Department of Astronomy, The University of Arizona, 933 N Cherry Ave, Tucson, AZ, 85719}
\email[show]{agaspar@arizona.edu}  

\author[orcid=0000-0002-0834-6140,sname='Leisenring',gname='Jarron M.']{Jarron M.\ Leisenring}
\affiliation{Steward Observatory and the Department of Astronomy, The University of Arizona, 933 N Cherry Ave, Tucson, AZ, 85719}
\email[]{jarronl@arizona.edu}  

\author[orcid=0000-0002-9977-8255,sname='Wolff',gname='Schuyler Grace']{Schuyler Grace Wolff}
\affiliation{Steward Observatory and the Department of Astronomy, The University of Arizona, 933 N Cherry Ave, Tucson, AZ, 85719}
\email[]{sgwolff@arizona.edu}  

\author[orcid=0000-0002-6964-8732,sname='Lawson',gname='Kellen']{Kellen Lawson}
\affiliation{NASA Goddard Space Flight Center, Exoplanets and Stellar Astrophysics Laboratory, Code 667, Greenbelt, MD 20771, USA}
\email[]{kellen.d.lawson@nasa.gov} 

\author[orcid=0000-0003-2303-6519,sname='Rieke',gname='George H.']{George H.\ Rieke}
\affiliation{Steward Observatory and the Department of Astronomy, The University of Arizona, 933 N Cherry Ave, Tucson, AZ, 85719}
\email[]{grieke@arizona.edu}  

\author[orcid=0000-0003-0777-7392,sname='Deng',gname='Dingshan']{Dingshan Deng}
\affiliation{Lunar and Planetary Laboratory, The University of Arizona, Tucson, AZ 85721, USA}
\email[]{dingshandeng@arizona.edu}  

\author[orcid=0000-0002-7893-6170,sname='Rieke',gname='Marcia J.\']{Marcia J.\ Rieke}
\affiliation{Steward Observatory and the Department of Astronomy, The University of Arizona, 933 N Cherry Ave, Tucson, AZ, 85719}
\email[]{mrieke@arizona.edu}  

\author[orcid=0000-0003-4623-1165,sname='Sefilian',gname='Antranik A.\']{Antranik A.\ Sefilian}
\affiliation{Steward Observatory and the Department of Astronomy, The University of Arizona, 933 N Cherry Ave, Tucson, AZ, 85719}
\email[]{asefilian@arizona.edu}  

\author[orcid=0000-0002-5627-5471,sname='Beichman',gname='Charles']{Charles Beichman}
\affiliation{Jet Propulsion Laboratory, California Institute of Technology, Pasadena, CA}
\email[]{charles.a.beichman@jpl.nasa.gov}  

\author[orcid=0000-0002-4248-5443,sname='Lovell',gname='Joshua B.\']{Joshua B.\ Lovell}
\affiliation{Center for Astrophysics, Harvard \& Smithsonian, 60 Garden Street, Cambridge 02138, MA USA}
\email[]{joshua.lovell@cfa.harvard.edu}  

\author[orcid=,sname='Krist',gname='John']{John Krist}
\affiliation{Jet Propulsion Laboratory, California Institute of Technology, Pasadena, CA}
\email[]{john.e.krist@jpl.nasa.gov}  

\author[orcid=0000-0001-7591-2731,sname='Ygouf',gname='Marie']{Marie Ygouf}
\affiliation{Jet Propulsion Laboratory, California Institute of Technology, Pasadena, CA}
\email[]{marie.ygouf@jpl.nasa.gov}  

\author[orcid=0000-0002-3414-784X,sname='Llop-Sayson',gname='Jorge']{Jorge Llop-Sayson}
\affiliation{Jet Propulsion Laboratory, California Institute of Technology, Pasadena, CA}
\email[]{jorge.llop.sayson@jpl.nasa.gov}  

\author[orcid=0000-0001-5966-837X,sname='Bryden',gname='Geoffrey']{Geoffrey Bryden}
\affiliation{Jet Propulsion Laboratory, California Institute of Technology, Pasadena, CA}
\email[]{geoffrey.bryden@jpl.nasa.gov}  

\author[orcid=,sname='Stark',gname='Christopher C.\']{Christopher C.\ Stark}
\affiliation{NASA Goddard Space Flight Center, Exoplanets and Stellar Astrophysics Laboratory, Code 667, Greenbelt, MD 20771, USA}
\email[]{christopher.c.stark@nasa.gov}  

\author[orcid=0000-0002-8382-0447,sname='Chen',gname='Christine H.\']{Christine H.\ Chen}
\affiliation{Space Telescope Science Institute, 3700 San Martin Drive, Baltimore, MD 21218, USA}
\email[]{cchen@stsci.edu}  

\correspondingauthor{Andr\'as G\'asp\'ar}




\begin{abstract}

Debris disks, composed of rocks, boulders, planetesimals, and the dust produced in their collisions, present the 
most readily observable components of mature planetary systems. They also serve as valuable 
diagnostic tools, enabling studies of planetary dynamical interactions and
mineral compositions. Observed from optical to radio wavelengths, each band reveals
unique information about the dust populations. Optical and near-infrared 
observations are specifically sensitive to light scattered off the surfaces of the
micron-sized particles. Here, we present results from the
JWST/NIRCam GTO program 2780, designed to observe five disk systems previously identified to be exceptionally 
bright at optical wavelengths (HD~10647, HD~32297, 
HD~61005, HD~107146, and HD~181327) with six filters using the NIRCam coronagraphs. 
The NIRCam data complement previous shorter-wavelength images of these same systems.
They reveal scattered light from the disks and from the extended halos of tiny grains 
under the influence of radiative forces, at high resolution and signal to noise. All 
the systems show evidence for water ice, although it can have differing radial distributions 
and tends to show stronger signatures in the halos. In the two cases we could analyze, 
the scattering phase function in the disks resembles the behavior of dust in the 
Solar System with evidence for enhanced forward scattering in the halos, consistent with 
the latter being composed of tiny grains. MIRI images for two systems are more 
centrally concentrated than the shorter wavelength ones, suggesting a role for 
dragged-in larger grains.
\end{abstract}

\keywords{\uat{Debris Disks}{363} --- \uat{JWST}{2291} --- \uat{Coronagraphic imaging}{313} --- \uat{Circumstellar dust}{236}}


\section{Introduction}

In 1983, the launch of the Infrared
Astronomical Satellite (IRAS) opened the thermal infrared window to our Universe.
During its calibration, it was found that one of its six standard stars, Vega,
is two orders of magnitude brighter in the far-infrared than expected from its stellar photosphere 
\citep{aumann84}. The ``Vega phenomenon'', as it became known, was hypothesized 
early-on to originate from either a shell or disk of dust. Soon thereafter, many other
similar systems were  identified in the IRAS observations  \citep[e.g.,][]{aumann85,gillett86,backman93}.
The disk around $\beta$ Pictoris, one of the original ``Fab Four'' disks 
(the others being Vega, Fomalhaut, and $\epsilon$ Eridani), is so massive that it was easily identified in
ground-based optical wavelength coronagraphic observations as well \citep{smith84}; and even through 
an eyepiece (Brad Smith, priv.\ communication). While the number of disks identified in
spatially unresolved thermal emission rose toward a hundred \citep[e.g.,][]{1998ApJ...497..330M}, other than $\beta$ Pictoris, 
they remained elusive in scattered light. This limited their further study until they 
were detected over a broader range of the spectrum.

When they became available, panchromatic observations revealed various properties of these systems. Thermal infrared
observations in multiple bands detected the black-body emission of the grains, modified according to grain 
absorption characteristics as well as the underlying size-distribution and minimum grain size
cutoff. Optical/near-infrared light, on the other hand, originates
from the central star and scatters off the surfaces of the smallest grains (from sub-$\micron$ to
a few $\micron$ in size), much like sunlight scattering in smoke, which consists of similar size
particles. The amount of light received depends on the optical properties of the dust (its reflectance, a.k.a.\ albedo), the
scattering angle, and, critically, on the minimum grain sizes present and the size-distribution of the particles.
Therefore, scattered light observations are excellent diagnostic tools to understand the compositions
of the disks. 

With the launch of the Hubble Space Telescope (HST), the number of spatially resolved
debris disks observed in scattered light 
rose dramatically. The Advanced Camera
for Surveys (ACS) coronagraph provided filtered images, enabling broad-band spectral characterization, while
the still-operational Space Telescope Imaging Spectrograph (STIS) coronagraph is able to image low-surface-brightness components given its unfiltered imaging setup. Further huge advances in scattered light 
imaging have been obtained with advanced coronagraphs on large ground-based telescopes, e.g, with the Gemini
Planet Imager \citep{macintosh2014} and SPHERE \citep{beuzit2019}. The Atacama Large Millimeter Array (ALMA) 
has provided unique insights  to the inherent structure of the parent body structures for many  disks 
\citep[e.g.,][and references therein]{2026A&A...705A.195M}.

Given that different dust populations
and characteristics dominate thermal emission and light scattering, not all thermal infrared bright
disks are prominent in scattered light. The Vega disk was only recently detected with HST by a deep
imaging program \citep{2024AJ....168..236W}, the $\epsilon$ Eridani disk is still 
undetected \citep{wolff23,2024AJ....168..169K}, and the Fomalhaut disk is barely visible 
and only in a narrow ring, resulting in a high enough surface density to be detected \citep{2005Natur.435.1067K,gaspar20}.
None of these three systems were detected in near-infrared ($\ge 1.82~\micron$) scattered light by JWST 
\citep[][]{2024AJ....167...26Y,2025AJ....169...17B,2025arXiv250808463L}, 
alluding to  characteristics of the size-distribution of the particles and to the inclinations of the disk systems. Sources where smaller
dust particles are more abundant and/or that are viewed at a favorable inclination with more forward scattering are more readily detectable 
at shorter wavelength scattered light \citep{schneider14}. 

In this paper, we present preliminary results from our JWST/NIRCam Guaranteed Time Observing 
(GTO) program 2780 (PI: G\'asp\'ar, Co-PI: Leisenring), aimed at imaging some of
the brightest and spatially most extended circumstellar debris disks in scattered light at multiple 
near-infrared wavelengths. The primary goal of the program was to study the material composition of 
the dust in the systems via multi-filter near-infrared observations, most importantly water-ice content. 
This paper showcases the dataset.
In Section \ref{sec:obs}, we present the observations, detailing the lessons learned from this early
GTO program. Sections \ref{sec:reds} and \ref{sec:miris} discuss the reduction methods 
we employed for the NIRCam and MIRI datasets (also taken for two of the sources), while 
Section \ref{sec:results} presents the results from the program for the individual targets. Section \ref{sec:compare} introduces a comparative study. Finally, Section \ref{sec:summary} 
presents a summary of our findings and technical details of the image are provided in Appendix A.

\section{Observations}
\label{sec:obs}

Of the 900 hours of GTO time awarded to NIRCam PI Marcia Rieke, 57.5 hours were reserved for the
scattered light debris disks program (2780; PI G\'asp\'ar, Co-PI Leisenring), placing it among the largest JWST GTO programs. 
The observations, originally submitted prior to launch, were designed to utilize the simultaneous 
short- and long-wavelength coronagraphic observing capabilities of NIRCam. Due to the delayed 
commissioning of this observing mode, the execution of the program was postponed to Cycle 2. 
The five targets were all observed using the same NIRCam coronagraphic configuration: the target was placed on the MASK335R round coronagraphic mask and imaged simultaneously with the shortwave (SW) and longwave (LW) modules through a total of six filters (F182M, F210M, F250M, F300M, F335M, and F444W) with the SUB320A subarray.
The SW channel (F182M and F210M) has a pixel scale of $\sim$31~mas pixel$^{-1}$ with a $\sim$10\arcsec\ field of view (FoV) while the LW channel (F250M, F300M, F335M, and F444W) has a pixel scale of $\sim$63~mas pixel$^{-1}$ with a $\sim$20\arcsec\ FoV.
The science targets were observed at a single position centered on the coronagraphic mask and acquired at two roll angles offset by $\sim$10$^{\circ}$ to obtain a rotational dither and facilitate PSF subtraction through Angular Differential Imaging (ADI). 
Corresponding reference star observations were subsequently observed using the 5-point diamond small grid dither (SGD) pattern to subtract the stellar PSF through Reference Differential Imaging (RDI).
In addition to the NIRCam observations, two of the targets (HD~10647 and HD~107146) were observed with the MIRI detector using the F2100W filter (non-coronagraphic observation) and one of the targets (HD~32297) was observed with NIRSpec using fixed slit spectroscopy under a shared observing time program (with PI: C.\ Chen).  
For the MIRI SUB256 array observations, background and PSF images were also obtained. In Table \ref{tab:JWSTobs}, we detail the observing sequences of our GTO program 2780.

\begin{figure*}[!t]
    \centering
    \includegraphics[width=0.885\linewidth]{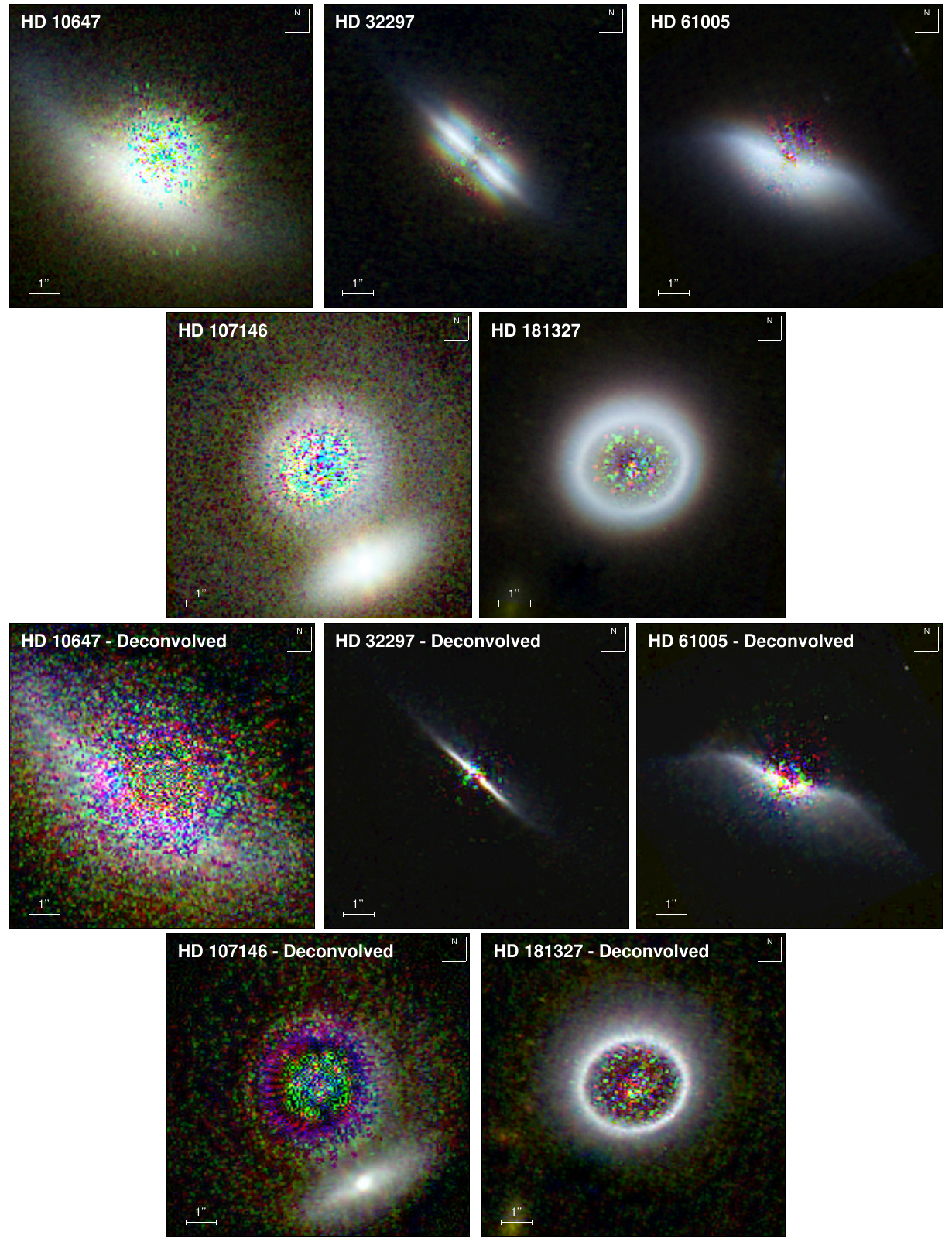}
    \caption{The color combined, NIRCam coronagraphic images (observed and deconvolved) 
    of the five debris disks in our program. Filters were averaged in pairs and assigned 
    colors according to their wavelengths (F182M+F210M: blue; F250M+F300M: green; F335M+F444W: red) and their fluxes
    were normalized by the stellar photosphere to produce reflectance colors. In each panel, scale bars in the bottom-left corner indicate $1''$, with North up and East to the left. The images are post-processed using the technique outlined in 
    Section \ref{sec:reds}.}
    \label{fig:allims}
\end{figure*}

Four out of the five observing sequences experienced technical issues, with three of them requiring
repeat observations. During the observation sequence of HD~10647 on 10/28/2023, one of the mirror
segments was struck by a micrometeoroid. The impact occurred early in the sequence and the 
change in the wavefront remained stable. The PSF observations, the second roll observation of the target,
and the majority of the first roll were executed following the impact; the few pre-impact images were discarded.
The observations of HD~32297 (02/26/2024) 
were executed following a major tilt event, affecting three segments located on one of the wings, and were 
therefore granted repeated executions. The first observing sequence of HD~61005 (04/21/2024) resulted in 
the target observations being skipped due to guide-star acquisition failure. Since contemporaneous PSFs are 
necessary for coronagraphy, the entire sequence was repeated. Finally, the observation of HD~181327 (09/13/2023) 
was a partial failure, due to the target acquisition algorithm locking onto a hot pixel in the second roll of the 
target sequence. We were granted a repeat of this second failed roll and a paired new PSF observation.

\begin{table*}[t]
\begin{center}
\caption{JWST observations of the GTO 2780 program.\label{tab:JWSTobs}}
\begin{tabular}{lllllll}
Target                 & Instrument/Filters                    & Readout    & N$_{\rm group}$ & N$_{\rm int}$ & Dither               & Time (s)\tablenotemark{$\dagger$} \\
\tableline\tableline
\multicolumn{7}{c}{HD~10647 -- Date-Obs: 10/28/2023}\\
\tableline
BCKGND                 & MIRI/F2100W/SUB256                    & FASTR1     & 5               & 300           & EXT/4pt              & 2155 \\
HD~10647 (Roll 1)      & NIRCam/F(182,210,250,300,335)M,F444W  & MEDIUM8    & 10              & 18            & None                 & 1906 \\
HD~10647               & MIRI/F2100W/SUB256                    & FASTR1     & 5               & 300           & EXT/4pt              & 2155 \\
HD~10647 (Roll 2)      & NIRCam/F(182,210,250,300,335)M,F444W  & MEDIUM8    & 10              & 18            & None                 & 1906 \\
$\iota$ Hor (PSF Ref.) & NIRCam/F(182,210,250,300,335)M,F444W  & MEDIUM8    & 10              & 3             & 5pt Dmnd             & 1588 \\
$\iota$ Hor (PSF Ref.) & MIRI/F2100W/SUB256                    & FASTR1     & 5               & 300           & EXT/4pt              & 2155 \\
\tableline\tableline
\multicolumn{7}{c}{HD~32297 -- Date-Obs: 09/12/2024 (failed on 02/26/2024)}\\
\tableline
HD~32297               & NIRSpec/S200A1/G395H/F290LP           & NSRAPID    & 6               & 5             & 12                   & 656 \\
HD~32297 (Roll 1)      & NIRCam/F(182,210,250,300,335)M,F444W  & MEDIUM8    & 10              & 16            & None                 & 1694 \\
HD~32297 (Roll 2)      & NIRCam/F(182,210,250,300,335)M,F444W  & MEDIUM8    & 10              & 16            & None                 & 1694 \\
HD~30365 (PSF Ref.)    & NIRCam/F(182,210,250,300,335)M,F444W  & MEDIUM8    & 10              & 3             & 5pt Dmnd             & 1588 \\
\tableline\tableline
\multicolumn{7}{c}{HD~61005 -- Date-Obs: 11/16/2024 (failed on 04/21/2024)}\\
\tableline
HD~61005 (Roll 1)      & NIRCam/F(182,210,250,300,335)M,F444W  & MEDIUM8    & 10              & 16            & None                 & 1694 \\
HD~61005 (Roll 2)      & NIRCam/F(182,210,250,300,335)M,F444W  & MEDIUM8    & 10              & 16            & None                 & 1694 \\
HD~56161 (PSF Ref.)    & NIRCam/F(182,210,250,300,335)M,F444W  & MEDIUM8    & 10              & 3             & 5pt Dmnd             & 1588 \\
\tableline\tableline
\multicolumn{7}{c}{HD~107146 -- Date-Obs: 02/03/2024}\\
\tableline
BCKGND                & MIRI/F2100W/SUB256                    & FASTR1     & 10              & 150           & EXT/4pt              & 1976 \\
HD~107146 (Roll 1)    & NIRCam/F(182,210,250,300,335)M,F444W  & MEDIUM8    & 10              & 18            & None                 & 1906 \\
HD~107146             & MIRI/F2100W/SUB256                    & FASTR1     & 10              & 150           & EXT/4pt              & 1976 \\
HD~107146 (Roll 2)    & NIRCam/F(182,210,250,300,335)M,F444W  & MEDIUM8    & 10              & 18            & None                 & 1906 \\
HD~111398 (PSF Ref.)  & NIRCam/F(182,210,250,300,335)M,F444W  & MEDIUM8    & 10              & 3             & 5pt Dmnd             & 1588 \\
HD~111398 (PSF Ref.)  & MIRI/F2100W/SUB256                    & FASTR1     & 10              & 150           & EXT/4pt              & 1976 \\
\tableline\tableline
\multicolumn{7}{c}{HD~181327 -- Date-Obs: 10/11/2023 (partial failure on 09/13/2023)}\\
\tableline
HD~181327 (Roll 1)      & NIRCam/F(182,210,250,300,335)M,F444W  & MEDIUM8    & 10              & 16            & None                 & 1694 \\
HD~181327 (Roll 2)      & NIRCam/F(182,210,250,300,335)M,F444W  & MEDIUM8    & 10              & 16            & None                 & 1694 \\
HD~180134 (PSF Ref.)    & NIRCam/F(182,210,250,300,335)M,F444W  & MEDIUM8    & 10              & 3             & 5pt Dmnd             & 1588 \\
\end{tabular}
\tablenotetext{\dagger}{SW integrations (F182M, F210M) are doubled relative to the LW (F250M, F300M, F335M, F444W) integrations -- whose values are given in the table -- due to the simultaneous observing sequence and how the filters were paired.}
\end{center}
\end{table*}

In the following section, we describe the high-level reduction methods we employed on this challenging coronagraphic dataset. In the appendix, we give further details on the custom algorithm we used to carry out the reductions. The final, 6-color combined images of our five targets are  shown in Figure \ref{fig:allims}, showcasing the observed and deconvolved images. 
All of our reduced and processed data are available for download at 
\dataset[10.5281/zenodo.21683353]{https://doi.org/10.5281/zenodo.21683353}, 
while pipeline data can be downloaded from MAST: \dataset[10.17909/6a5r-qf97]{https://doi.org/10.17909/6a5r-qf97}.

\section{NIRCam Data Reductions}
\label{sec:reds}

The raw JWST images were reduced to stage 2 level products (\texttt{calints.fits}) using the JWST pipeline \citep[v1.18.0;][]{bushouse} as implemented within the \texttt{spaceKLIP}\footnote{\url{https://github.com/spacetelescope/spaceKLIP}} environment \citep[v2.2.1;][]{2022SPIE12180E..3NK,2023ApJ...951L..20C}. We employed the default settings provided in the spaceKLIP tutorials, except for the $1/f$ noise subtraction which we set to ``median'' instead of ``savgol''. 
For extended sources, the default Savitzky–Golay filtering results in a striated background that interprets faint disk structure as a noise component, and this oversubtraction only becomes apparent later during post-processing. 
The stage 1 reductions flag and correct effects such as saturation, bias levels, column and row offsets,
non-linearity, dark current, cosmic rays, and spatial $1/f$ noise.
The stage 2 reductions follow up with observation level corrections: background subtraction, flat-field corrections, outlier corrections, and flux calibration. Stage 2 also applies an updated World Coordinate System (WCS) to the images. We refer readers to the JWST pipeline and spaceKLIP documentations, their example tutorials, and the numerous papers \citep[e.g.,][]
{2023AJ....166..150L,2024AJ....167..183M,2024AJ....167..181W,2024ApJ...967L...8L} using the packages for further details. 
We also employed the built-in \texttt{ImageTools} package of spaceKLIP to perform median level subtractions and to clean/flag the bad 
pixels in the images using multiple methods, based on the data quality flags, temporal variations across the 
integrations, and sigma clipping. We do not employ additional steps to identify and clean outliers since the spaceKLIP algorithms provide reliable bad pixel identification and cleaning functionality beyond the default pipeline. The reduction steps up to this point 
were handled identically for all images; the post-processing steps deviated from each other following this stage.

\begin{figure*}
    \centering
    \includegraphics[width=1.0\linewidth]{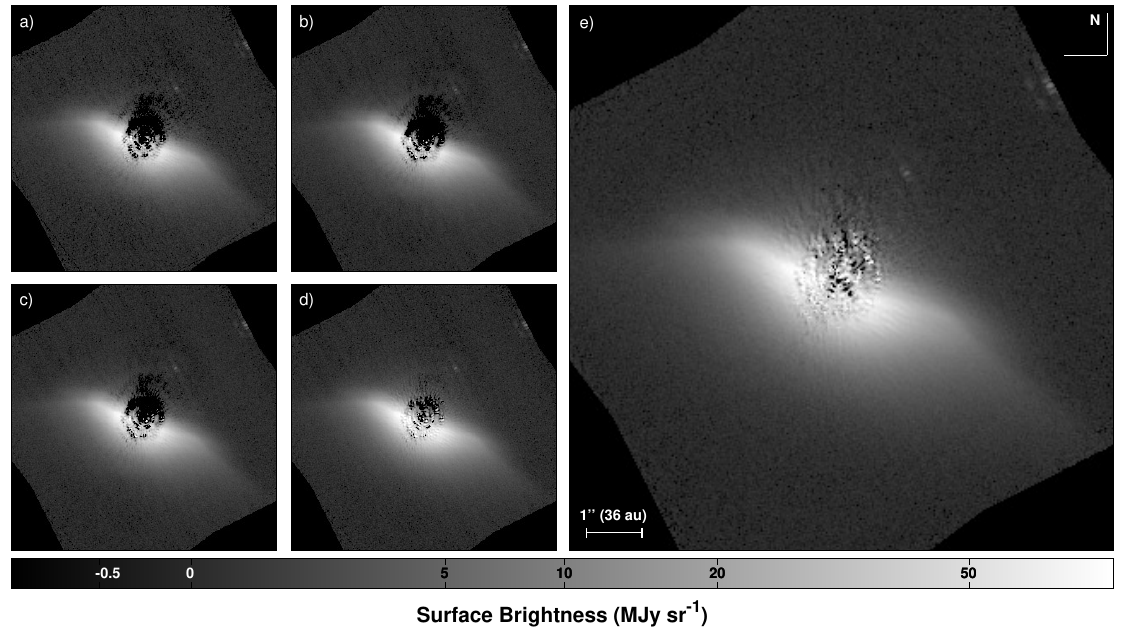}
    \caption{Comparison of the results of the various post-processing techniques for the HD~61005 F182M dataset. The methods shown are:
    a) SpaceKLIP, b) Winnie ``basic'' RDI, c) Winnie HPFRDI, 
    d) Winnie MCRDI, and e) Custom radially weighted linearly combined
    Reference Differential Imaging (this work). The backgrounds of each resulting image had slightly
    different median values, so the following offsets were applied to bring them to the levels achieved
    in our custom reduction shown in panel (e): 
    a) +1.77 MJy sr$^{-1}$, b) +0.36 MJy sr$^{-1}$, c) +0.37 MJy sr$^{-1}$, d) +0.38 MJy sr$^{-1}$. All images are in logarithmic scaling between -0.6 and 80 MJy sr$^{-1}$.}
    \label{fig:comps}
\end{figure*}

In stage 3 post-processing, the corongraphic data are aligned, WCS is updated to account for stellar positioning behind the mask, the reference PSFs are subtracted, and the subtracted images
are combined. There are mathematical variations on how these last steps can be executed, each method with unique advantages, 
depending on the astrophysical object. The observations obtained by the program were among the first NIRCam coronagraphic 
data on extended disks; therefore, we decided to test various post-processing methods for image retrieval.
In the following subsections, we introduce the commonly used
statistical and classical post-processing methods we tested: KLIP \citep{2012ApJ...755L..28S,2023A&A...679A..18R}
as employed by SpaceKLIP, ``basic'' Reference Differential Imaging (RDI) -- linear PSF combination -- 
as employed by Winnie \citep{2023AJ....166..150L}, high-pass filtered RDI (HPFRDI) as employed by Winnie, and model constrained
RDI \citep[MCRDI;][]{2022ApJ...935L..25L} as employed by Winnie.
We also introduce a custom post-processing algorithm that we adopted as the method of 
choice, details of which can be found in Appendix~\ref{sec:postproc}.
Unique steps for particular datasets are mentioned where they were employed. The final images calculated using each 
process for the HD~61005 dataset at F182M (the wavelength with the highest spatial resolution
and most challenging PSF residual subtractions) are  presented in Figure~\ref{fig:comps}. 
To quantify the levels of oversubtraction in the core ($\le 1^{\prime\prime}$ for F182M)
and the remnant noise in the background following each post-processing method, we measured the mean levels above the disk plane in a semi-circular region 
around the core and in a square box at the edge of the field (Table~\ref{tab:pmethods}). The hybrid method we introduced has a higher value than either of the other methods in the core, the Model-Constrained Reference Differential Imaging (MCRDI)\footnote{implemented using the \texttt{winnie} package (\url{https://github.com/kdlawson/Winnie}}
approaches zero, while all other methods severely oversubtract. Given the extended brightness of the disk, we expect a positive signal 
in the core, even after PSF subtraction. The MCRDI method has the lowest noise level in the core. The background
is nearest to zero with the lowest noise levels with the hybrid method. After testing the various post-processing
methods, we decided to employ the custom hybrid of the classic and linear combined RDI methods 
for all datasets to produce the scientific analysis quality products.

\subsection{SpaceKLIP} \label{subsec:space_KLIP}

The \texttt{spaceKLIP} package  employs the Karhunen–Lo\`eve image projection (KLIP) statistical image post-processing method 
\citep{2012ApJ...755L..28S,2023A&A...679A..18R}, with specific routines for JWST observations as well as tools to 
calculate the JWST contrast curves. The KLIP algorithm optimizes post-processing of 
coronagraphic datasets for directly imaged exoplanets (i.e.\ point sources) but can be used to retrieve extended
features as well. We used the 2.2.1 developmental version of \texttt{spaceKLIP} for the reductions and post-processing tests presented in this paper.

\begin{table}[!t]
\begin{center}
\caption{Mean flux values (in MJy sr$^{-1}$) in the core ($\le 1^{\prime\prime}$ for F182M) and
background for each post-processing method showcased in Figure \ref{fig:comps}\label{tab:pmethods}.}
\begin{tabular}{lrr}
Method              & Core                 & Background \\
\tableline\tableline
Hybrid (this paper) & $6.69\pm11.31$       & $0.005\pm0.039$  \\
Winnie MCRDI        & $1.39\pm5.100$        & $0.380\pm0.058$  \\
SpaceKLIP           & $-10.03\pm11.41$     & $1.770\pm0.267$  \\
Winnie HP           & $-6.41\pm9.720$       & $0.370\pm0.051$  \\
Winnie RDI          & $-9.94\pm12.49$      & $0.360\pm0.222$
\end{tabular}
\end{center}
\end{table}

The stage 3 post-processing with spaceKLIP, following the reduction steps presented previously, was executed 
along the standard steps described in the tutorials. One of the most critical steps in coronagraphic data reduction
is determining the accurate position of the occulted central source relative to the coronagraphic mask and to the
reference observations. Best practices for this step vary based on the instrument and type of occulter used. 
Importantly, extended objects around the bright central sources present additional complications for automated 
centering algorithms, possibly resulting in systematic offsets in the solutions. SpaceKLIP determines the positions 
of the sources (target and PSF) by comparing them to STPSF models, generated for the epoch of observation 
and at various on- and off-axis positions using webbpsf\_ext\footnote{https://pypi.org/project/webbpsf-ext/}.
For the spaceKLIP reduction, we determined the stellar positions for each individual frame, which deviates 
from the recommendation to  only execute this for the first frame to save processing time.
The KLIP algorithm itself was executed using a single annulus without subsections and 50 principal component KL modes.
Panel (a) in Figure \ref{fig:comps} shows the basic spaceKLIP post-processing, without any additional fine-tuning.
Striping residuals are still present as well as significant over-subtraction residuals, which stem from the extended
nature of the science target. Note that the image had to be shifted by 1.77 MJy sr$^{-1}$ to bring it to the same background
level as our custom reduction shown in Figure \ref{fig:comps}(e). 

\subsection{The Winnie processing package}

Winnie is a python package developed specifically for the post-processing of coronagraphic observations of extended 
objects, fine-tuned for JWST NIRCam round-mask datasets \citep{2023AJ....166..150L}. The package provides various 
flavors of reference differential imaging (RDI) techniques to optimize stellar PSF subtraction and improve characterization of surrounding extended disk structures and point source companions.
The processing takes, as input, images that have already been aligned by the
spaceKLIP imagetools code. Therefore, we used the processed, aligned, and padded products of SpaceKLIP introduced
in Section \ref{subsec:space_KLIP} as the input for Winnie, but co-added for increased processing speed. Below, we discuss the
various flavors of RDI that we tested with Winnie. Note that Winnie is an extremely flexible post-processing package
and we only show the results of the basic reductions provided by each method, without further fine-tuning of the steps,
such as using custom exclusion zones or alignments; with those, the results achievable
with Winnie are likely of much higher quality than presented here. Our goal is to present the results achievable out of
the box.

$(i)$ The simplest post-processing available with Winnie is a linear combination PSF reduction, where the constructed 
reference PSF is subtracted from the target images and then they are co-added, termed ``basic RDI'' 
by Winnie. Note that there is discrepancy in the literature on what basic (or ``classic'') RDI constitutes. Historically,
``classic'' RDI involved scaling all PSFs to match that of the target source, subtracting them, and then 
co-adding the residuals. This was a successful method for HST reductions \citep[e.g.][]{schneider14,gaspar20}, where
the PSFs are stable and less dependent on positioning behind the occulting mask. An update to this method, the locally 
optimized linear combination of images \citep[LOCI;][]{lafreniere07} developed for ground-based high-contrast image processing, 
allows the PSFs to be scaled within large ranges -- including in the negative -- and then combined, producing 
PSFs optimized to subtract semi-static speckles in the PSF cores. The original LOCI method produces 
PSFs and subtracted images optimized and subtracted in zones, while the basic RDI method employed by 
Winnie uses the LOCI algorithm with a single optimization and subtraction zone. Given the variation of
the PSF for NIRCam as a function of positioning behind the mask, allowing the PSFs to scale with arbitrary values (i.e.\ LOCI processing variants, with either single or multiple zones) yields superior results to ``classic'' RDI.
A reduction using the Winnie RDI technique is shown in panel (b) of Figure \ref{fig:comps}, with results similar 
to that given by SpaceKLIP. By providing optimization zones (excluding disk areas), the classical RDI reduction as 
performed by Winnie could be improved; here we show what it can do without fine-tuning.

$(ii)$ Winnie provides an improvement over the LOCI RDI methods for extended sources by also introducing a high-pass 
filtering method, which attempts to remove the extended disk flux for the PSF fitting step, by attenuating the low frequency 
features, thereby allowing a better fit to the PSF, which is comprised of sharp high-frequency signals. 
We show the results from the Winnie High-Pass Filtered RDI (HPFRDI) post-processing in panel (c) of 
Figure \ref{fig:comps}.

$(iii)$ The most complex processing method offered by Winnie is the Model-Constrained Reference Differential Imaging  \citep[MCRDI;][]{2022ApJ...935L..25L}.
The difficulty in removing the stellar contribution from the coronagraphic imaging of extended systems lies in differentiating
the two components (stellar and extended), especially if the extended source is compact. The instrumental PSF convolves 
with both, resulting in wider coronagraphic speckles. In addition, for NIRCam, the structure of the 
coronagraphic PSF depends greatly on the exact position of the occulted stellar source behind the coronagraph.
MCRDI simulates the complex coronagraphic image of the combined star plus extended disk system and performs a 
Levenberg-Marquardt fit of the simulation to the observations, by varying the model disk parameters. Apart from optimizing
the post-processing, this method also yields a robust preliminary model of the circumstellar disk. In panel (d) of Figure \ref{fig:comps}, we show the Winnie
MCRDI reduction of the HD~61005 F182M observation, and in Table \ref{tab:HD61005mcrdi}
we list the fitted disk parameters found by this reduction method. Note that
this is a difficult system to reduce with MCRDI, as the wings of the disk are 
not easily fit by a model parameterized by a single (or multiple) concentric disk ring(s).

\begin{table}[!t]
\begin{center}
\caption{Winnie MCRDI fits for HD~61005 \label{tab:HD61005mcrdi}}
\begin{tabular}{ll}
Variable                                    & Fitted value \\
\tableline\tableline
Fiducial radius (r$_0$)                     & 49.91~au   \\
Disk aspect ratio at r$_0$ (h$_0$)          & 0.1547    \\
Disk density inner slope ($\alpha_{\rm in}$)     & 5.76     \\
Disk density outer slope ($\alpha_{\rm out}$)    & -2.37    \\
Position Angle (PA)                         & 68.11$^{\circ}$ \\
Eccentricity (e)                            & 0.045          \\
Inclination ($\iota$)                       & 80.14$^{\circ}$\\
Vertical profile index                      & 3.29 \\
g$_1$ Henyey-Greenstein parameter           & 0.833          \\
g$_2$ Henyey-Greenstein parameter           & 0.399         \\
wg1 (weight for g$_1$)                      & 0.69 
\end{tabular}
\end{center}
\end{table}

\subsection{Custom hybrid classic-LOCI RDI}
\label{sec:cRDI}

\begingroup
\setlength{\tabcolsep}{10pt}
\begin{deluxetable*}{llllll}
\tablewidth{0.99\textwidth}
\tablecaption{System Parameters of the Survey Targets \label{tab:props}}
\tablehead{
\colhead{Parameter/Star} & \colhead{HD~10647} & \colhead{HD~32297} & \colhead{HD~61005} & \colhead{HD~107146} & \colhead{HD~181327}
}
\startdata
Age (Myr)                                       & 700$^{+300}_{-240}$ [1]    &  15-30 [7,8,9]              &  40-100 [13,32]        & 100-300 [18,19]   & 18.5$^{+2.0}_{-2.4}$ [29] \\
Stellar mass (M$_{\odot})$                      & 1.15$^{+0.01}_{-0.02}$ [1] &  1.65$\pm$0.1 [9]           &  0.96$\pm$0.01  [14]   & 1.09 [26]         & 1.36$\pm$0.02 [30]  \\
Stellar luminosity (L$_{\odot})$                & 1.41 [2]                   &  6.16$\pm$1.4 [9]           &  0.583$\pm$0.048 [13]  & 1.1 [25]          & 0.44$\pm$0.02 [30]  \\
Stellar T$_{\rm eff}$ (K)                       & 6207$\pm$11 [1]            &  8000$\pm$150[9]            &  5500$\pm$50 [13]      & 5850 [26]         & 6500 [30]  \\
Spectral-Type                                   & F9V [2]                    &  A6V [9]                    &  G8Vk [15]             & G2V [20]          & F5/F6V [27,28]  \\
$\rm{[Fe/H]}$                                   & -0.027$\pm$0.008 [1]       &  -0.7[10]                   &  0.01$\pm$0.04         & -0.04 [21]        & 0.05$\pm$0.06 [30]  \\
Distance (pc)                                   & 17.35$\pm$0.01 [3]         & 129.73$\pm$0.54[3]          & 36.45$\pm$0.02[3]      & 27.47$\pm$0.02[3] & 47.78$\pm$0.07[3]  \\
\hline
\multicolumn{6}{c}{Disk parameters} \\
\hline
Warm Disk R$_{\rm in}$ (au; $^{\prime\prime}$)  & 3; 0.17 [4]                & 1.1; 0.01 [11]              & -                      & 40; 1.46 [23]     & -  \\
Warm Disk R$_{\rm out}$ (au; $^{\prime\prime}$) & 10; 0.58 [4]               & -                           & -                      & 60; 2.18 [23]     & -  \\
Cold Disk R$_{\rm in}$ (au; $^{\prime\prime}$)  & 34; 1.96 [5]               & 78.5; 0.61 [12]             & 42; 1.15 [7]           & 100; 3.6 [23]     & 75; 1.56 [16] \\
Cold Disk R$_{\rm peak}$ (au; $^{\prime\prime}$)& 81.6; 4.7 [5]              & 110; 0.85 [11]              & 49.91; 1.37 [16]       & 120; 4.4 [23]     & 80; 1.67 [16]   \\
Cold Disk R$_{\rm out}$ (au; $^{\prime\prime}$) & 134; 7.72 [5]              & 122; 0.94 [12]              & 67; 1.84 [7]           & 140; 5.1 [23]     & 85; 1.78 [16] \\
Inclination ($^{\circ}$)                        & 76.7$\pm$1.0 [5]           & $83.6^{+4.6}_{-0.4}$ [12]   & $80\fdg14$ [16]        & 18.5$\pm$2 [24]   & 28.54$\pm$0.31 [16] \\
Position angle ($^{\circ}$)                     & 57$\pm$1.0 [5]             & 47.9$\pm$0.2 [12]           & 68.11 [16]             & 58.5$\pm$5 [25]   & 100$\fdg$39$\pm$0$\fdg$63   \\
Total L$_{\rm IR}$/L$_{\rm \ast}$ [$\times10^{-4}$]   & 3.07$\pm$0.08 [5,6]  & 62.9 [11]                   & 20 [17]                & 12 [19,22,25]     & 20 [31]  \\
\hline
\enddata
\tablecomments{References: [1] \cite{carvalho25}, [2] \cite{marmier13}, [3] \cite{gaia22}, [4] \cite{schuppler16}, [5] \cite{lovell21}, [6] \cite{moor06}, [7] \cite{macgregor18}, [8] \cite{kalas05}, [9] \cite{rodigas14b}, [10] \cite{gebran16}, [11] \cite{donaldson13}, [12] \cite{macgregor18}, [13] \cite{desidera2011}, [14] \cite{dasilva21}, [15] \cite{gray2006}, [16] This work, [17] \cite{hines2007}, [18] \cite{isaacson2010}, [19] \cite{matra2025},
[20] \cite{harlan1970}, [21] \cite{gaspar2016}, [22] \cite{thebault2023}, [23] \cite{marino2018}, [24] \cite{schneider14}, [25] \cite{ardila2004}, [26] \cite{watson11}, [27] \cite{nordstrom2004}, [28] \cite{2021ApJS..254...42B}, [29] \cite{2020AA...642A.179M}, [30] \cite{reggiani24}, [31] \cite{2012AA...539A..17L}, [32] \cite{isaacson2010}
}
\end{deluxetable*}
\endgroup

We also execute a custom post-processing sequence on the datasets, based on a hybrid method, that incorporates 
elements of both the classical and the LOCI RDI methods. The hybrid method combines the advantages of the classical 
method, where all PSFs are scaled at the same positive scaling -- resulting in a high SNR reduction outside of 
the PSF cores, and that of the linear combination of PSFs, where they can have 
arbitrary and varying coefficients and are combined into a single PSF -- resulting in a high SNR reduction 
inside the core. Unlike for LOCI, the best solution with our method is not locally optimized in concentric 
sections and produces a radially smooth PSF. We give a detailed description of the hybrid method in the Appendix 
and only a brief overview in this section.

The post-processing steps followed the general stage 1 and 2 reductions performed with spaceKLIP, 
as for the other methods, as well as its median subtraction and bad pixel identification/cleaning procedures. We 
did not calculate the alignments or shift the images with space\-KLIP; instead individual PSF offsets were determined 
by manually minimizing subtraction residuals by eye in small fractional pixel steps 
(as opposed to numerical automated methods) and separate masks were created for de-striping, for the 
calculation of the coefficients for image combination, and PSF background sources.

For the hybrid method, we place constraints on the sum of the coefficients to limit oversubtraction and solved for
the coefficients via minimization of the integrated median absolute deviation curves, instead of
the traditional least-squares solution of the subtracted image. We also place lower bounds
on the median flux levels in radial bins during these fits, thereby avoiding the over-subtraction
that is common when post-processing coronagraphic observations of extended sources. Finally, we smooth
the coefficients to equal values toward the edge of the field of view, thereby lowering the per pixel
noise of the final image outside of the PSF core. This additional step reveals exquisite details on faint extended sources compared
to the standard LOCI algorithm. Finding a solution with these constraints requires an iterative minimization
search, instead of simply solving for the minimum via matrix inversion as is done with LOCI \citep{lafreniere07}. 
This can be computationally intensive, depending on the dataset. The hybrid reduction produces results with lower 
PSF subtraction residuals and virtually no over-subtractions. The hybrid classic/LOCI RDI reduction of the 
F182M HD~61005 dataset is shown in panel (e) of Figure \ref{fig:comps}.

\begin{figure}
    \centering
    \includegraphics[width=0.99\linewidth]{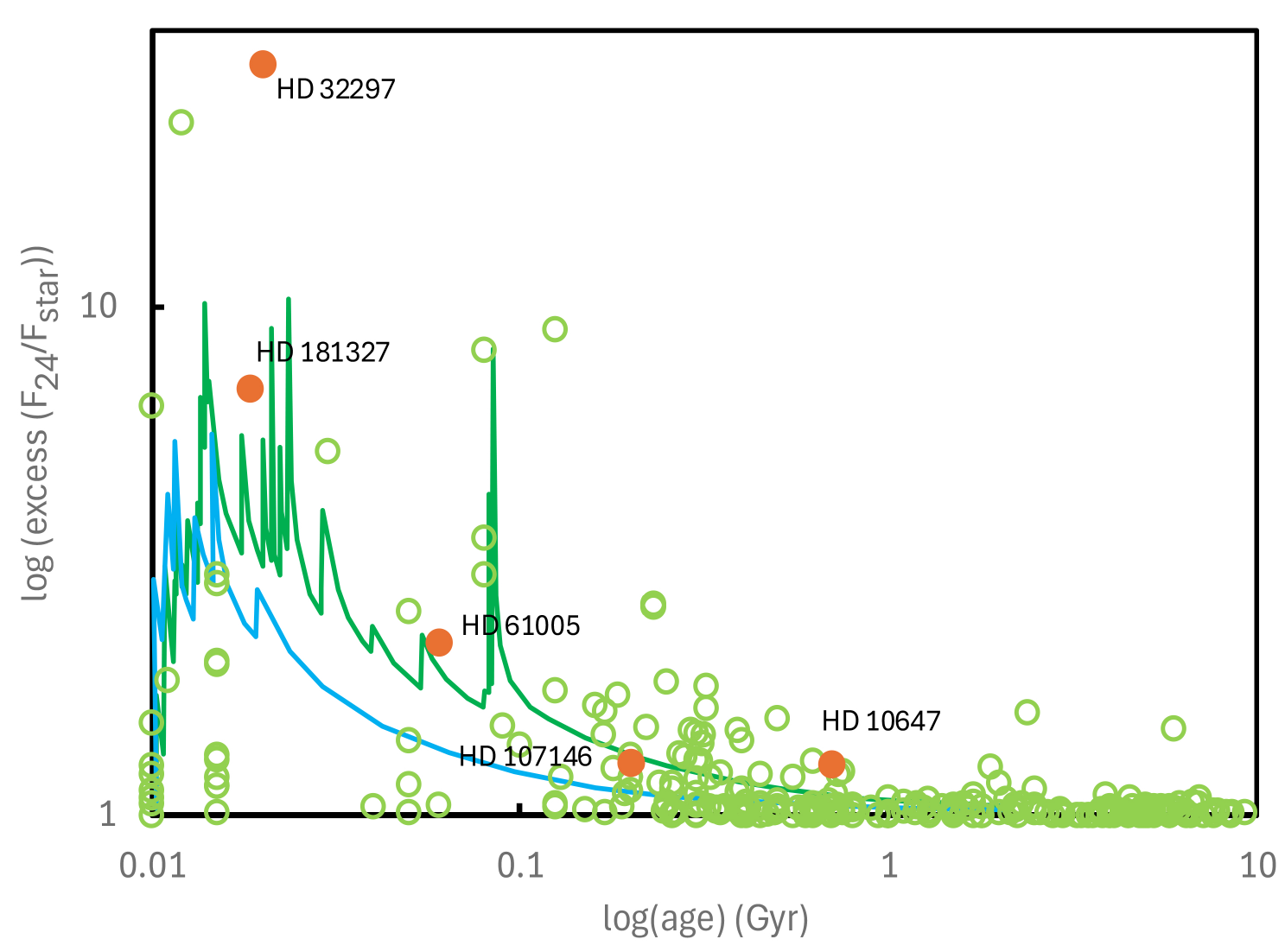}
    \caption{Debris disk excesses vs.\ age. The excesses (open circles) are in terms of the ratio of the observed 
    24 $\micron$ flux to that expected from the stellar photosphere, with data from \citet{su06, gaspar13}. 
    The disks in this study are shown as filled circles (data from \citet{sierchio14,lovell21} and WISE), 
    and two theoretical models of debris disk evolution are shown by lines \citep{genda2015}. To have 
    such bright scattered light excesses has imposed the selection bias that these five disks are both 
    young and that the warm dust signatures are brighter than typical for their ages.}
    \label{fig:overview}
\end{figure}

The images presented in Figure \ref{fig:allims} and the ones we analyze in Section \ref{sec:results} were 
post-processed using the custom method presented in this subsection. The images were deconvolved utilizing 
the built-in function within Winnie \citep{2023AJ....166..150L}. Using STPSF, Winnie produces a grid of representative JWST/NIRCam coronagraphic PSFs sampling the variations in PSF morphology across the field of view and normalized such that an infinite aperture would measure a total flux of 1 at the exit pupil.  These PSFs are combined with coronagraph transmission maps (where each pixel's value indicates the fractional flux throughput for a point source falling on that pixel; generated with WebbPSF-ext) to enable accurate modeling of the throughput and blurring induced by the JWST/NIRCam optics \citep{2023AJ....166..150L}. The deconvolution procedure used in Winnie is a variation of the Richardson-Lucy deconvolution algorithm that has been altered to account for the significantly shift-variant NIRCam coronagraphic PSF. These modifications are described in \citet{2025ApJ...994..199M} and demonstrated in the context of disk analysis in \citet{2025ApJ...994..199M} and \citet{2026arXiv260723992L}. This procedure is carried out for 50--100 iterations, with the number of iterations being tuned for each observation to strike a balance between the sharpening of astrophysical features and the amplification of background noise (see analysis in \citealt{2026arXiv260723992L}). While some residual (plausibly chromatic) blurring is expected for this range of iterations, any chromatic impact on extracted surface brightness measurements will be small compared to the chromatic blending of nearby resolution elements in the non-deconvolved images. Moreover, for the purpose of assessing the presence of broad spectral features, chromatic biasing in either case is likely irrelevant.

\begin{figure*}[!t]
    \centering
    \includegraphics[width=0.775\linewidth]{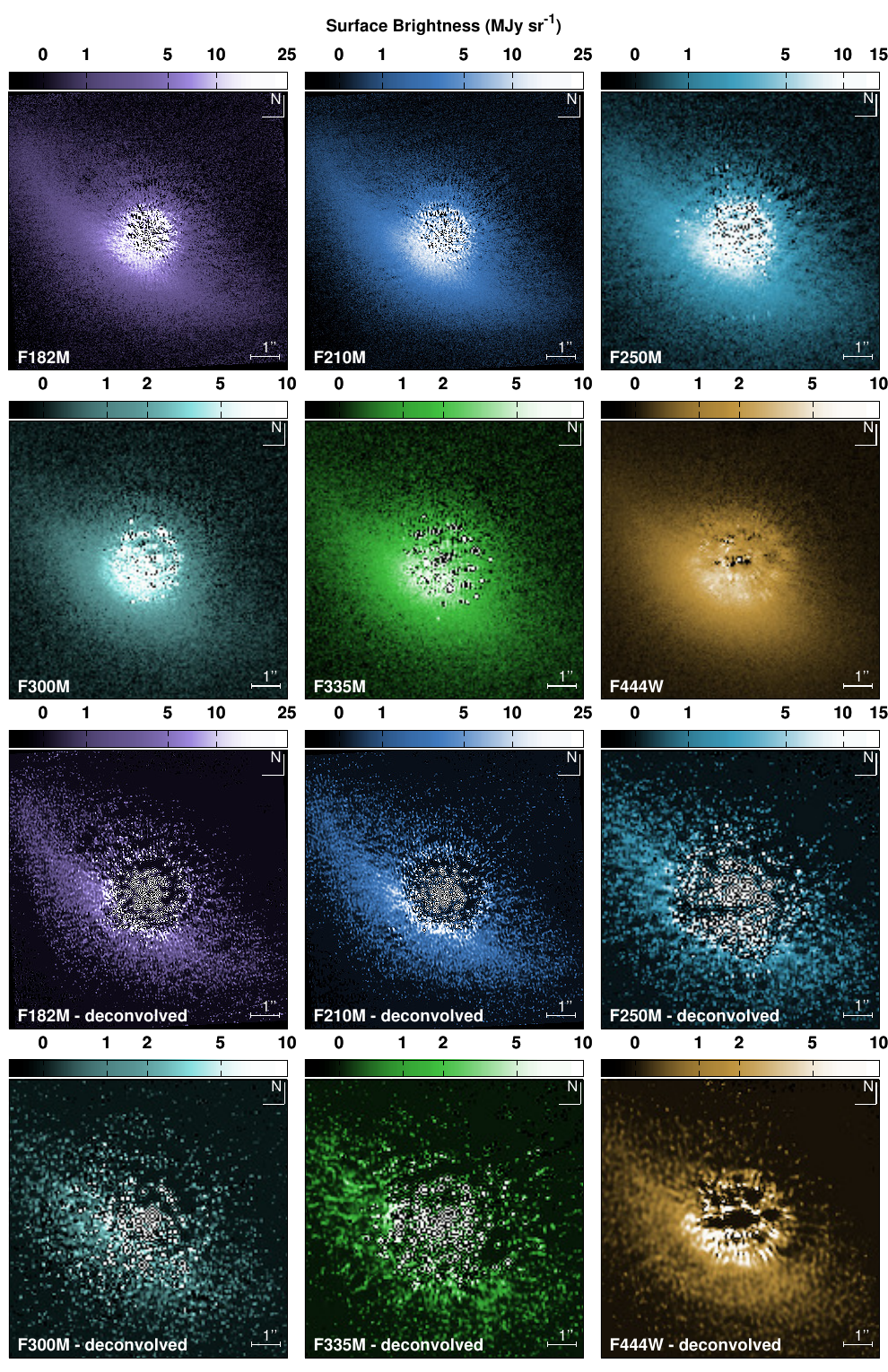}
    \caption{The six-color observations of the HD~10647 (q$^1$ Eri) system. 
    Images are displayed in logarithmic scaling in units of MJy sr$^{-1}$ with levels displayed
    for each filter in their respective colorbars. Analysis, for this system, was performed on the observed 
    (not deconvolved) images.}
    \label{fig:hd10647_gallery}
\end{figure*}

\section{MIRI Data Reductions}
\label{sec:miris}

Two of the sources in the survey, HD~107146 and HD~10647, were also observed with MIRI, using direct imaging, with
the F2100W filter. The observations included the acquisition of a background image and that of a reference PSF (see Table \ref{tab:JWSTobs} for details),
all executed using the 4pt EXTENDED dither pattern integrating with the SUB256 subarray.
Unlike for the GO 1193
observations of Fomalhaut, Vega, and $\epsilon$ Eridani, we did not perform a rotational dither, in order to maximize 
integration time. The targets in this program were also fainter than those in our previous program (GO 1193), therefore we were not affected by saturation in the central pixels.
The MIRI background observations were the first to be executed in each sequence to ensure a clean 
image void of latent imprints. 

The reduction and post-processing of these observations followed along the now ``standard'' methods that we 
have described in great detail in the Vega \citep{su24} and $\epsilon$ Eridani \citep{wolff25} papers. For brevity, we
refer readers to those papers, especially the Appendix of \cite{wolff25}. The uncalibrated datasets were processed with the 
1.12.5 and 1.13.4 versions of the JWST pipeline \citep{bushouse} for the HD~10647 and HD~107146 observations, respectively, with the majority of the settings kept at default values. We did turn off
dark current corrections, and used the background images (observed identically to the target and PSF observations) as the dark correction. The custom post-processing steps, identical to those carried out for $\epsilon$ Eridani \citep{wolff25}, were executed using IDL/IDP3 and IRAF. 

\section{Results}
\label{sec:results}

The five targets observed with our program span a range of stellar and system properties, allowing a comparative study of
disk scattered light properties. In Table \ref{tab:props}, we list the attributes of the targets in our survey.
Figure~\ref{fig:overview} compares their thermal mid-infrared excesses with those of other debris systems. Their 
disks are all relatively bright at 24 $\mu$m for their ages, suggesting that their bright systems of scattered 
light may result from recent collisional activity. 
In what follows, we present an overview of the results for each of the five disks in our sample.

\subsection{HD~10647}

\subsubsection{Background}

HD~10647 (q$^{1}$ Eridani, HR 506, HIP 7978) is an older \citep[$\sim 700~{\rm Myr}$;][]{carvalho25} 
nearby ($d = 17.35 \pm 0.01$ pc) F9V main sequence star 
that hosts both a warm and a cold debris belt and at least one giant planet. These characteristics make it an archetype for 
studying the coupling between planetary architectures and circumstellar debris in Solar-type systems. 
There are only 14 known debris disks located closer to us than HD~10647 and of those, only four have been 
spatially resolved in scattered light observations (Fomalhaut, Vega, AU Mic, and HD~207129), of which 
only HD~207129 is also a solar-type star. The debris disk system is amongst the brightest in fractional 
infrared luminosity \citep[$L_{\mathrm{IR}}/L_\star > 3\times10^{-4}$;][]{lovell21,moor06}, an unusual 
property at its advanced age.

In 2003, Mayor et al.\ announced the possible detection of a planet in the system, HD~10647b, 
at the XIX$^{\rm th}$ IAP Colloquium based on CORALIE data. In 2006, \citet{butler06} reported a 
long-period ($P\sim1000$\,days) Jovian planet orbiting at 2.03\,au, based on 28 radial velocity points
acquired with UCLES on the Anglo-Australian Telescope, with an $M\sin i$ of 0.93\,M$_{\jupiter}$.
\citet{marmier13} presented a total of 108 Doppler measurements with CORALIE, determining the planet
to be orbiting at 2.015$\pm$0.011\,au with a period of 989.2$\pm$8.1 days and having a mass of 
$M\sin \iota \approx 0.94 \pm 0.08\,M_{\jupiter}$. In 2024, \cite{bisht24}
analyzed 330 archival radial velocity measurements from multiple instruments and refined the orbital
parameters of the planet, deriving a period of 992.10$\pm$1.45 days and semi-major axis of 2.02$\pm$0.01\,au,
and a slightly reduced planetary mass of $M\sin i \approx 0.90 \pm 0.04\,M_{\jupiter}$.\footnote{Note that the value
of $M_{\rm DA} \approx 1.07 \pm 0.05\,M_{\jupiter}$ given by \cite{bisht24} for the ``disk aligned'' mass was determined with a multi-variable MCMC fit to the
observed radial velocities, including the disk inclination, but is the median value of a highly skewed solution distribution (priv.\ communication), therefore we decided to quote the \cite{marmier13} solution in Table \ref{tab:props}.} 
Adopting the value determined by \cite{marmier13} and taking into account the inclination
of the system, HD~10647b has a mass of $M \approx 0.975 \pm 0.08\,M_{\jupiter}$. 

\begin{figure}[!t]
    \centering
    \includegraphics[width=0.9\linewidth]{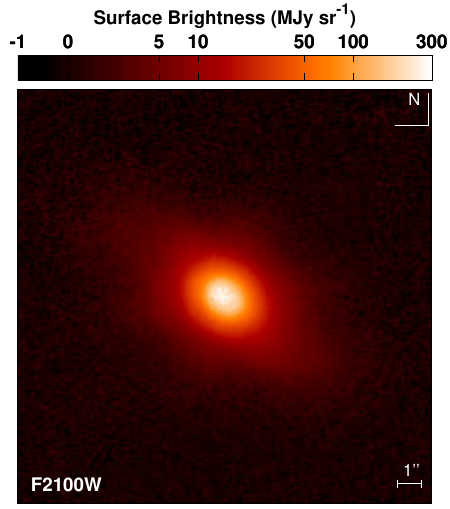}
    \caption{The MIRI F2100W observation of the HD~10647 (q$^1$ Eri) system. A PSF scaled to
    0.19 Jy was subtracted to obtain this image, slightly higher than the 0.165 Jy photospheric
    brightness of the star (determined via SED fitting), to produce a clean reduction, hinting at 
    an additional spatially unresolved central component.}
    \label{fig:hd10647_miri}
\end{figure}

The star’s far-infrared excess was first detected by \textit{IRAS} \citep{stencel91,silverstone00,zuckerman04a} and later 
confirmed by \textit{Spitzer}/MIPS \citep{trilling08}, revealing strong 24 and 70\,$\mu$m excesses indicative 
of an extended circumstellar disk, while Spitzer/IRS 5.5–35\,$\mu$m spectra \citep{chen06} measured excess flux 
in the 30--34\,$\mu$m band but not at 8--13.5 $\mu$m. The first spatially resolved images of the disk were obtained in scattered 
light with \textit{HST}/ACS, initially presented in a conference proceeding by \cite{stapelfeldt07} and later analyzed 
in detail by \cite{lovell21}. Far-infrared observations by \textit{Herschel}/PACS \citep{liseau10} revealed a broad -- 40 au wide -- inclined 
belt at a stellocentric distance of roughly 85 au, although at a relatively low angular resolution of 6$^{\prime\prime}$ at 70\,$\mu$m.
Subsequent spectral energy distribution (SED) modelling by \citet{schuppler16} showed that a two-component 
architecture provides a superior fit to the data, consisting of a cold outer belt ($\sim$75--125\,au)  
and a warm inner component ($\sim$3--10\,au), with about three orders of magnitude more dust mass in the outer belt than
in the inner. \cite{kennedy14} also inferred additional infrared excess from the inner regions of the system,
with a radius of $\sim$10\,au assuming blackbody grains.

\begin{figure*}[!t]
    \centering
    \includegraphics[width=0.99\linewidth]{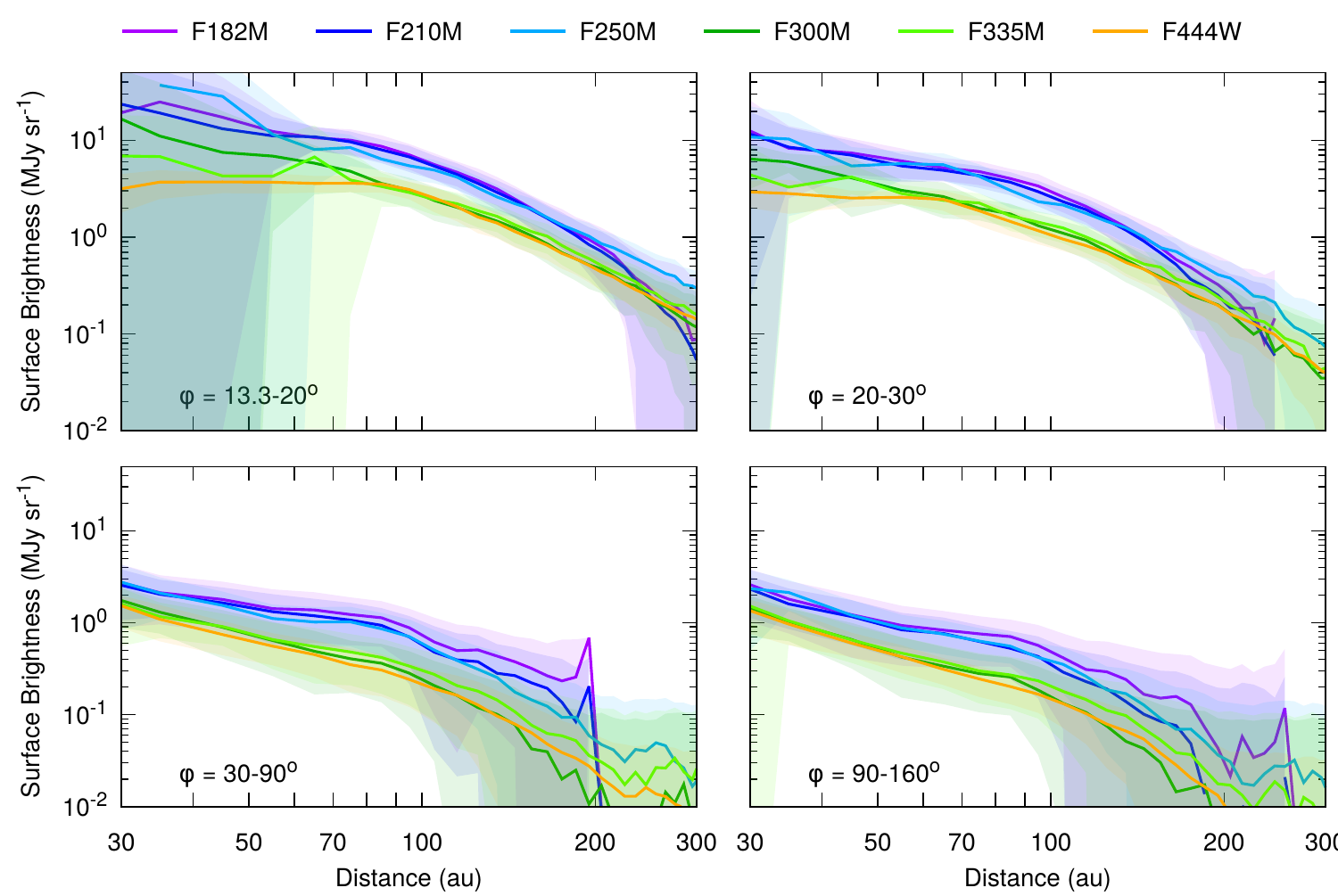}
    \caption{Deprojected median radial surface brightness profiles (with 1 $\sigma$ per-pixel 
    errors represented by shaded regions) for the HD~10647 system at various scattering angle directions  $\varphi$ (see 
    Figure \ref{fig:hd10647_phi}). The angular separation towards $\varphi \sim 90\deg$ direction is considerably
    larger than towards the forward- and back-scattering angles, implying that the main-belt is better resolved.}
    \label{fig:hd10647_SBs}
\end{figure*}

\begin{figure}[!t]
    \centering
    \includegraphics[width=1.0\linewidth]{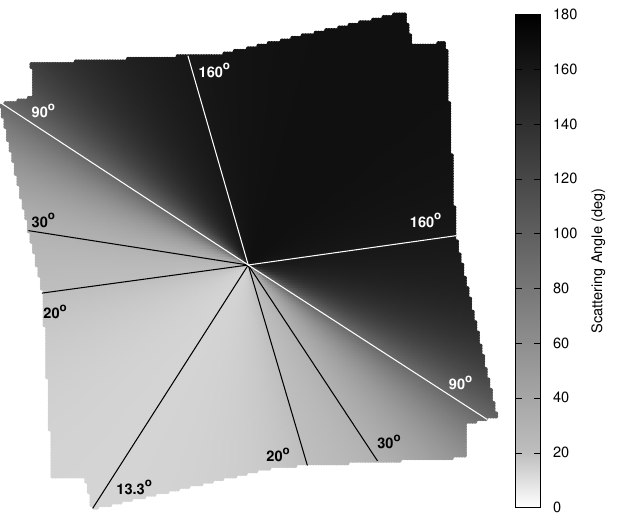}
    \caption{The scattering angles in the HD~10647 system.}
    \label{fig:hd10647_phi}
\end{figure}

High-resolution \textit{ALMA} observations by \citet{lovell21} at 0.86\,mm and 1.3\,mm resolved the outer belt with 
unprecedented detail, measuring a radial extent from 34 to 134\,au and peak emission at $81.6 \pm 0.5$\,au, with 
inclination $76.7\degr \pm 1.0\degr$ and position angle $54.2\degr \pm 0.5\degr$. The continuum images reveal a 
significant azimuthal asymmetry, with the southwest ansa both brighter and slightly closer to the star, possibly 
indicative of gravitational sculpting by an unseen perturber. Dynamical modelling by \cite{lovell21} suggests that an additional 
planet of mass $8\,M_\oplus$--$11\,M_{\jupiter}$ near $\sim$60\,au could reproduce the observed offset and 
brightness asymmetry. Notably, \citet{lovell21} also detect a tentative 0.86\,mm excess within $\sim$10\,au 
($70 \pm 22\,\mu$Jy), consistent with a compact warm dust population coincident with the inner belt inferred from 
the SED. 

Because of the highly inclined nature of HD~10647, it was analyzed both parametrically and non-parametrically by \citet{2026A&A...705A.196H} to investigate the disk’s radial structure, and parametrically by \citet{2026A&A...705A.197Z} to investigate the disk’s vertical structure (in addition to its radial structure), as part of the ALMA survey to Resolve exoKuiper belt Substructures (ARKS) program.
In general, these ALMA analyses imply that the disk is radially broad and vertically extended, and can be well-described as two overlapping radial Gaussian distributions, with best-fit radii and widths ($R_i {\pm} \sigma_i$) of $89.2{\pm}12.0\,$au and $168{\pm}48\,$au respectively \citep{2026A&A...705A.196H}, and with two vertical aspect ratios (corresponding to the narrower and broader radial Gaussian components respectively) of $h_{\rm HWHM1}{=}0.052$ and $h_{\rm HWHM2}{=}0.225$ \citep{2026A&A...705A.197Z}. 
These two vertical populations result in the conclusion of \citet{2026A&A...705A.197Z} that HD~10647 has two distinct dynamical dust populations, with the lower aspect ratio (dynamically cooler component) corresponding to the dominant component by mass (${\approx}88\%$) and the higher aspect ratio (dynamically hotter component) corresponding to the sub-dominant component by mass (${\approx}12\%$). 
Based on the inner–edge steepness, \citet{2026A&A...705A.196H} additionally reported the upper-limit mass and semi-major orbital axis for a planet that could plausibly sculpt the disk inner edge, as $2.5\,M_J$ inside 48\,au.
Further analysis by the ARKS team resulted in this source being identified as being on an asymmetric orbit based on the integrated major-axis flux density being raised towards the south-west \citep{2026A&A...705A.200L}, in keeping with the earlier results of \citet{lovell21}. 
Although the ARKS asymmetry result is noted as being less pronounced in comparison to the earlier result of \citet{lovell21} given the inclusion of new higher-resolution data, this remains significant \citep[see further discussion in][]{2026A&A...705A.200L}.

\begin{figure*}[!t]
    \centering
    \includegraphics[width=0.49\linewidth]{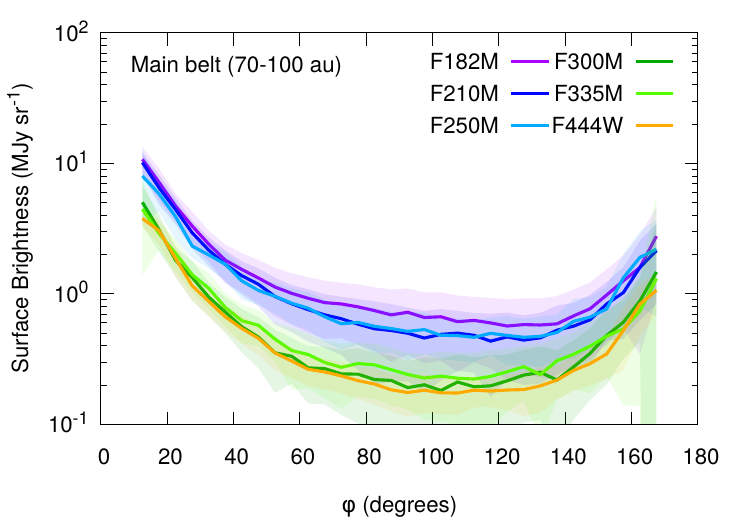}
    \includegraphics[width=0.49\linewidth]{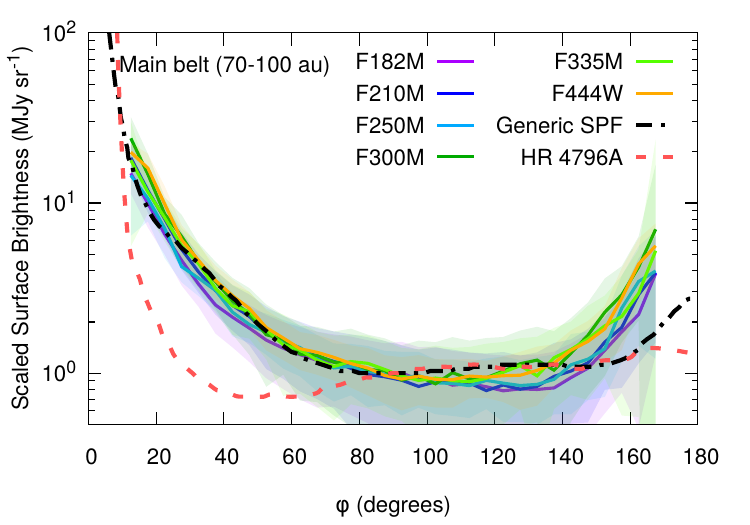}
    \caption{\textit{Left:} The measured scattering phase distributions for the HD~10647 system determined within the main belt. \textit{Right:}
    SPF curves scaled to 1 MJy sr$^{-1}$ surface brightness at $\varphi=90^{\circ}$ overploted with a generic SPF compiled
    by \cite{2024MNRAS.528.6959H} as well as the observed SPF of HR4796A \citep{2017A&A...599A.108M}. The generic SPF is produced by 
    averaging the SPFs of a number of solar system objects \citep[see details in][]{2024MNRAS.528.6959H}. 
    The scattering properties of the dust in this system are very similar to the generic solar system SPF; deviations at 
    large angles are likely due to measurement errors in our dataset, as the stellar core becomes the dominant source 
    of signal and noise compared to the low levels expected for backscattering. Interestingly, the SPF of HR 4796A, another
    bright scattered light disk, shows lower forward scattering at angles where both systems have data,
    relative to their fluxes normalized at $\varphi=90^{\circ}$. This does not exclude HR 4796A from having a brighter SPF
    peak at $\varphi<13\fdg3$}.
    \label{fig:hd10647_Phis}
\end{figure*}

\subsubsection{JWST results}
\label{sec:hd10647analysis}

The NIRCam observations, shown in Figure \ref{fig:hd10647_gallery}, reveal the disk at all six observed wavelengths, although it is
certainly brighter and detected with a higher SNR at shorter wavelengths. The disk is highly inclined with strong forward scattering towards
the south-east direction. The MIRI observations, shown in Figure \ref{fig:hd10647_miri}, showcase a bright central core
while also revealing it to be extended towards the outer regions. The HD~10647 inner disk is seemingly similar to the
Fomalhaut \citep{gaspar23}, Vega \citep{su24}, and $\epsilon$ Eridani \citep{wolff25} disks, with an extended inner disk of 
micron-sized dust particles.

Analysis for the HD~10647 system was carried out on the observed (not the deconvolved) images, as the low surface brightness
and spatially extended nature of the system does not provide an ideal case for deconvolution, only introducing noise
during the additional processing step. We measured the surface brightness of the HD~10647 disk in the NIRCam dataset 
assuming the system architecture and orientation determined by the ALMA dataset \citep[][see also Table \ref{tab:props}]{lovell21}. 
In Figure \ref{fig:hd10647_SBs}, we showcase the deprojected, median radial surface brightness values with their standard deviation 
measured within various scattering angle regions ($\varphi)$. We do this because the scattering efficiency is heavily dependent on the
scattering angles, shown in Figure \ref{fig:hd10647_phi}, and we want to investigate whether the 
surface brightness profiles are scattering angle and color dependent. The profiles are, in fact, very similar
in all four directions, with the three shorter wavelength bands (F182M, F210M, and F250M) yielding a factor of a few
larger brightness values. The main belt at $\sim$80~au is most easily identified in the $\varphi=30^{\circ}-90^{\circ}$
bin. 

\begin{figure*}[!t]
    \centering
    \includegraphics[width=0.99\linewidth]{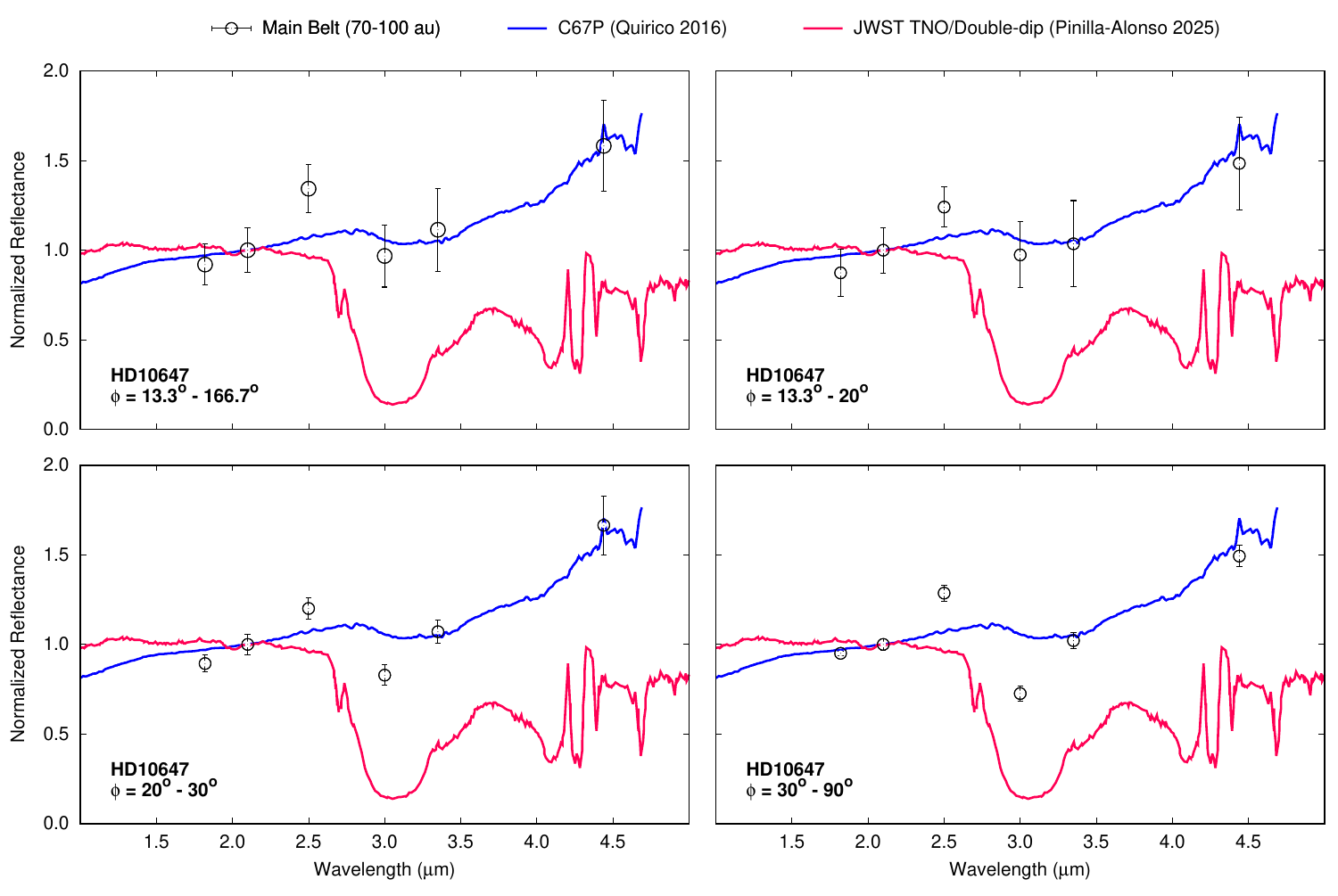}
    \caption{The integrated and normalized (at 2.1 $\micron$) reflectance spectra of the HD~10647 main belt (70-100 au), 
    scaled by deprojected $R^2$, compared to that of known Solar System objects (see Section \ref{sec:hd10647analysis}).
    The {\it Top Left} panel shows the reflectance spectra using the entire scattering angle range, while the other panels shows the same restricted to narrower angles. 
    The spectra closely follows that of comet C67 in the solar system, with a surprising peak at 2.5 $\micron$ and an absorption feature at 
    3 $\micron$ at the expected location for water ice. The larger scattering angle areas ($\varphi>20^{\circ}$) 
    avoid the higher noise core region, therefore are better characterized.
    \label{fig:hd10647_spectra}}
\end{figure*}

\begin{figure}[!t]
    \centering
    \includegraphics[width=0.99\linewidth]{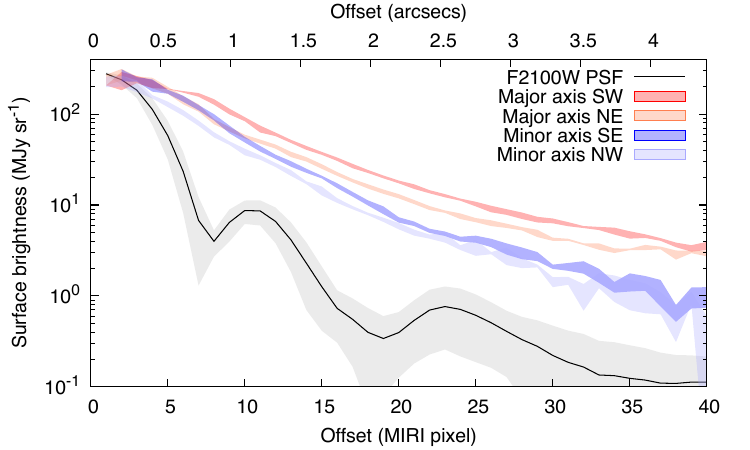}
    \caption{The radial surface brightness profile of the HD~10647 disk along its major and minor axes 
    as observed by MIRI with the F2100W filter. The disk is clearly brighter and more extended
    towards its SW direction.}
    \label{fig:hd10647_miri_radial}
\end{figure}

In Figure \ref{fig:hd10647_Phis}, we show the scattering phase function (SPF) for the main belt (70-100 au).
The SPF was determined by averaging the surface flux brightness values of all pixels located within the main
belt region (70-100 au) in $\Delta\varphi=5^{\circ}$ bins (based on scattering angles shown in Figure \ref{fig:hd10647_phi}),
weighted with the per-pixel standard deviations. The
errors of the SPF in each bin are equal to the standard deviations of the pixels in each spatial bin.
Given the inclination angle of the system ($\iota=76.7^{\circ}$), the residual speckles in the stellar core, 
and the low efficiency of backscattering, measurements above $\varphi\sim140^{\circ}$ are of low quality. 
The {\it left panel} in the figure shows the absolute measurement values, while the 
{\it right panel} shows them normalized to 1 MJy sr$^{-1}$ at
$\varphi=90^{\circ}$. We also show a generic SPF compiled by \cite{2024MNRAS.528.6959H} -- merging observations
of several solar system objects, such as Saturn’s D68 and G rings \citep{hedman15}, Jupiter’s ring \citep{2004Icar..172...59T} 
and various comets \citep[see details in][]{2024MNRAS.528.6959H} --  to compare the values in the HD~10647 system 
to the average dust scattering properties in the Solar System. We also present the SPF of the disk surrounding HR 4796A \citep{2017A&A...599A.108M},
one of the best studied systems, which shows significant deviation in the forward scattering ranges (10$^{\circ}$-80$^{\circ}$)
both from the solar system generic SPF and that of HD~10647, with the latter two in good agreement with each other. 
Interestingly, the normalized scattering function profiles for HD~10647 seem to be wavelength independent.

\begin{table}[!t]
\setlength{\tabcolsep}{2pt}
\begin{center}
\caption{Photometry of the HD~10647 disk. 
\label{tab:hd10647phot}}
\begin{tabular}{lcccccc}
\hline\hline
                   & F182M  & F210M & F250M & F300M & F335M & F444W \\
\hline
F$_{\nu,\star}$\tablenotemark{a} (Jy) & 15.43  & 12.70 & 9.49  & 6.92  & 5.66  & 3.31 \\
\hline
\multicolumn{7}{c}{$\varphi = 13\fdg3 - 166\fdg7$}\\
\hline
F$_{\rm sc}$ ($\mu$Jy)      & 592 & 530 & 532 & 280 & 263 & 219 \\
$\sigma_{\rm sc}$ ($\mu$Jy) & 74  & 65  & 53  & 50  & 54  & 35 \\
\hline
\multicolumn{7}{c}{$\varphi = 0^{\circ} - 20^{\circ}$}\\
\hline
F$_{\rm sc}$ ($\mu$Jy)      & 238 & 224 & 208 & 119 & 103 & 87 \\
$\sigma_{\rm sc}$ ($\mu$Jy) & 36  & 28  & 19  & 22  & 24  & 15 \\
\hline
\multicolumn{7}{c}{$\varphi = 20^{\circ} - 30^{\circ}$}\\
\hline
F$_{\rm sc}$ ($\mu$Jy)      & 84  & 77  & 69  & 35  & 37  & 34  \\
$\sigma_{\rm sc}$ ($\mu$Jy) & 4.5 & 4.4 & 3.4 & 2.4 & 2.2 & 3.3 \\
\hline
\multicolumn{7}{c}{$\varphi = 30^{\circ} - 90^{\circ}$}\\
\hline
F$_{\rm sc}$ ($\mu$Jy)      & 122 & 106 & 102 & 42 & 48 & 41   \\
$\sigma_{\rm sc}$ ($\mu$Jy) & 1.7 & 3.4 & 3.5 & 2.5 & 2.1 & 1.7 \\
\hline
\end{tabular}
\end{center}
\tablecomments{Photometry of the Main Belt (70-100 au) disk around HD~10647. The ``{\it sc}'' scatter values are scaled by $(R/R_0)^2$, where $R_0$ is the disk center at 85 au. As the disk is narrow, these values are within 1\% of the nominal values.}
\tablenotetext{a}{Stellar photosphere values.}
\end{table}

We next construct a broadband reflectance spectrum (${\rm L}_{\rm disk}/{\rm L}_{\ast}$) using the six filter observations 
in \mbox{Figure \ref{fig:hd10647_spectra}}, showcasing it totaled for the entire scattering angle range (multiplied by 
an appropriate $R^2$ map) in the first panel and then for various other scattering angle ranges in the other panels.
Total photometric errors are calculated following the methodology described in Appendix \ref{app:A_pt_3} and measurements
are tabulated in Table~\ref{tab:hd10647phot}. We compare these reflectance measurements to the well characterized 
spectra of comet C67 \citep{quirico16} and the bulk average spectra of ``double-dip''-type trans-Neptunian objects \citep[TNOs;][]{pinilla25}.
We must note that there is a reasonable question whether reflectance spectra of large bodies, such as TNOs, are directly comparable to 
that of small dust particles, which is why we are including measurements of comets, which are shrouded by small volatiles.
Including comparisons with the TNOs is for qualitative purposes only.
The broadband spectra show an increase at 4.44 $\micron$.
The measurements around HD~10647 are most similar to those of cometary bodies. It lacks a ``significant'' water-ice dip at
3 $\micron$ overall, but ice seems to be detected at scattering angles above 20$^{\circ}$. We should note that the solar system
observations will generally be obtained at larger scattering angles, with the TNOs measured at $\varphi\approx180^{\circ}$.
The most curious feature is a reflectance bump at 2.5 $\micron$, the origin of which is unknown.

Finally, in Figure \ref{fig:hd10647_miri_radial}, we present the measured radial surface brightness profiles of the
HD~10647 debris disk at 21 $\micron$ using the MIRI observations (Figure \ref{fig:hd10647_miri}). To produce the 
observations, the observed PSF was scaled to a total brightness of 190.6 mJy, which is higher than the photospheric
brightness of the central star at 165 mJy. This discrepancy results from a spatially unresolved central component, which
convolves with the stellar PSF, with a brightness of 25.6 mJy. The remaining disk component is extended from the inner regions
well out to 100 au with a total integrated flux of 21.7 mJy, similar to that of the inner unresolved component. Much like 
for the observations of Fomalhaut \citep{gaspar23} and Vega \citep{su24} we see an inner disk region filled with dust 
for HD~10647, highlighting the prevalence of dust transport due Poynting-Robertson drag \citep{burns79} around this 
luminous solar-type star. In Figure \ref{fig:hd10647_miri_radial}, we plot the scaled 
surface brightness of the observed F2100W PSF with a black line, and compare it to the radial surface brightness of the disk
along its major and minor axes. Just like for the ALMA observations \citep{lovell21, 2026A&A...705A.200L}, the SW side of the disk,
along the major axis, is brighter than its NE side and the SE side is brighter than the NW side along the minor axes 
in the regions outwards of $0\farcs5$. 

Figure~\ref {fig:hd10647_miri} underestimates the spatial extent
of the debris disk excess because the disk of HD~10647 is cold, with relatively subdued emission at 24 $\mu$m compared 
with its total luminosity. In the 273 FGK members of the combined DEBRIS and DUNES samples, HD~10647 has the {\it largest}
fractional excess \citep{sibthorpe2018}. Its fractional excess is at least four times larger than those of all but one star,
HD~69830, which is a site of ongoing collisional activity \citep{beichman05b}. This reinforces the conclusion that 
HD~10647 owes the brightness of its disk to a recent collision of planetesimals and the aftermath.  

\begin{figure}[!t]
    \centering
    \includegraphics[width=0.99\linewidth]{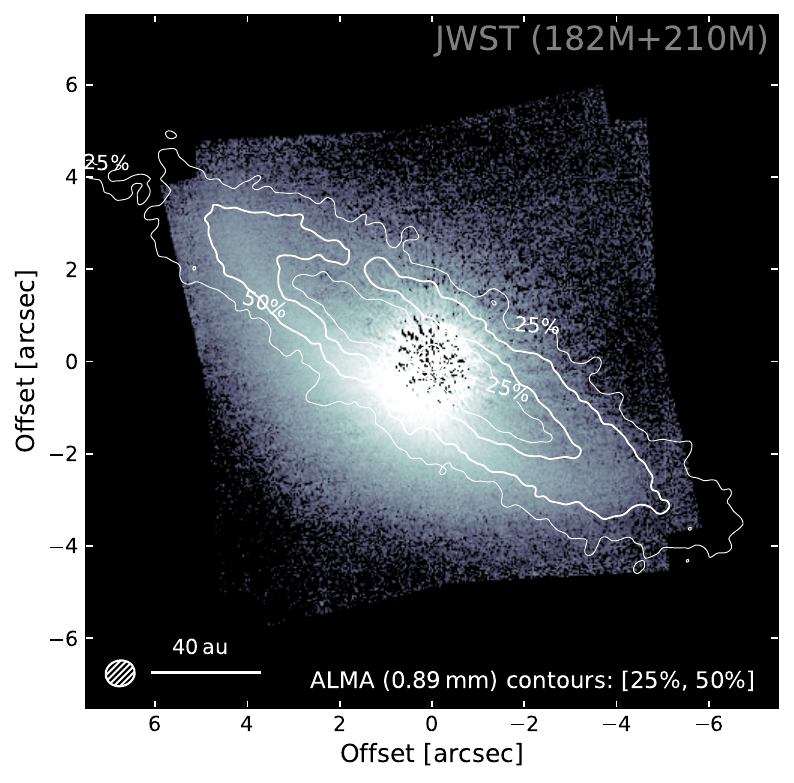}
    \caption{The JWST NIRCam SW observations (F182M and F210M combined to increase SNR) of HD~10647, with
    ALMA surface brightness contours from the ARKS observations \citep{2026A&A...705A.195M} overlaid. The ALMA
    peaks are co-located with the NIRCam brightness peaks.}
    \label{fig:hd10647_alma}
\end{figure}

The asymmetries in the disk provide clues to the nature of this event. The peak surface brightnesses in the ALMA images 
are at identical radii from the star to the NE or SW \citep{lovell21,2026A&A...705A.200L}, suggesting that the parent bodies orbit in an
approximately circular ring. The HST image at 0.6 $\mu$m shows significantly greater extent along the major axis to 
the NE than to the SW \citep{lovell21}. In Figure \ref{fig:hd10647_alma}, we compare the locations of the ALMA brightness
peaks to the combined F182M and F210M NIRCam image, showing good agreement between the two.

At 21 $\mu$m (Figure~\ref{fig:hd10647_miri_radial}), the NE apex is not larger but slightly smaller 
than the SW one; it does not show an extended structure such as observed at optical scattered light with HST. The dominant force acting on the very small grains and that distinguishes sharply in size in this regime 
is radiation pressure. For a star similar to HD~10647, grains are expected to transition from being strongly 
affected by radiation forces to dominantly under gravitational forces as they increase in size above 1 - 4 $\mu$m 
\citep[depending on composition, e.g.,][]{arnold2019}. This transition must be responsible for the change in
behavior with wavelength. There is no imprint of the planet at 2 au in the innermost structure of the disk at 21 $\mu$m, perhaps because our beam width is $\sim$ 17 au.

An independent set of grain size estimates can be derived from the scattering behavior. Based on the analogy 
with Saturn ring particles, \citet{hedman15} argued that the scattered light is likely dominated by grains 
containing modest amounts of water ice, with minimum sizes of a few $\mu$m, which is just above the blowout 
limit. In this regime, radiation pressure places the grains on highly eccentric but still bound orbits, allowing 
them to contribute to the observed scattered-light signal. In the mean time, grains smaller than this are removed 
from the system too rapidly to make a significant contribution to our images.

The continuous production of grains near the original collision site is consistent with the relatively cold far
infrared SED of this system. That is, the data suggest the collision of two bodies far from the star, with 
loosely held layers of water ice mixed with minerals, as is typical in the Kuiper Belt of the Solar System \citep{pinilla25}. 
Most extreme debris systems under study have collisions closer to the star and the collisional products are 
apparent at wavelengths of 10 $\mu$m and shorter, but this is not the case for HD~10647.

\begin{figure*}[!t]
    \centering
    \includegraphics[width=0.7877\linewidth]{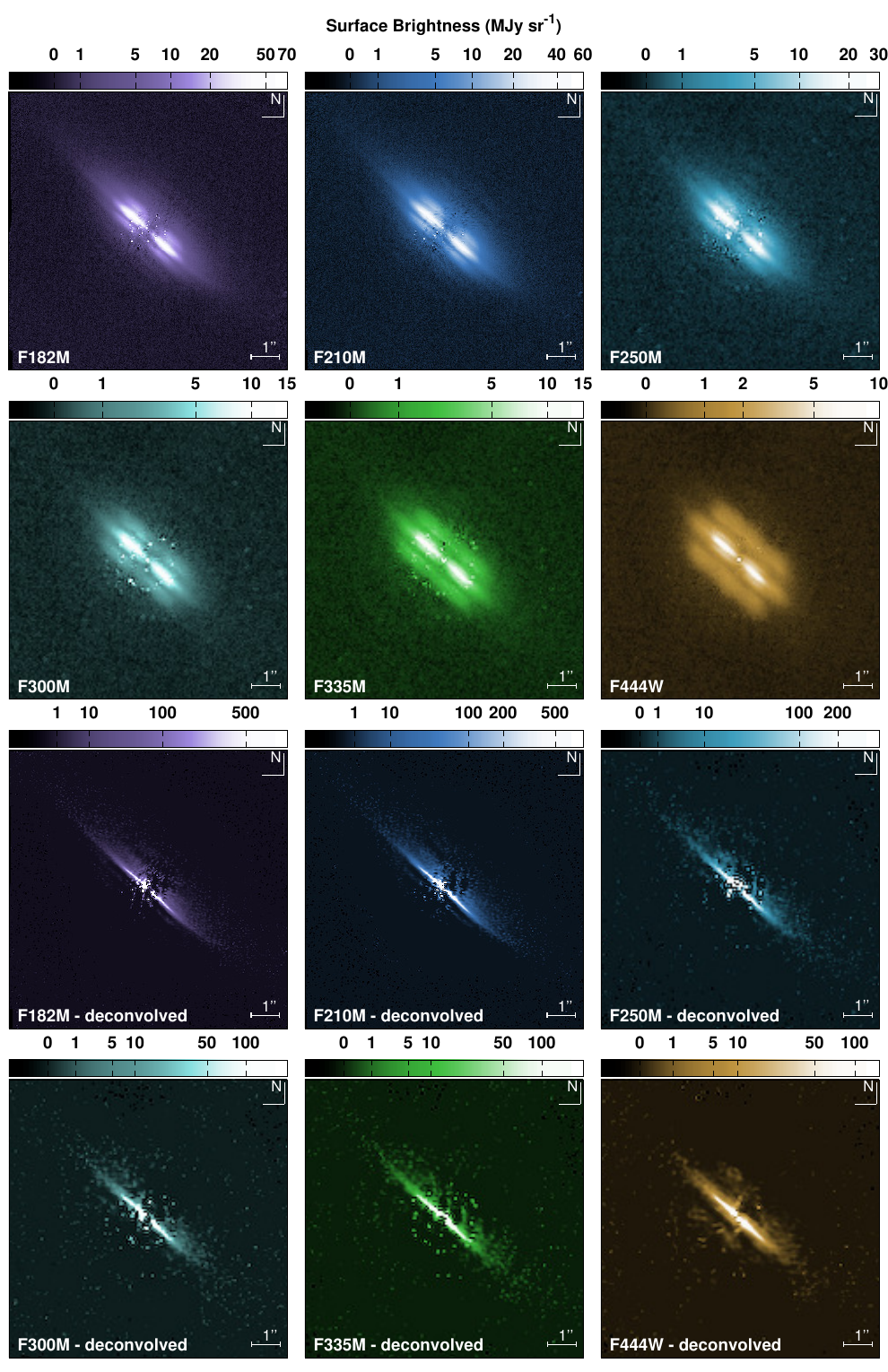}
    \caption{The six-color observations of the HD~32297 system. Images are displayed in logarithmic scaling with levels displayed
    for each filter in their respective colorbars.}
    \label{fig:hd32297_gallery}
\end{figure*}

\begin{figure*}[!t]
    \centering
    \includegraphics[width=0.99\linewidth]{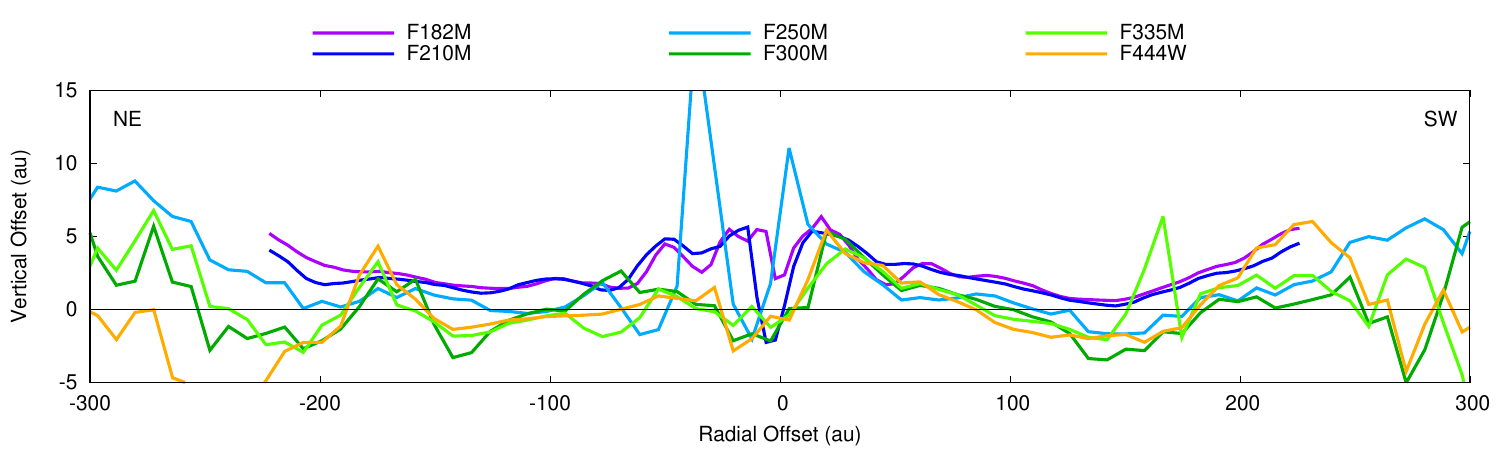}
    \caption{The location of the disk peak brightness with respect to the best fitting mid-plane orientation. 
    The short wavelength observations trend towards the N direction, possibly due to a minor disk inclination, resulting
    in enhanced forward scattering. The disk flares to the N direction for both lobes outside of 200 au radius.}
    \label{fig:hd32297_flares}
\end{figure*}

\subsection{HD~32297}

\subsubsection{Background}

HD~32297 is a young (15-30 Myr) and early-type star (A6V) at a distance of 129 pc and surrounded by a massive and extended edge-on 
debris-disk (see Table \ref{tab:props}) that is extremely bright in scattered light.  We showcase our 
NIRCam images of the system in Figure \ref{fig:hd32297_gallery}. The PSF of JWST greatly broadens 
the edge-on disk of this system, therefore, the deconvolved images -- also presented in Figure 
\ref{fig:hd32297_gallery} -- reveal a thinner disk architecture than natively observed with the 
observatory. 

\begin{figure*}[!t]
    \centering
    \includegraphics[width=0.99\linewidth]{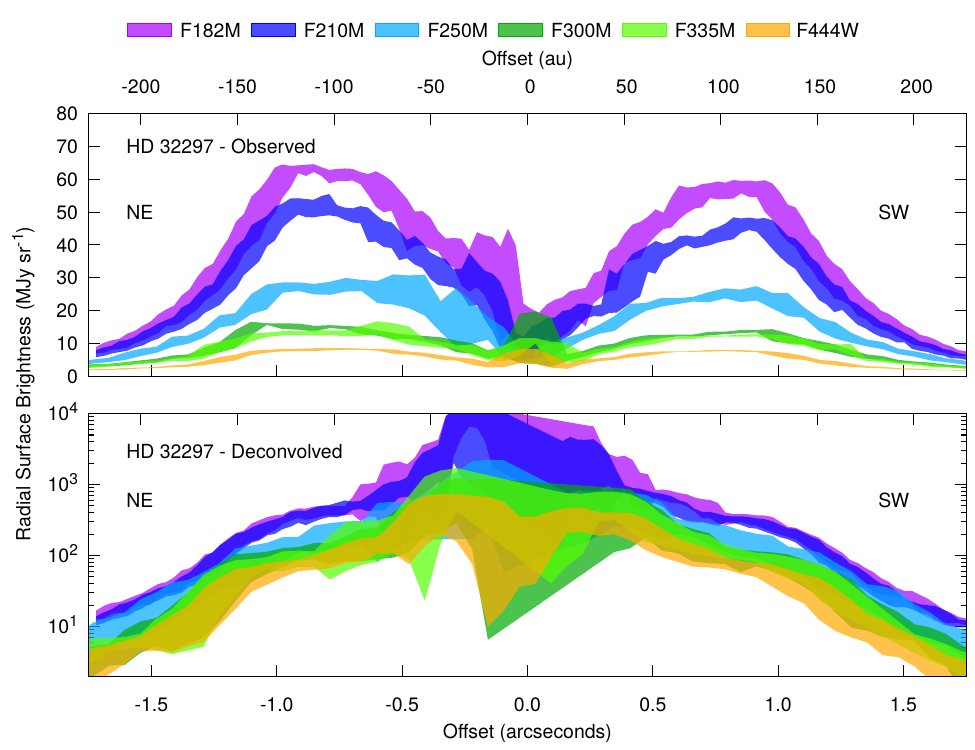}
    \caption{Peak radial surface brightness of the HD~32297 system measured across its edge-on disk plane at all six 
    NIRCam wavelengths, for both the observed (top) and deconvolved images (bottom). The SW lobe of the disk appears 
    to be about 9\% fainter than the NE side.}
    \label{fig:hd32297_radial}
\end{figure*}

Based on its large IRAS infrared excess, the system was targeted in an HST Near-Infrared Camera and 
Multi-Object Spectrometer (NICMOS) corongraphic survey \citep{schneider05}. 
The observations in the F110W filter provided the first resolved images of the  disk, revealing a large 
($\ge 3\farcs3$; 400 au) spatial extent in near-IR scattered light and a brightness asymmetry towards its SW side.
Follow-up ground-based observations at optical wavelengths \citep[R filter 647 nm;][]{kalas05} showed the disk to 
be even more extended, with nebulosity detected out to 15$^{\prime\prime}$ (1680 au). Asymmetries detected at this time hinted at 
interactions with the local Taurus molecular cloud. The system was revisited by HST, first using NICMOS \citep{debes09}
and then STIS \citep{schneider14}. The NICMOS observations also found an asymmetry in the radial surface brightness
profiles, with increasing offsets towards the longer wavelengths. A break in these surface
brightness profiles appears at 104-115 au in the SW lobe and at 80-97 au for the NE lobe, further highlighting the 
asymmetry of the system. \cite{kalas05} and \cite{debes09} argue that the asymmetry of the disk results from 
interactions with the local interstellar medium and also measure a warping of the disk (a change in PA between 
the two lobes of the disk). 

The new NICMOS observations expanded the panchromatic imaging with 1.6 and 2.05 $\micron$ images, revealing a mostly 
neutral color, hinting at $\sim \micron$ size grains. Coronagraphic imaging, using a four quadrant phase mask, with the
Palomar 5 m Hale telescope at K$_{\rm s}$ band \citep{mawet09} reached a $0\farcs4$ ($\sim 50~{\rm au}$) inner working angle and showed
the disk to have a gray color -- in agreement with the NICMOS observations.  
The STIS observations \citep{schneider14} present the deepest optical scattered light observations of 
the system to date and provide unparalleled understanding of the morphology of the system.
They connect the core of the disk seen with NICMOS to the extended nebulosity observed from the 
ground to $\sim 1560~{\rm au}$. A bright ``swept'' feature is prominent  ``above'' 
the disk plane towards the NW direction. The disk plane curves towards this direction also with the SW 
lobe being truncated relative to the NE side.

There are additional groundbased observations  at near-infrared wavelengths. \cite{boccaletti12} imaged 
the system at H and K$_{\rm s}$ bands using the  coronagraph of the Nasmyth Adaptive Optics System and 
Near-Infrared Imager and Spectrograph (NACO) at the Very Large Telescope (VLT), with the goal of detecting 
morphological signatures of planets. They  determined the disk to have a near edge-on 
(inclination $\sim 88^{\circ}$).  Observations with the Large Binocular Telescope Interferometer (LBTI)/LMIRcam
at L$^{\prime}$ (3.8 $\micron$) characterized the disk  at these longer  wavelengths \citep{rodigas14b}. They 
searched for planets, unsuccessfully, and found tentative indications of  water-ice in the disk grains. 
The HD~32297 system has been  observed polarimetrically by NIRC2 on Keck \citep{esposito14}, HICIAO on 
Subaru \citep{asensio-torres16}, with the Gemini Planetary Imager (GPI) at H-band \citep{duchene20}, 
and with SPHERE on the VLT \citep{bhowmik19}. These latter observations pushed the inner-working angle 
very close to the star at $0\farcs15$ and favored a `gray-to-blue' color for the disk, suggesting the 
presence of sub-blowout-size particles. 

Thermal mid- and far-IR observations were obtained by \cite{moerchen07} and \cite{fitzgerald07}, using 
respectively the Thermal Region Camera and Spectrograph (T-Recs) on Gemini-S and MICHELLE/Gemini-N. 
They reveal a spatial extent of $\sim$ 150 au. The mid-IR color temperature of 186$\pm$15 K, based 
on the T-Recs measurements at the location of the main belt, is significantly hotter than the blackbody 
temperature, hinting at emission by micron-sized particles. 
\cite{donaldson13} obtained Herschel PACS and SPIRE observations from 63 to 500 $\micron$ and modeled 
them along with with shorter wavelength observations to provide evidence for an inner component within the habitable 
zone of the system. This study provided a detailed analysis of the grain material composition, which they 
found to be dominated by water ice with a 90\% porosity and with sizes above 2 $\micron$. They derived 
constraints on the belt structures, and an updated disk fractional luminosity 
($L_{\rm inner}/L_{\ast} = 6.9\times10^{-4}$ and $L_{\rm main}/L_{\ast} = 5.6\times10^{-3}$, 
assuming $L_{\ast} = 5.3~L_{\odot}$), higher than previous estimates. 

The Combined Array for Research in Millimeter-wave Astronomy \citep[CARMA;][]{maness08} and then ALMA \citep{macgregor18}
observations characterized the main planetesimal belt at millimeter wavelengths. The 1.3 mm ALMA data places the main belt
within 78.5$\pm$8.1 and 122$\pm$3 au, with an extended halo stretching out to 440$\pm$32 au. Models have united the
scattered light and radio observations \citep{olofsson22}, showing that porous grains with a range of $\beta$ values (defining the ratio of the radiative to gravitational forces acting on the dust particles) are
necessary, with the halo populated by $\micron$-sized particles.

As with HD~10647, HD~32297’s highly-inclined (near edge-on $i{=}88.48^\circ$) orientation led to its analysis in both ARKS papers II and III \citep{2026A&A...705A.196H,2026A&A...705A.197Z}.
The ARKS analyses found this system to be radially broad, and vertically narrow, best-fit by a single dust distribution comprising a radial double power-law (describing the inner and outer edges of the disk) with a ‘critical radius’ and power-law slopes of 105.1\,au, and $\alpha_{\rm in}{=}110$ and $\alpha_{\rm out}{=}{-6.6}$ \citep[i.e., both steep; see][]{2026A&A...705A.196H}, and with a vertical Lorentzian profile, with a $h_{\rm HWHM}{=}0.0098$ \citep[i.e., narrow;][]{2026A&A...705A.197Z}. \citet{2026A&A...705A.197Z} infer that HD~32297 cannot have been self-stirred within the age of the system, and discuss planet-disk stirring scenarios to stir the disk within the relatively young age of the system.
Based on the inner–edge steepness, \citet{2026A&A...705A.196H} additionally reported the upper–limit mass and semi-major axis of a planet that could plausibly sculpt the disk inner edge, as 1.0$M_J$ inside 74\,au. A second planet constraint is provided based on the tentative non-parametrically modelled gap at 79.9\,au which is consistent with a $2.0\,M_J$ mass planet centered on this location.  
Although the star was too faint to be detected in the ARKS data, \citet{2026A&A...705A.197Z} and \citet{2026A&A...705A.200L} show the system to be offset from the phase center of the data, which at the resolution of the data, is significantly shifted from the Gaia–based location of the star, resulting in the inference of \citet{2026A&A...705A.200L} that this system is eccentric.
Whilst the system has not been modelled as eccentric at these wavelengths yet, \citet{2026A&A...705A.200L} demonstrate that the offsets are consistent with an eccentricity $e_f{>}0.03$ given the uncertain pericenter direction of the offset, which appear to agree with GPI analyses by \citet{duchene20} and \citet{2024ApJ...961..245C} who measured modest eccentricities of 0.05 and 0.04 respectively (with pericenter directions towards the south-east) in HD~32297’s disk.

Finally, the system is  one of the few debris disks with detections of atomic and molecular gas. In 2007, \cite{redfield07}
observed Na I absorption towards the system, in excess to that measured at neighboring stars, concluding that the disk itself must carry gas up to a total mass of 0.3 M$_{\oplus}$.  
Early Herschel SPIRE data detected  [C II] at 158 $\micron$. Based on generic assumptions, \cite{donaldson13} estimate a lower limit of 0.1~M$_{\oplus}$ for
the total gas mass, given this [C II] detection, confirming the estimate by  \cite{redfield07}. 
Various CO molecules were also measured in the system by ALMA over the years \citep{macgregor18,moor19,macmanamon2026}.
All of these results point to the gas in the system being secondary in origin \citep{worthen24}, with a continuous 
dynamic recycling of CO and high collisional activity.

\begin{figure}[!t]
    \centering
    \includegraphics[width=0.99\linewidth]{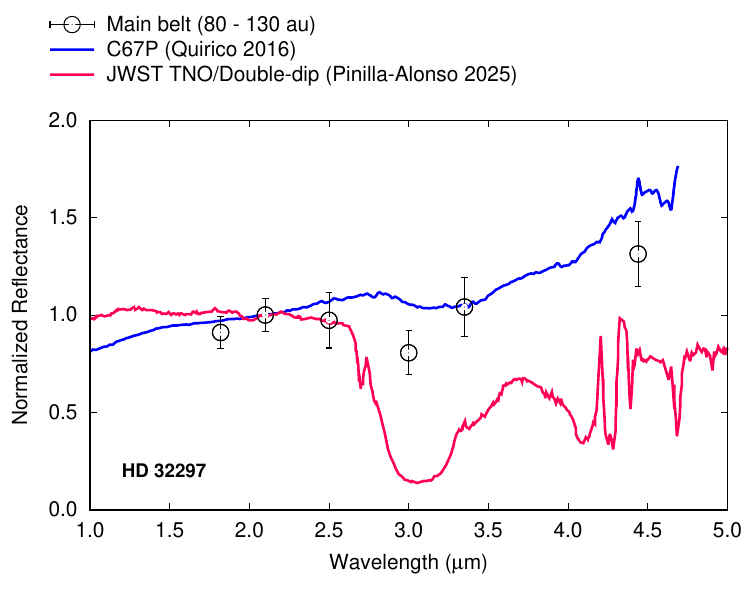}
    \caption{The normalized (at 2.1 $\micron$) reflectance spectra-photometry of HD~32297,
    compared to the observed reflectance spectra of comet C679P Churyumov–Gerasimenko \citep{quirico16} 
    and the average of the ``double-dip'' class of TNO objects \citep{pinilla25} 
    using the deconvolved images. The spectrum is quite flat and 
    similar to that of comet C67P, an inner Solar System object. 
    A 3.0 $\micron$ dip, indicative of the presence of water ice appears present at low significance.}
    \label{fig:hd32297_spec}
\end{figure}

\subsubsection{JWST Results}

Our JWST/NIRCam observations, shown in Figure \ref{fig:hd32297_gallery}, contribute an array of images at high spatial 
resolution and at new wavelengths to enable compositional studies. The edge-on disk is clearly resolved at all six
wavelengths, with the highest resolution and per-pixel signal-to-noise ratio provided at the two short wavelength 
filters. The NE lobe of the disk can be tracked to over $4\farcs3$ at a PA of $\sim 46\fdg8$, while the SW lobe 
is truncated at $3\farcs2$ and oriented at a PA of $\sim 227\fdg3$. Fitting the full extent of the disk through the location
of the central star, we determine the PA of the system at $47\fdg44$. The brightest parts of the disk are ``above''
the midplane at shorter wavelengths (see Figure \ref{fig:hd32297_flares}), indicative of the system being slightly 
inclined with enhanced forward scattering from the arc pointing towards our vantage point. The disk is clearly flared
towards the north direction at all wavelengths, likely a result of interactions with the interstellar medium 
\citep{kalas05,debes09}. 

\begin{figure}[!t]
    \centering
    \includegraphics[width=0.99\linewidth]{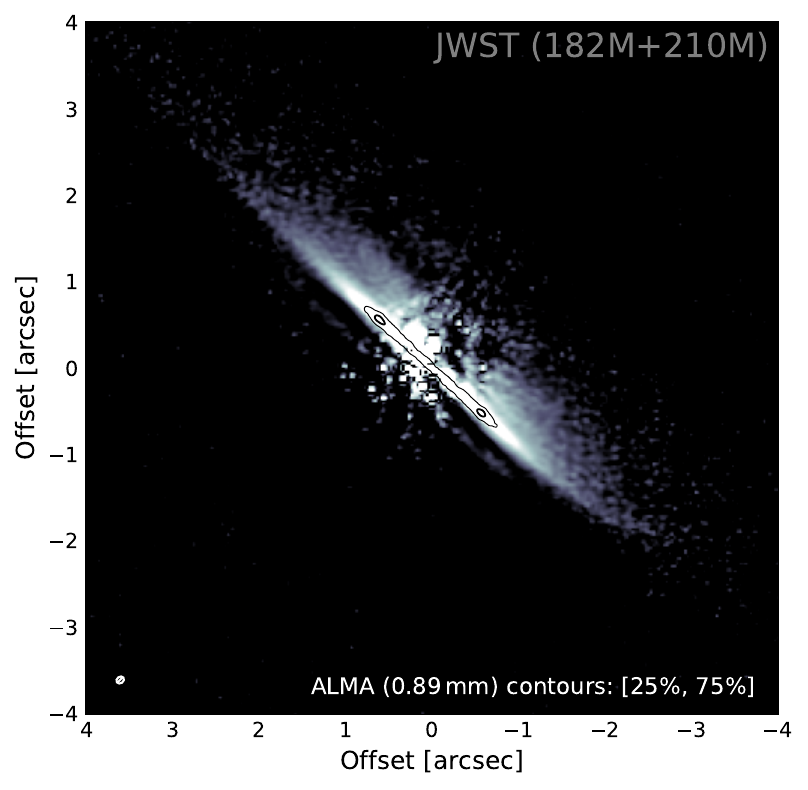}
    \caption{The JWST NIRCam SW observations (F182M and F210 combined to increase SNR) of HD~33297, with
    ALMA surface brightness contours from the ARKS observations \citep{2026A&A...705A.195M} overlaid.}
    \label{fig:hd32297_alma}
\end{figure}

Determining the surface brightness profile of the edge-on disk of HD~32297 is complicated by the coronagraphic 
transmission profile. The NIRCam coronagraph provides a continuous reduction of transmission as the center of the
occulting mask is approached. For the MASK335R round mask, the transmission reaches $\sim$50\% at $\sim 0\farcs67$
and 100\% at $\sim 1\arcsec$. The bulk of the disk emission for this system is within $1\farcs5$, therefore we must
factor in the losses due to the coronagraphic mask to establish more accurate surface brightness measurements. An
additional complication arises from the convolution of this compact disk with the broad and wavelength dependent
JWST PSF, which is clearly visible in the parallel bands above and below the disk mid-plane, tracing the 
first Airy-ring. The FWHM of the PSF varies by a factor of 2.3 between our shortest and longest wavelength observations,
considerably expanding the apparent width of the compact disk. We produced deconvolved images at each observed 
orientation, and then de-rotated and combined them. In Figure \ref{fig:hd32297_gallery}, we also show the 
deconvolved images for HD~32297.  The procedure reveals the true thin nature of the disk, as seen previously 
for example with HST/STIS \citep{schneider14} and ALMA \citep{2026A&A...705A.196H}. In Figure \ref{fig:hd32297_alma}, we show
the ALMA surface brightness contours from the ARKS survey, overlaid on
a combined F182M+F210M NIRCam image.

\begin{figure}[!t]
    \centering
    \includegraphics[width=0.99\linewidth]{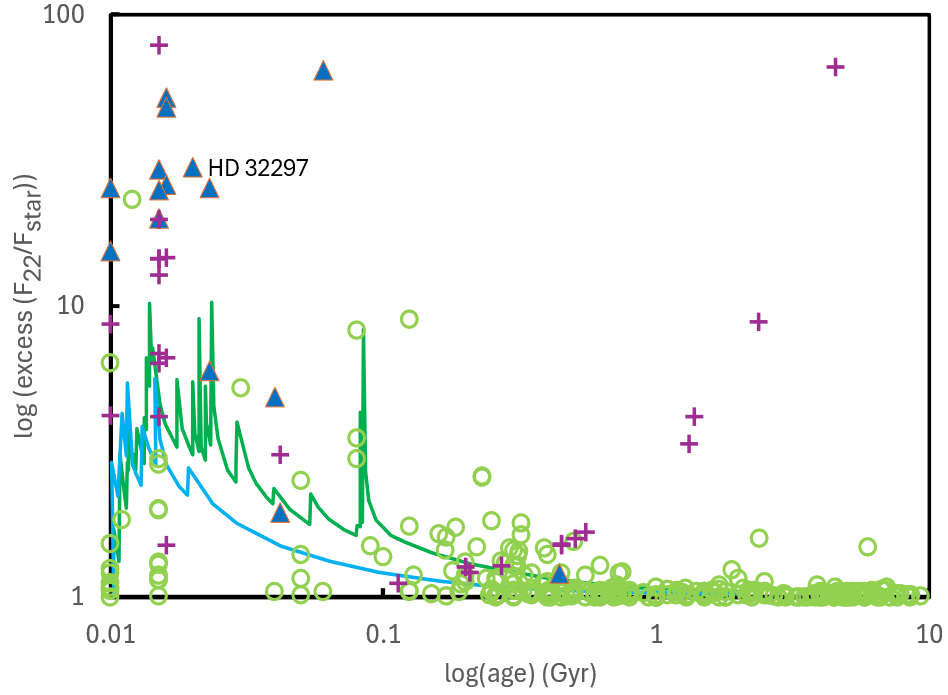}
    \caption{Similar to Figure~\ref{fig:overview}, except now we plot the CO-detected systems  
    (filled blue triangles, from \citet{macmanamon2026}). We also show the systems with demanding CO upper limits  
    $<$ 100 mJy km s$^{-1}$ as pluses. Ages are from \citet{sierchio14,gaspar13,macmanamon2026}, and 
    22 $\mu$m excesses are estimated from 2MASS and WISE or MIPS.}
    \label{fig:codetections}
\end{figure}

We measure the disk surface brightness across the midplane by finding the location of the midplane via fitting
a Gaussian to pixel values in $0\farcs2$ wide bands across
(see Figure \ref{fig:hd32297_flares}), and averaging the flux values in a 3$\times$3 pixel centered on
the peak locations. In Figure \ref{fig:hd32297_radial}, we show the measured surface brightness for both the 
observed and deconvolved images, highlighting the importance of performing a deconvolution for this 
particular observation. The disk brightness seemingly decreases towards the center in the observed
images, while it continues to increase in the deconvolved frames. The disk profiles in the deconvolved images are
in agreement with  the ALMA observations \citep{macgregor18}, which placed the main belt between 78.5 and 122 au. While the 
deconvolved NIRCam observations increase in brightness within $0\farcs3$, due to the uncertain nature of
the deconvolution algorithm as it addresses bright speckle residuals, we will not analyze the signal within
this region.

\begin{table}[!t]
\setlength{\tabcolsep}{1pt}
\begin{center}
\caption{HD~32297 Photometry\label{tab:hd32297phot}}
\begin{tabular}{lcccccc}
\hline\hline
                   & F182M  & F210M & F250M & F300M & F335M & F444W \\
\hline
F$_{\nu, \star}$\tablenotemark{a} (mJy) & 814  & 652 & 474 & 356  & 285  & 171 \\
\hline
F ($\mu$Jy)         & 704 & 619 & 438 & 273 & 282 & 213  \\
$\sigma$ ($\mu$Jy)  & 64 & 52 & 64 & 38 & 41 & 27  \\
\hline
\end{tabular}
\end{center}
\tablecomments{Photometry of the HD~32297 main belt disk 
($80 - 130$~au; $0\farcs6 - 1^{\prime\prime}$), measured within a disk width of $0\farcs15$.}
\tablenotetext{a}{Stellar photosphere values.}
\end{table}

We measure the normalized reflectance spectra of the disk using the deconvolved images in the main belt 
(80-130 au; $0\farcs6$-$1^{\prime\prime}$), within a disk width of $0\farcs15$ (see Table \ref{tab:hd32297phot} for photometric values).
Photometric error estimates are given using the methodology described in Appendix \ref{app:A_pt_3}. Given the edge-on orientation of the disk, our line-of-sight
measurements through the disk mid-plane naturally mix fluxes from various domains. Due to these effects, 
the flux measurements were not scaled by $R^2$. In Figure \ref{fig:hd32297_spec}, we show these measurements, compared
to various Solar System objects, like we did for HD~10647. 
The characteristic water ice  (or icy tholin \citep{brown2023}) dip at 3 $\mu$m is only marginally present. The scattering 
is dominated by materials that are relatively featureless over the range of our imaging, at least at our low effective spectral resolution. 

\begin{figure*}[!t]
    \centering
    \includegraphics[width=0.7877\linewidth]{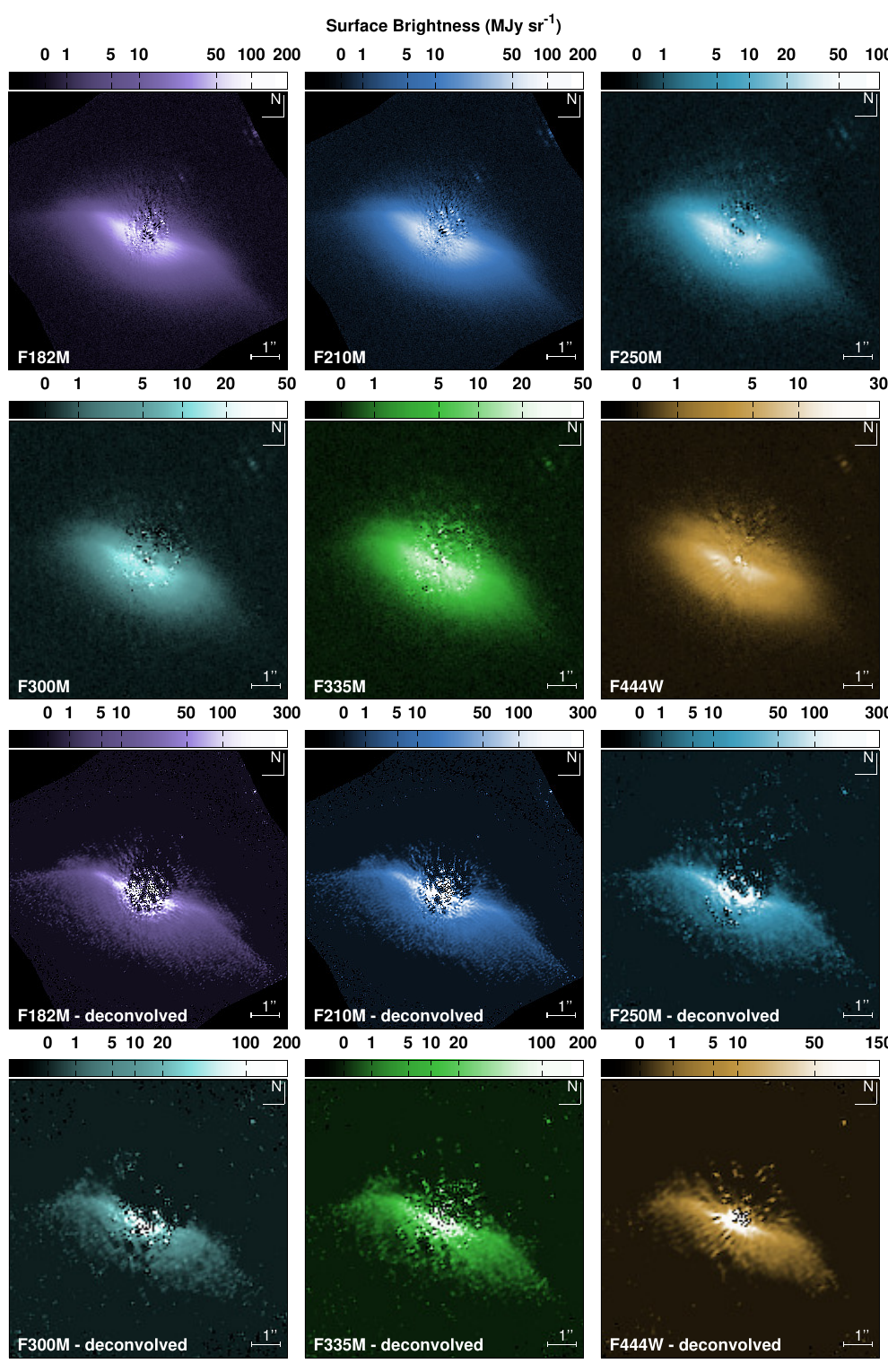}
    \caption{The six-color observations of the HD~61005 system. Images are displayed in logarithmic scaling with levels displayed
    for each filter in their respective colorbars.}
    \label{fig:hd61005_gallery}
\end{figure*}

The morphology of this system has been modeled by \citet{jones2023}. They favor a giant impact as the primary cause of the morphology. 
However, they are unable to account accurately for the tilts of the wings, their secondary wing is brighter than observed, and the 
model does not yield the kink in the primary wing. They also do not reproduce the bright central disk seen with ALMA and in our images 
in Figure~\ref{fig:hd32297_gallery}. However, they are optimistic that reasonable refinements of their model 
(e.g., some level of interaction with the ISM) can resolve the remaining discrepancies.

Figure~\ref{fig:codetections} shows the debris disk excesses 
above the stellar photospheres at 22 or 24 $\mu$m vs age, highlighting disks with gaseous $^{12}$CO(J=3-2) detections and deep non-detections. The detections tend to lie in the very strongest excesses and among young systems, and 
HD~32297 is among them as shown.  This poses an attractive scenario in which large, volatile-rich protoplanets have collided to produce 
the profuse grain system, and the release of the CO gas accompanies that of icy grains. One might, therefore, expect the grains around 
this star to be icy. However, Figure~\ref{fig:hd32297_spec} shows it has only subtle hints of the 3 $\mu$m absorption due to ice. For 
this system, the attractive scenario where the extreme debris disk behavior results from breakup of a volatile object that contributes 
an enormous population of icy grains seems not to be the case. It can be contrasted with HD~61005 (next section), which shows a 
definitive ice absorption, although it is undetected in CO bands.

\subsection{HD~61005}
\subsubsection{Background}

HD~61005 is a G8V star \citep{gray2006} located at 36.5 pc \citep{bailer2018}. Its age has been estimated to be 
$\sim$ 100 Myr using multiple standard methods, e.g., Ca II HK \citep{henry1996,white2007,  schroder2009, isaacson2010}. 
Ages of $\sim$ 40 Myr have also been proposed from its rotation rate \citep{weise2010} and from a likely membership 
in the Argus Association \citep{desidera2011}. We plot it at 60 Myr in Figure~\ref{fig:overview}. The system was 
identified as having a prominent debris disk in the FEPS
program \citep{hines2007} and independently from IRAS data \citep{rhee2007}.

\citet{hines2007} also presented a NICMOS coronagraphic image revealing a highly asymmetric debris disk,
which they dubbed ``the Moth.'' They suggested that interaction of the circumstellar debris disk with the 
ISM might be responsible for the unusual extended features of the system. More detailed imaging plus polarimetry 
obtained with HST/ACS were reported by \citet{maness2009}. They were able to separate the image into a nearly 
edge-on disk centered on the star and dust being stripped from this disk as the system passes through an 
interstellar cloud. The stripping scenario has been confirmed theoretically by \citet{pastor2017}. 

\citet{olofsson2016} reported images and polarimetry with VLT/SPHERE in a number of near-infrared bands. The high 
resolution of the SPHERE images definitively confirmed the presence of a nearly edge-on debris disk. They also 
obtained an image with ALMA at 1.4 mm, which is of lower resolution but appears to be dominated by this edge-on disk.  
They model the images and the SED, resulting in estimates that the minimum grain size in the disk is 
$\sim$ 1.9 $\mu$m, with a steep size distribution with power law index 3.84. They suggest that the grains are 
porous and predominantly amorphous silicates (e.g., astrosil) with a small addition of water ice, from modeling 
the SED. The disk appears more extended in the near infrared than with ALMA, which they suggest reflects the effects
of radiation pressure on the smaller grains. \citet{macgregor18} find consistent results with a higher resolution 
ALMA image. Their model indicates that the debris system has a broad belt between 42 and 67 au with a ``halo'' (still
thin and in the disk plane) extending further out. The latest ARKS results \citep{2026A&A...705A.196H} observe this halo
out to $\sim 126$ au.
\citet{esposito2016} find a similar belt inner edge from angular 
differential imaging in the near infrared. It is unclear whether the halo is part of the wings of the ``Moth'' or an extension
of the planetesimal disk, although the straightness of the contours suggests the latter. \citet{macgregor18} put
the disk inclination at $85.6^\circ \pm 0.1^\circ$.  \citet{absil2021} use interferometry in H-band (1.6 $\mu$m to 
find an excess over the photosphere of the star of $0.81 \pm 0.12$ \%. 

As with HD~10647 and HD~32297, ARKS analysis of HD~61005 was presented in both \citet{2026A&A...705A.196H} and \citet{2026A&A...705A.197Z}. 
Those studies suggest that HD~61005 hosts a relatively narrow belt with a narrow vertical distribution and a halo, that can be well-modelled as a double Gaussian radial distribution with best-fit radii and widths ($R_i {\pm} \sigma_i$) of $69.7{\pm}7.3\,$au and $94.0{\pm}32.3\,$au \citep{2026A&A...705A.196H}, and either a single vertical Lorentzian profile with vertical aspect ratio $h_{\rm HWHM}{=}0.0143$, or a double Gaussian vertical profile with vertical aspect ratios of $h_{\rm HWHM1}{=}0.014$ and $h_{\rm HWHM2}{=}0.117$ corresponding to the two Gaussian distributions respectively \citep{2026A&A...705A.197Z}.
\citet{2026A&A...705A.197Z} further show that if there are two vertical dust populations, that the dust population with the lower aspect ratio (the dynamically cooler component) contains ${\approx}90\%$ of the disk mass, with the other ${\approx}10\%$ made up by the higher aspect ratio (dynamically hotter component).
Unlike the other sources in this work that were analysed by ARKS, no planet constraints are placed on this system based on the emission morphologies.

The asymmetries in the HD~61005 disk have caused a number of investigators to suggest that it might be the site of
a massive planetesimal collision \citep{esposito2016, olofsson2016, macgregor18}. However, the detailed modeling of \citet{jones2023} cannot find a fit to its morphology under that assumption and prefers the hypothesis that its extended structure arises through interaction with the interstellar medium.
HD~61005 is assessed as asymmetric based on its ARKS data in \citet{2026A&A...705A.200L}, with a significant enhancement in the integrated sub-millimeter emission to the south of the disk, i.e. on the same side that the scattered–light halo extends.
\citet{2026A&A...705A.200L} remained agnostic as to the cause of the asymmetry.

\begin{figure}[!t]
    \centering
    \includegraphics[width=0.99\linewidth]{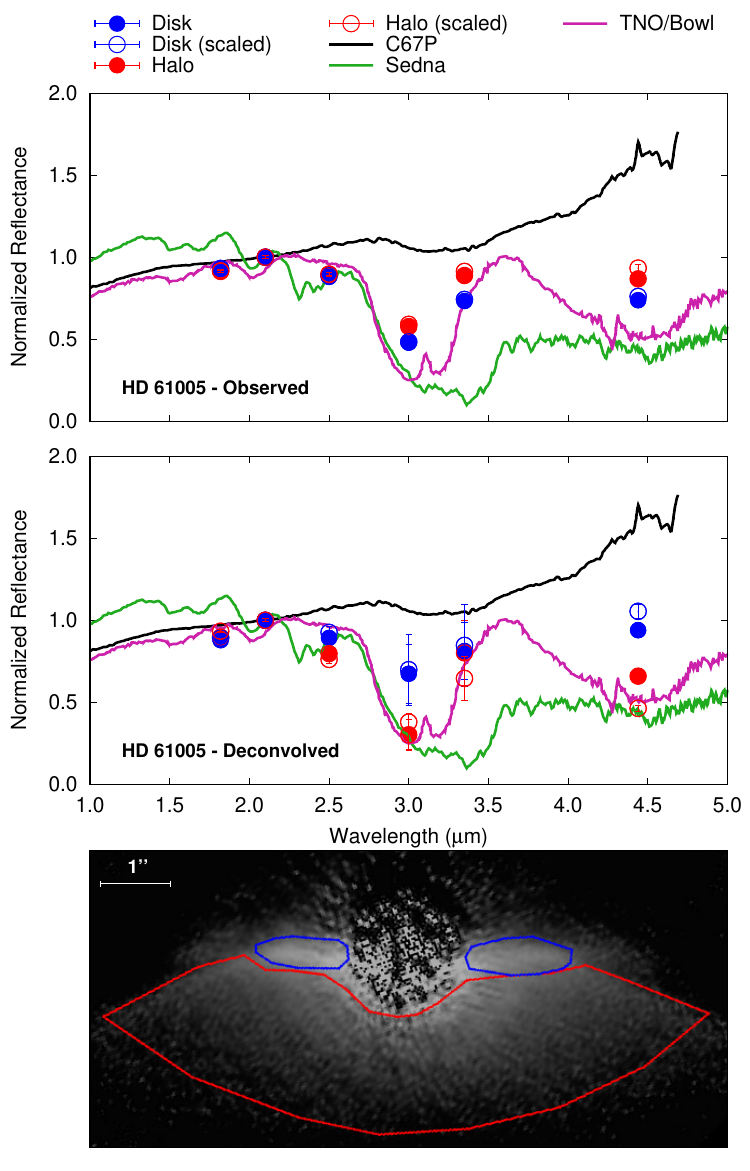}
    \caption{The normalized (at 2.1 $\micron$) and deprojected $R^2$ scaled reflectance spectra of 
    HD~61005, determined for the main disk and the extended halo towards the South (shown in the bottom panel), measured using 
    both the observed and the deconvolved images. This system shows strong water absorption at 
    3 $\micron$, with photo-spectra resembling TNO/Bowl-type objects. The deconvolution algorithm introduces 
    artifacts and noise patterns that complicate the analysis of the reflectance spectra, therefore we present 
    measurements using both reductions. The reflectance behavior in the 4.4 $\mu$m band for the 
    halo, in the deconvolved image, resembles that of water-rich grains, e.g., the rings of Saturn or of Sedna, 
    but in the disk there is a different grain composition indicated by significantly higher reflectance. 
    }
    \label{fig:hd61005_spectra}
\end{figure}

\begin{figure}[!t]
    \centering
    \includegraphics[width=0.99\linewidth]{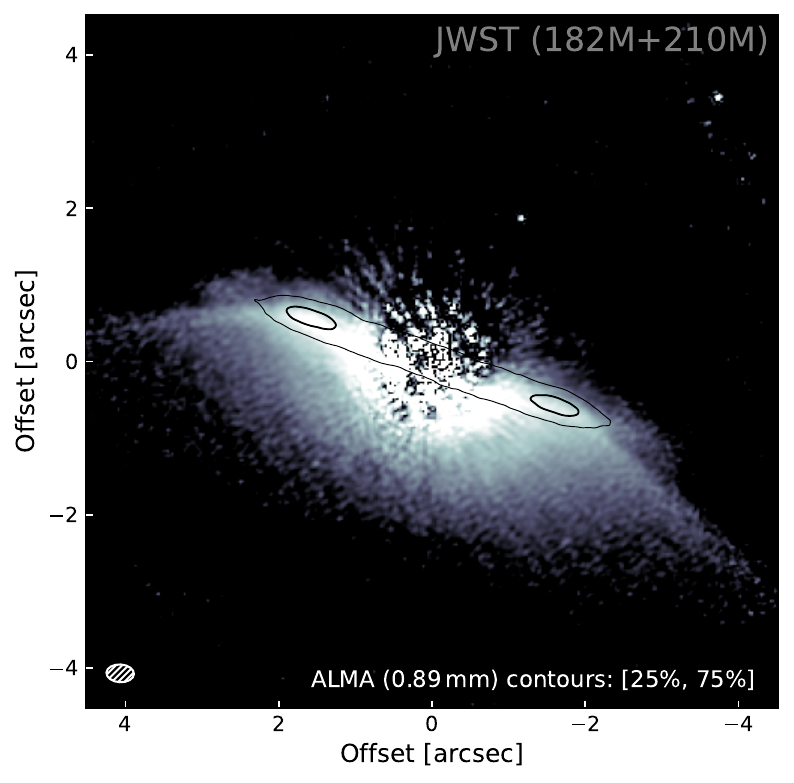}
    \caption{Contours from the ALMA observations obtained within the ARKS program \citep{2026A&A...705A.195M}, overlaid
    on the combined F182M and F210M JWST/NIRCam observations for HD~61005.}
    \label{fig:hd61005_alma}
\end{figure}

\subsubsection{JWST Results}

\begin{table}[!t]
\setlength{\tabcolsep}{2pt}
\begin{center}
\caption{HD~61005 Photometry \label{tab:hd61005phot}}
\begin{tabular}{lcccccc}
\hline\hline
                   & F182M  & F210M & F250M & F300M & F335M & F444W \\
\hline
F$_{\nu,\star}$ (Jy)\tablenotemark{a} & 2.24 & 1.78 & 1.32 & 0.98 & 0.80 & 0.45 \\
\hline
\multicolumn{7}{c}{Disk - deconvolved}\\
\hline
F$_{\rm sc}$ ($\mu$Jy)      & 1304 & 1161 & 797 & 446 & 442 & 309 \\
$\sigma_{\rm sc}$ ($\mu$Jy) & 19   & 18   & 31  & 138 & 130 & 12  \\
\hline
F ($\mu$Jy)                 & 1200 & 1085 & 718 & 403 & 396 & 258 \\
$\sigma$ ($\mu$Jy)          & 19   & 19   & 32  & 107 & 84 & 12   \\
\hline
\multicolumn{7}{c}{Halo - deconvolved}\\
\hline
F$_{\rm sc}$ (mJy)          & 49.46 & 42.12 & 23.80 & 8.79 & 12.22 & 4.93 \\
$\sigma_{\rm sc}$ (mJy)     & 0.49  & 0.36  & 0.89  & 1.17 & 2.59  & 0.20 \\
\hline
F ($\mu$Jy)                 & 1855 & 1642 & 970 & 273 & 592 & 274 \\
$\sigma$ ($\mu$Jy)          & 18   & 13   & 34  & 84  & 145 & 8   \\
\hline

\multicolumn{7}{c}{Disk - observed}\\
\hline
F$_{\rm sc}$ ($\mu$Jy)      & 859 & 733 & 482 & 195 & 245 & 141 \\
$\sigma_{\rm sc}$ ($\mu$Jy) & 6.6 & 5.7 & 6.4 & 3.8 & 4.2 & 2.2 \\
\hline
F ($\mu$Jy)                 & 776 & 664 & 433 & 175 & 218 & 124 \\
$\sigma$ ($\mu$Jy)          & 6.5 & 5.5 & 6.6 & 4.1 & 4.4 & 2.1 \\
\hline
\multicolumn{7}{c}{Halo - observed}\\
\hline
F$_{\rm sc}$ (mJy)          & 64.1  & 55.5 & 36.8 & 18.0 & 22.8 & 13.1 \\
$\sigma_{\rm sc}$ (mJy)     & 0.65  & 0.57 & 0.42 & 0.33 & 0.35 & 0.30 \\
\hline
F ($\mu$Jy)                 & 2099 & 1827 & 1216 & 580  & 730  & 401 \\
$\sigma$ ($\mu$Jy)          & 17.1 & 15.5 & 13.7 & 10.1 & 10.2 & 7.1 \\
\hline
\end{tabular}
\end{center}
\tablecomments{Photometry of the HD~61005 disk and halo. 
The ``{\it sc}'' scatter values are scaled by $(R/R_0)^2$, where $R_0$ 
is set at 70 au.}
\tablenotetext{a}{Stellar photosphere values.}
\end{table}

Our NIRCam observations, shown in Figure \ref{fig:hd61005_gallery}, reveal the ``Moth'' and its extended dust 
structure at all six wavelengths. The disk (or its extended halo), extending further out from the break-off 
point of the Moth's wing, can be traced at 1.82, 2.1, and even 2.5 $\mu$m out to $\sim$ 3'' (110 au), with 
the feature more pronounced in the deconvolved images. The wings extend for $\sim 3\farcs4$ (120 au) from the 
break-off point, which itself is located at $\sim 1\farcs4$ (51.1 au) from the star, near the center 
of the main belt, as seen by ALMA \citep{macgregor18}. The break-off point also coincides with the fiducial radius
of the disk model obtained from our MCRDI fit ($r_0 = 49.91$ au) during our tests with Winnie 
(see Table \ref{tab:HD61005mcrdi}). The results would indicate that the dust particles present within the wings
originate from the main belt and that the dust size distributions differ in the extended halo within the plane of the
disk and that in the wings.

To compare the composition and dust sizes present in the dust populations within the disk and wing/halo, 
we measured the normalized reflectance spectra of each component, scaled by the deprojected $R^2$ values,
as well as using unscaled values. Due to the increased speckle noise at longer wavelengths for this system when
performing the deconvolution, we measure the spectra for both observed and deconvolved images.
In Figure \ref{fig:hd61005_spectra}, we show these measurements compared to various solar system objects. 
Error calculations for the photometry followed the description in Appendix \ref{app:A_pt_3}.
The system reveals an absorption feature at 3 $\mu$m in both components with both processing methods. 
At 4.4 $\mu$m, we prefer the deconvolved image because at the lower resolution, the components are less-well 
separated in the observed one.
In the halo, the downturn in reflectance at 4.5 $\micron$ is as expected for water ice. However, in the disk 
the relatively higher reflectance at this wavelength, combined with the deep 3 $\micron$ feature, is suggestive 
of a mixture of ice and processed organic material (e.g., tholin-like components; \citealt{brown2023}).

Collisional processing in the parent belt can 
produce small fragments that are subsequently transported outward, where they contribute to the observed ``wings'' 
as they interact with the surrounding ISM. These observations are consistent with an ISM-interaction scenario for 
the formation of the wing/halo, which would selectively place smaller  -- blowout size ($\sim \mu$m) -- particles either
on bound ($e<1$) or unbound ($e\ge1$) trajectories. Recent Chandra/ACIS-S observations of the stellar atmosphere 
\citep{2026ApJ...999..125L} are also consistent with the ISM interaction model. This scenario naturally produces a spatial
segregation of grain sizes, such that the halo is dominated by smaller grains, which could enhance or reveal spectral 
features that are muted in the larger-grain population of the main disk.
Unfortunately, the nearly edge-on inclination of 
the system precludes scattering phase function studies in the HD~61005 system.

Finally, in Figure \ref{fig:hd61005_alma}, we show brightness contour levels measured in the ALMA ARKS observations
\citep{2026A&A...705A.195M}, overlaid on the combined F182M+F210M short wavelength JWST/NIRCam observations.
The scattered light disk profile bifurcates from the location of the ALMA planetesimal ring, with the halo
component extending outward from the disk plane \citep[called the ``vestigal wing'' by][]{jones2023}, 
while the ``wings of the moth'' reveal a sharp edge emanating from the same planetesimal ring towards the south-east. 

\subsection{HD~107146}

\begin{figure*}
    \centering
    \includegraphics[width=0.7877\linewidth]{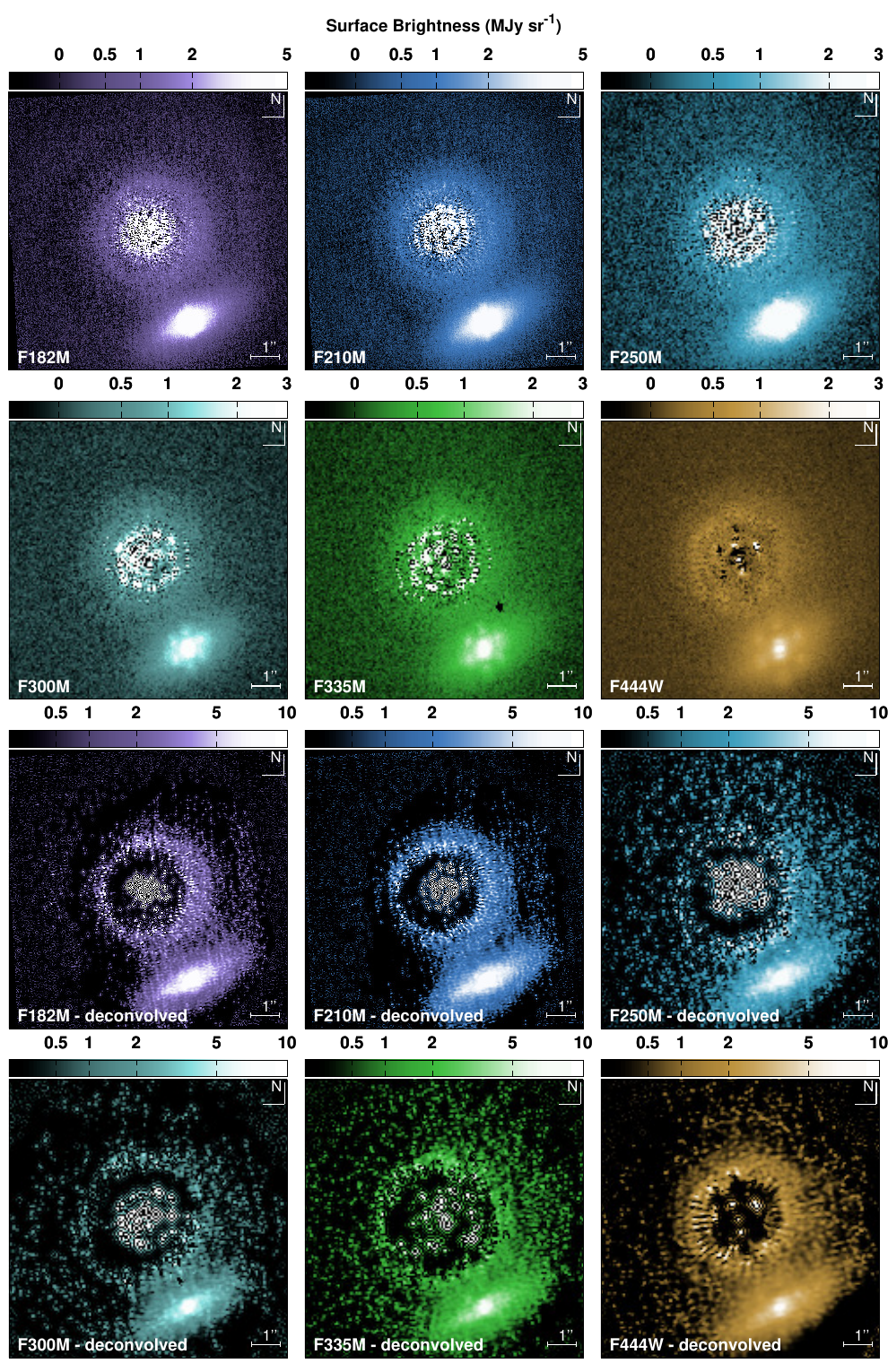}
    \caption{The six-color observations of the HD~107146 system. Images are displayed in logarithmic scaling with levels displayed
    for each filter in their respective colorbars.}
    \label{fig:hd107146_gallery}
\end{figure*}

\subsubsection{Background}

HD~107146 is a G2V star \citep{harlan1970,white2007} at a distance of 27.44 pc \citep{bailer2021}. There 
is a range of age estimates; we adopt 100 $-$ 300 Myr from \citet{isaacson2010,white2007,stanford2020, matra2025} 
and X-ray data from ROSAT (see \citet{sierchio14}). The metallicity of the star, [Fe/H] = -0.04 \citep{gaspar2016}, is close to the solar value. 

\begin{figure*}[!t]
    \centering
    \includegraphics[width=0.99\linewidth]{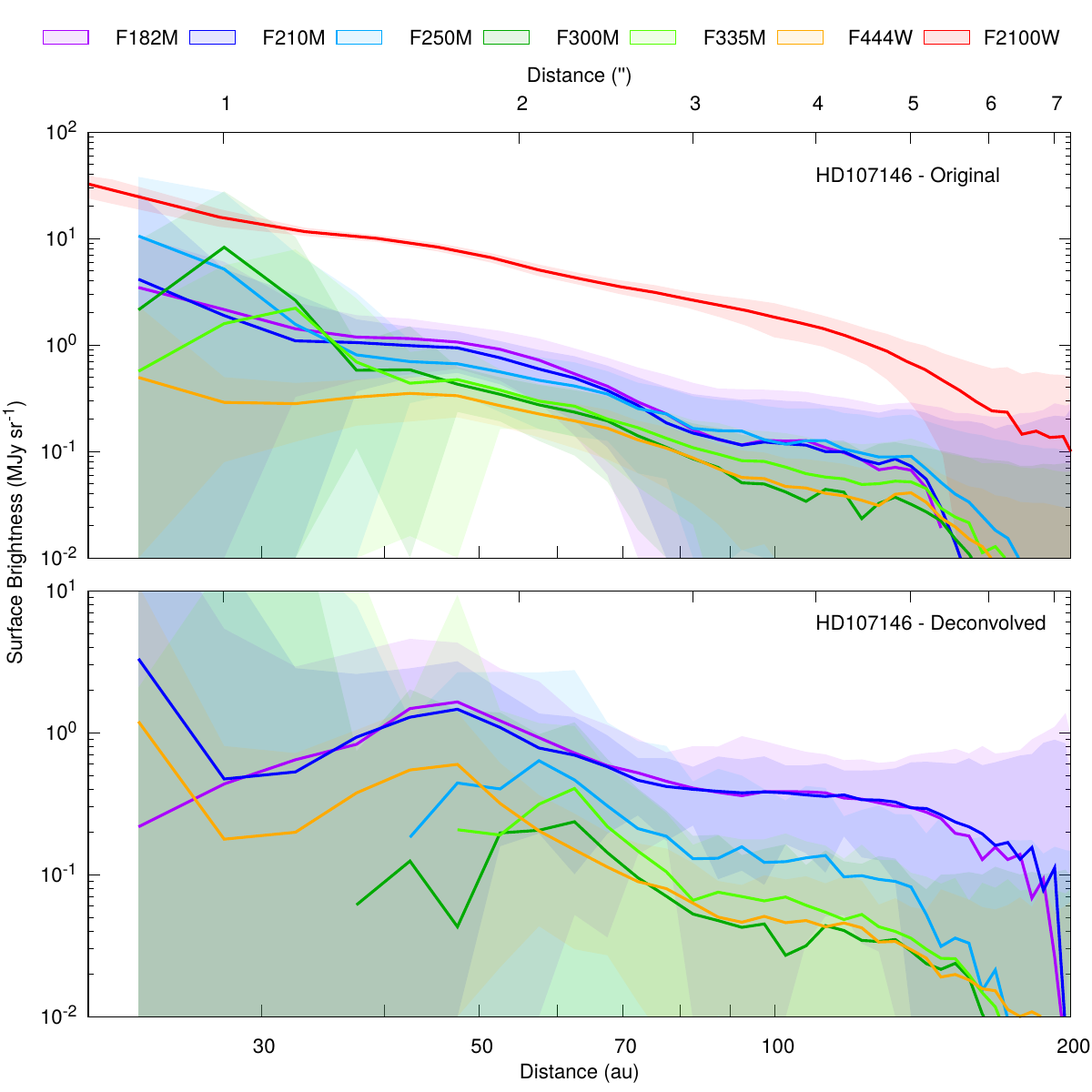}
    \caption{The radial surface brightness profile of the HD~107146 system for 
    the observed and deconvolved images. The two belts, located
    between 35-55 and 90-130 au are identified at all wavelengths, most clearly at the 
    the shorter wavelengths though. The relatively lower reflectance at 3.0 $\micron$, likely
    due to the presence of water-ice, is noticeable throughout the disk. The background galaxy was
    masked for the extraction of the median surface brightness values.}
    \label{fig:hd107146_SB}
\end{figure*}

An excess in this star at 60 $\mu$m is apparent in IRAS data \citep{abrahamyan2015}. However attention was 
not drawn to this until significantly later \citep{silverstone00}.  The SED was observed through the 
sub-mm by \citet{williams2004}. It demonstrates that the fractional luminosity of the disk is 
relatively large, $L_{\rm disk}/L_{\ast}\sim 1 \times 10^{-3}$ \citep{thebault2023,matra2025}.

The debris disk is one of the first to be imaged in scattered light \citep[with HST/ACS;][]{ardila2004,ertel11}. 
The system is ``infamous'' for having a bright background galaxy in its close proximity, first identified in the 2004 HST observations; the background galaxy
is currently passing behind the outer ring of the system. The disk is nearly face-on and therefore of low surface 
brightness, making ground-based followup imaging at optical/near-IR
wavelengths a challenge \citep{esposito2020,engler2025}.
The ALMA image \citep{marino2018,matra2025,2026A&A...705A.195M,2026A&A...705A.196H} provides a baseline for the 
ring structure of the system parent bodies. Initial analysis of ALMA observations revealed two belts 
(40-60 and 100-140 au) with a single gap in the system \citep{marino2018}. These ALMA observations also revealed
extended emission interior to these regions with a peak at 20 au. A mixed non-parametric and parametric modeling of the archival ALMA 
dataset \citep{2023MNRAS.522.6150I} resolved the broad gap structure as consisting of two narrower gaps, 
located at $56\pm1$ and $79\pm1$ au, therefore indicating a third narrower belt between
56 and 79 au.

\cite{2026A&A...705A.196H} analyzed new high resolution  data within the 
ALMA ARKS survey, separately fitting both non-parametric and parametric models to the dataset. Two non-parametric
and the parametric models all favor a three ring architecture for the system. 
\cite{2026A&A...705A.196H} determine the locations of these three rings
at $46.2^{+0.3}_{-0.5}$, $66.3^{+0.6}_{-0.7}$, and $118.07^{+0.12}_{-0.11}$ au,
with widths of $13.2^{+0.9}_{-1.2}$, $8.2^{+2.4}_{-1.9}$ and 
$58.7^{+0.6}_{-.4}$ au. This same study reports an upper limit of 
5 M$_{\jupiter}$ for a planet that could sculpt the inner edge of the disk,
orbiting at 32 au. 
The two non–parametrically modelled gap locations at 57.3\,au and 79.2\,au are also shown to be consistent with embedded planets with masses of $0.11\,M_J$ and $0.15\,M_J$ respectively.
While the disk structure strongly suggests that the gaps have
been cleared by at least one planet -- either embedded within the disk \citep[e.g.,][]{Friebe2022} or orbiting
completely interior to the disk's innermost edge \citep[e.g.,][]{yelverton18, Sefilian2021, Sefilian2023} -- none
have been detected so far. Upper limits on planets in this system, in the 1 - 3 M$_{\jupiter}$ range, 
have been placed by \citet{bendahan2026} and \cite{milli26} by combining direct imaging constraints 
with existing theoretical studies of planet--debris disk interactions. 

The HD~107146 system was also identified as asymmetric in the ARKS data by \citet{2026A&A...705A.200L}, who focused on ARKS-reprocessed ALMA Band 6 data. 
There it was shown by radially averaging that an azimuthal ‘arc’ asymmetry is present in the disk, which is most significant in the inner ring, but present too in the outer ring, with emission peaking at an angle of ${\approx}355^\circ$ (relative to $0^\circ$ due north). It is postulated that in addition to the proper motion offset \citep[shown in ARKS I;]{2026A&A...705A.195M} that this could be due to planet–disk interactions. Further discussion is presented (without modelling efforts) that describe the possibility that although a significant stellocentric offset was not found in the data, this disk could also be eccentric, with the azimuthal emission profile tracing density enhancements in an underlying eccentric ring \citep[as per the models of][]{2022MNRAS.510.2538L}.

\begin{figure}
    \centering
    \includegraphics[width=0.95\linewidth]{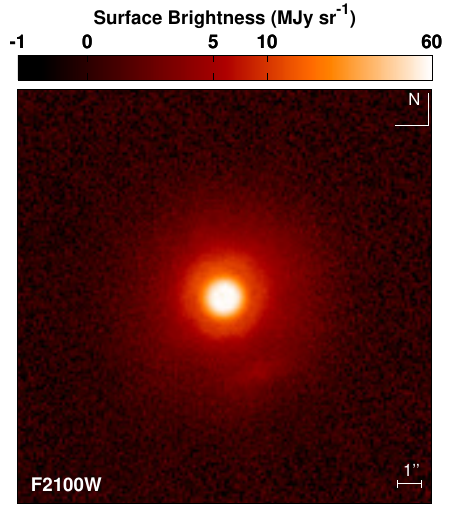}
    \caption{The MIRI F2100W observation of the HD~107146 system, shown
    in logarithmic scaling with the corresponding intensity levels indicated in the colorbar.}  
    \label{fig:hd107146_miri}
\end{figure}

 A substantially improved HST image, using the STIS coronagraph, is presented by \citet{schneider14}. 
 The image is dominated by a broad ring of similar dimensions to the ALMA image. The ring is brighter 
 to the SW, reflecting the  scattering properties of the grains and indicating that the ring is 
 probably inclined toward us in the SW. There is also a halo extending outward from the ring, likely
 to be scattered light from grains pushed into extreme elliptical orbits and/or escape trajectories 
 by radiation pressure. However, the drop-off in brightness of this structure is faster than 
 predicted by existing theoretical models and may be shaped by interactions with more distant bodies \citep{thebault2023}.

\subsubsection{JWST Results}

\begin{figure}[!t]
    \centering
    \includegraphics[width=0.99\linewidth]{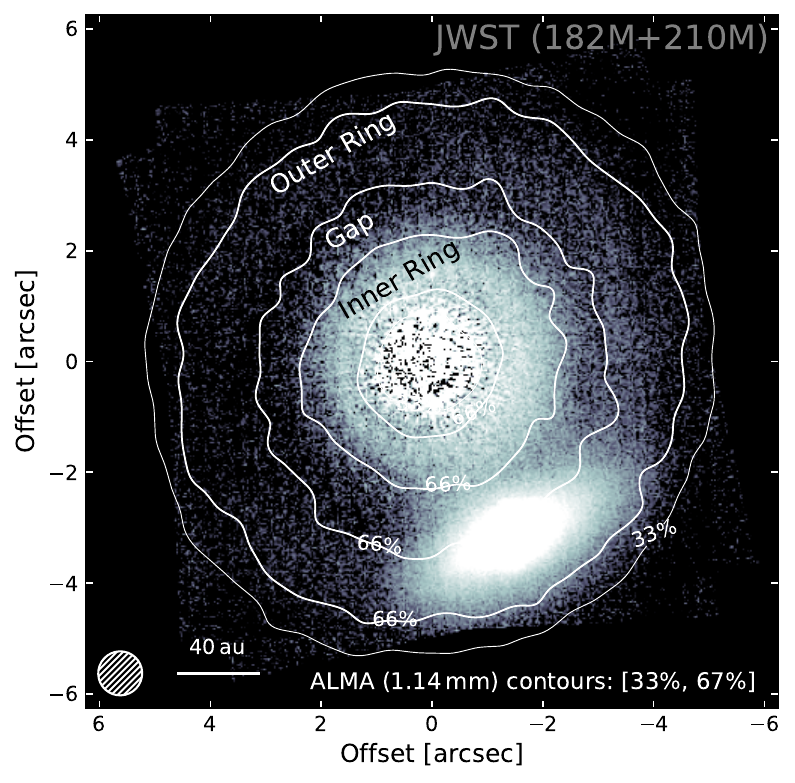}
    \caption{Contours from the ALMA observations obtained within the ARKS program \citep{2026A&A...705A.195M}, overlaid
    on the combined F182M and F210M JWST/NIRCam observations for HD~107146.}
    \label{fig:hd107146_alma}
\end{figure}

We present our six-color NIRCam images -- observed and deconvolved -- in Figure \ref{fig:hd107146_gallery}. The inner ring,
extending from $1\farcs3$ - $2\farcs0$ ($35-55$ au) is the brightest component at all NIRCam wavelengths and coincides with
the ALMA ring detected in the inner regions between $33-59$ au. This inner component is clearly extended towards the west 
direction at shorter wavelengths, up to $2\farcs35$ (64 au). While this component was detected in the HST/STIS images
\citep{schneider14}, it was not determined to be of astrophysical origin, due to circular artifacts at small inner 
working angles being easily produced via breathing within the HST optical train assembly and slight PSF color-mismatches. 
The intermediate belt at 66 au retrieved from the ARKS ALMA data via parametric modeling \citep{2026A&A...705A.196H} 
coincides with the azimuthally asymmetric extended outer edge of the inner disk in our NIRCam dataset. 
The outer ring is detected in the NIRCam observation, at very faint levels, with an inner edge at $\sim 3\farcs2$
(90 au) and an outer edge at $\sim 4\farcs8$ (130 au), in good agreement with the HST/STIS observations \citep{schneider14}
as well as the location where the bulk of the ALMA emission (see Figure \ref{fig:hd107146_alma}) originates \citep{marino2018}. 

In Figure \ref{fig:hd107146_SB}, we show the radial surface brightness profiles of the system as a function of 
deprojected disk radii at all six observed filters for both the observed and deprojected images. Estimating the sky background
levels is difficult for the two short wavelength filters, given the small FOV ($\sim 10^{\prime\prime}$) of the SUB320 
subarray and the spatially extended nature of the disk ($\sim 5^{\prime\prime}$), filling the entire subarray. 
Following deconvolution, the short wavelength filter images had negative background values; therefore, we applied individual
offsets to level the backgrounds to zero (see Figure \ref{fig:hd107146_SB} caption for details). The flux of the system
drops at wavelengths above 3 $\mu$m at all radii. The two main belts are also clearly resolved with MIRI with the F2100W
filter. The MIRI image, shown in Figure \ref{fig:hd107146_miri}, reveals a bright residual in the image core.
As for all extended sources with possible compact inner regions \citep[see e.g.,][]{wolff25}, the question remains 
whether the subtracted PSF -- scaled to produce a result that is free of residuals -- accounts for purely photospheric
emission or possibly additional excess. To investigate this question, we determined the 21 $\micron$ photospheric values 
of both the target HD~107146 ($\sim 54.7$ mJy) and the PSF HD~111398 ($\sim 53.6$ mJy) via fitting their available optical
and near-IR photometry with Kurucz stellar models \citep{castelli03}. We confirm the estimated photospheric flux
value for the PSF by measuring its total emission in our images ($\sim 56.63$ mJy). For the MIRI post-processing, we
applied a PSF scaling of 1.03, determined via by-eye scaling, which is in agreement with the 1.02 factor given by the
fitted photospheric estimates. Based on our MIRI image processing, the bright central residual core is of astrophysical
origin, as we did not under- or over-subtract the PSF, and is likely the inner ring revealed by ALMA observations at 
20 au ($0\farcs7 \approx$ 6.6 px). While beyond the scope of the present study, this finding may also motivate revisiting dynamical models of gap formation to assess whether they can account for both the observed gaps and the inner emission, which is not typically considered in gap-carving scenarios.

\begin{figure}[!t]
    \centering
    \includegraphics[width=0.99\linewidth]{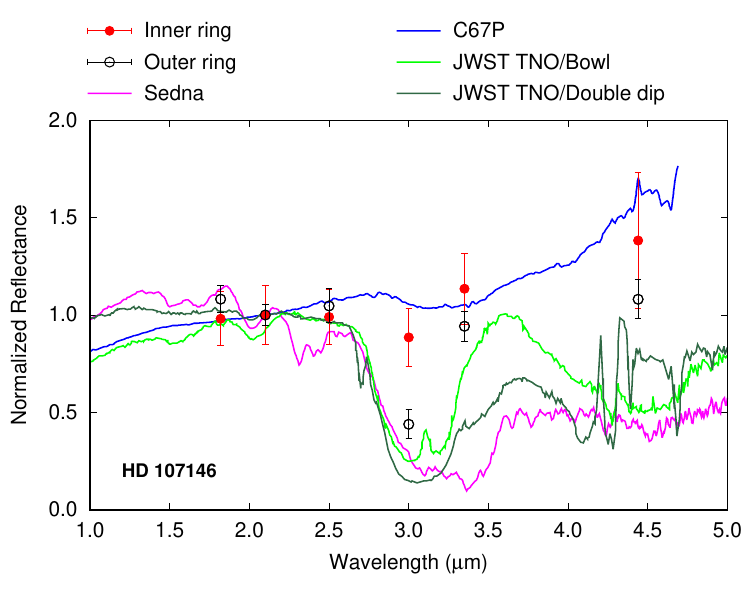}
    \caption{The normalized (at 2.1 $\micron$) and deprojected $R^2$ scaled reflectance spectra 
    of the solar-type star HD~107146, determined for the two main rings in the system, 
    measured using the observed images (not deconvolved). This system shows exceptionally strong 
    water absorption at 3 $\micron$ in the outer ring, with the photo-spectra 
    of the outer Kuiper-belt analog ring resembling that of Bowl or Double dip-type objects. The inner ring
    is similar to comet C67P. As with HD~61005, the reflectivity behavior in the 4.4 $\mu$m band for the 
    outer ring resembles that of water-rich grains, e.g., the rings of Saturn, but in the inner disk there is a 
    different grain composition indicated by significantly higher reflectance. That is, the scattering 
    grains in the inner ring are likely composed of minerals or icy tholins, whereas the outer ring appears to be dominated by  
    water-ice grains. 
    }
    \label{fig:hd107146_spectra}
\end{figure}

\begin{table}[!t]
\setlength{\tabcolsep}{2pt}
\begin{center}
\caption{HD~107146 Photometry \label{tab:hd107146phot}}
\begin{tabular}{lcccccc}
\hline\hline
                   & F182M  & F210M & F250M & F300M & F335M & F444W \\
\hline
F$_{\nu,\star}$ (Jy)\tablenotemark{a} & 5.26 & 4.24 & 3.14 & 2.32 & 1.91 & 1.07 \\
\hline
\multicolumn{7}{c}{Inner Ring ($35-55$ au)}\\
\hline
F$_{\rm sc}$ ($\mu$Jy)      & 238 & 195 & 143 & 95 & 100 & 68 \\
$\sigma_{\rm sc}$ ($\mu$Jy) & 34  & 29  & 21  & 16 & 16  & 17 \\
\hline
F ($\mu$Jy)                 & 197 & 163 & 119 & 82 & 85 & 56 \\
$\sigma$ ($\mu$Jy)          & 31  & 28  & 19  & 16 & 16 & 16 \\
\hline
\multicolumn{7}{c}{Outer Ring ($90-130$ au)}\\
\hline
F$_{\rm sc}$ ($\mu$Jy)       & 98  & 73  & 57  & 17.6 & 31.0 & 20.0 \\
$\sigma_{\rm sc}$ ($\mu$Jy)  & 6.3 & 3.8 & 4.7 & 3.0  & 2.5  & 1.9 \\
\hline
F ($\mu$Jy)                  & 96  & 72  & 56  & 17.8 & 31.0 & 20.1 \\
$\sigma$ ($\mu$Jy)           & 5.9 & 3.1 & 3.6 & 2.0  & 1.8  & 1.8  \\
\hline
\end{tabular}
\end{center}
\tablecomments{Photometry of the HD~107146 disk rings. 
The ``{\it sc}'' scatter values are scaled by $(R/R_0)^2$, where $R_0$ 
is set at 45 and 110 au, for the inner and outer ring, respectively. }
\tablenotetext{a}{Stellar photosphere values.}
\end{table}
 
Finally, we present the normalized reflectance spectra for the HD~107146 system in Figure \ref{fig:hd107146_spectra}, measured 
using the observed images (not deconvolved) appropriately scaled with the deprojected $R^2$. The background galaxy was 
conservatively masked during the measurements, taking into account its maximum extent at 4.44 $\micron$. The errors were calculated
following the methods described in Appendix \ref{app:A_pt_3}. The system reveals a very strong water-ice 
absorption feature in its outer ring, while the inner ring has a reflectance spectra more similar to inner solar system comets. 
The 3.35 and 4.44 $\micron$ emission increases compared with the shorter wavelength measurements, much like  that observed for inner solar system objects. 

\begin{figure*}
    \centering
    \includegraphics[width=0.7877\linewidth]{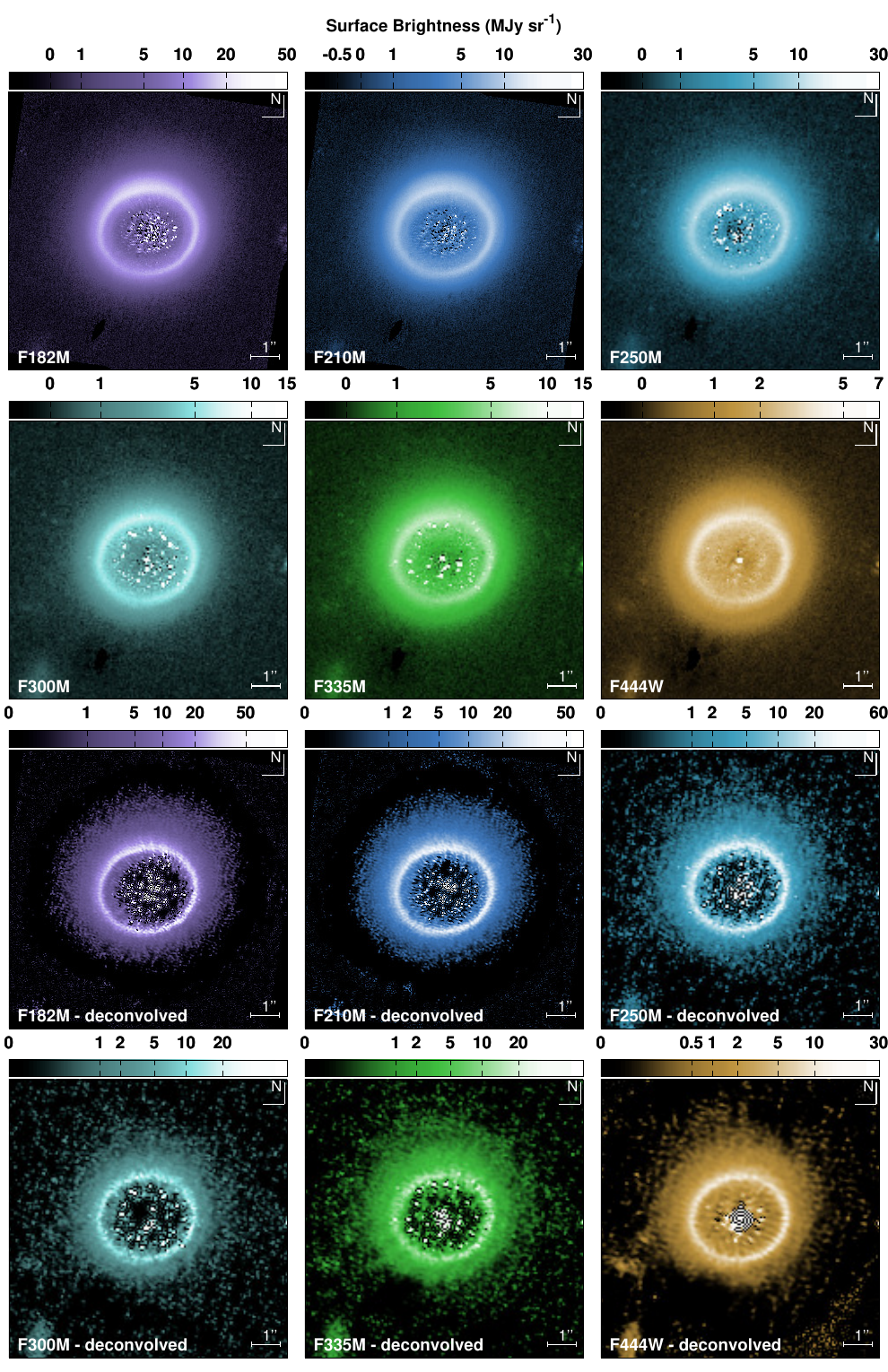}
    \caption{The six-color observations of the HD~181327 system. Images are 
    displayed in logarithmic scaling with levels displayed for each filter in their respective colorbars.}
    \label{fig:hd181327_gallery}
\end{figure*}

\subsection{HD~181327}

\begin{figure}
    \centering
    \includegraphics[width=0.99\linewidth]{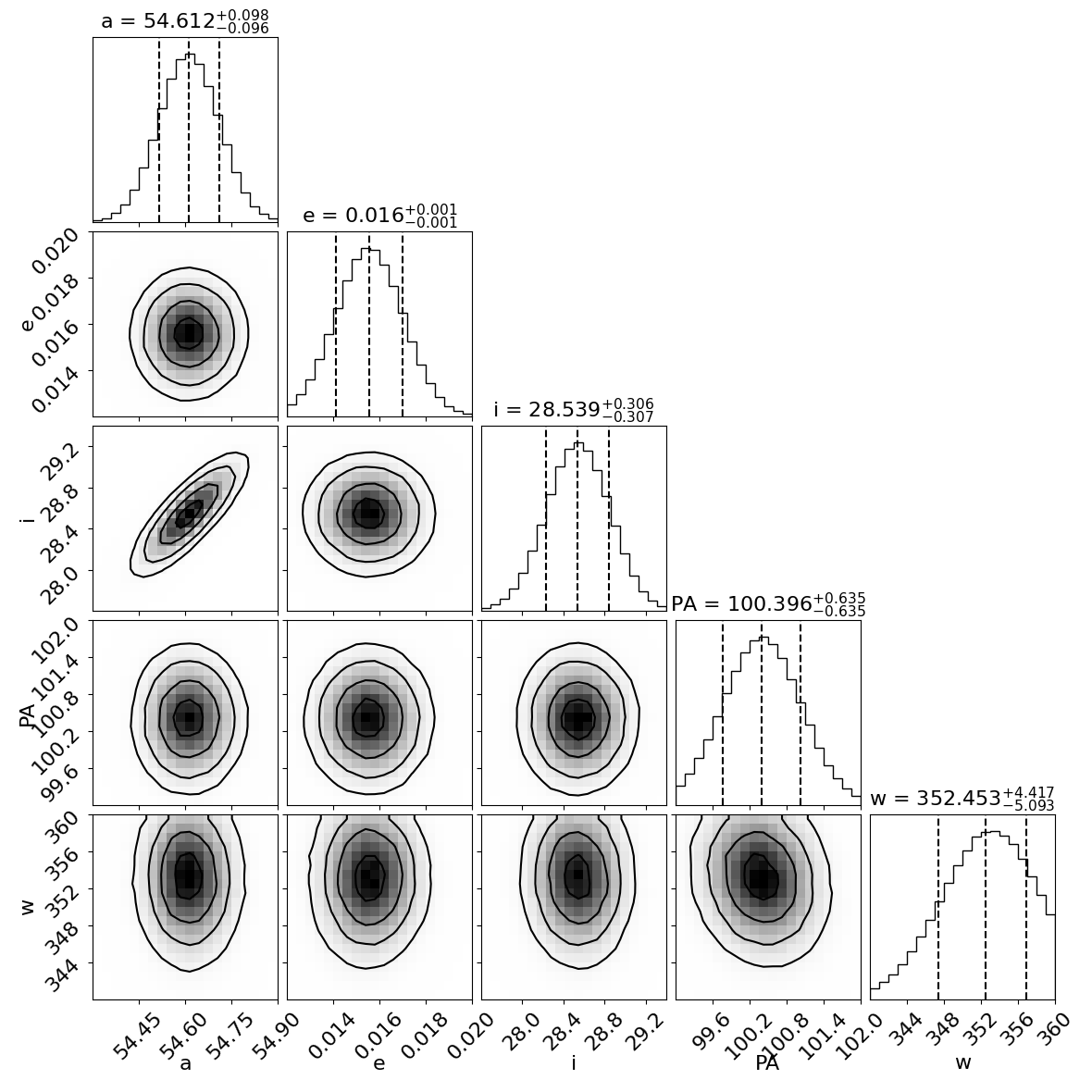}
    \caption{Orbital fits to the peak surface brightness values measured for the HD~181327 system using 
    the F182M deconvolved image. The semi-major axis (a) is in pixel values, while the orbital angles are in degrees.}
    \label{fig:hd181327orbfit}
\end{figure}

\subsubsection{Background}

HD~181327 is a F5/F6V-type star \citep{nordstrom2004, torres2006,  2021ApJS..254...42B} located at 47.78 pc \citep{gaia22} 
and identified as a member of the $\beta$ Pictoris moving group \citep{2004ARA&A..42..685Z, dasilva2009}
with a dynamical age estimated at  $18.5_{-2.4}^{+2.0}$ Myr \citep{2020AA...642A.179M} and an estimated age of $25 \pm 3$ 
Myr from lithium depletion \citep{messina2016}. It was first identified as a debris disk host based on its IRAS detection of far infrared excess emission \citep{1998ApJ...497..330M}.

The debris system is extensively studied through imaging. \citet{2006ApJ...650..414S} first resolved it with HST/ACS and 
HST/NICMOS coronagraphic observations. An improved reduction of the NICMOS image is presented by \citet{milli2024}, along with
new high resolution SPHERE/IRDIS observations (discussed later). These 
data showed a well-defined ring of material centered at $\sim$ 86 au (assuming a distance of 48.2 pc) and inclined about 32$^{\circ}$ 
from face-on. The ACS image also suggests a faint halo. The deepest scattered light detection was obtained using HST/STIS 
\citep{2014ApJ...789...58S}; their image shows the ring radius to be 90.5 $\pm$ 1.1 au (assuming a larger distance of 51.8 pc) 
at an inclination of $\sim$ 28.5$^{+2.0}_{-2.1}$ degrees. They also find a halo extending to $\sim$ 200 au as well as  large scale
asymmetries in the brighter part of the disk:  ``the NE side of the disk exhibits a peak in the surface brightness, 
approximately 30\% brighter than the SW side of the disk. There is also a NW–SE asymmetry, with the NW side of the 
disk approximately 10\% brighter than the SE side. Additionally, the disk appears more radially extended toward
the N than toward the S.'' The NICMOS and optical images all agree that the flux falls off rapidly inside the 90 au ring.

\begin{figure*}
    \centering
    \includegraphics[width=0.7\linewidth]{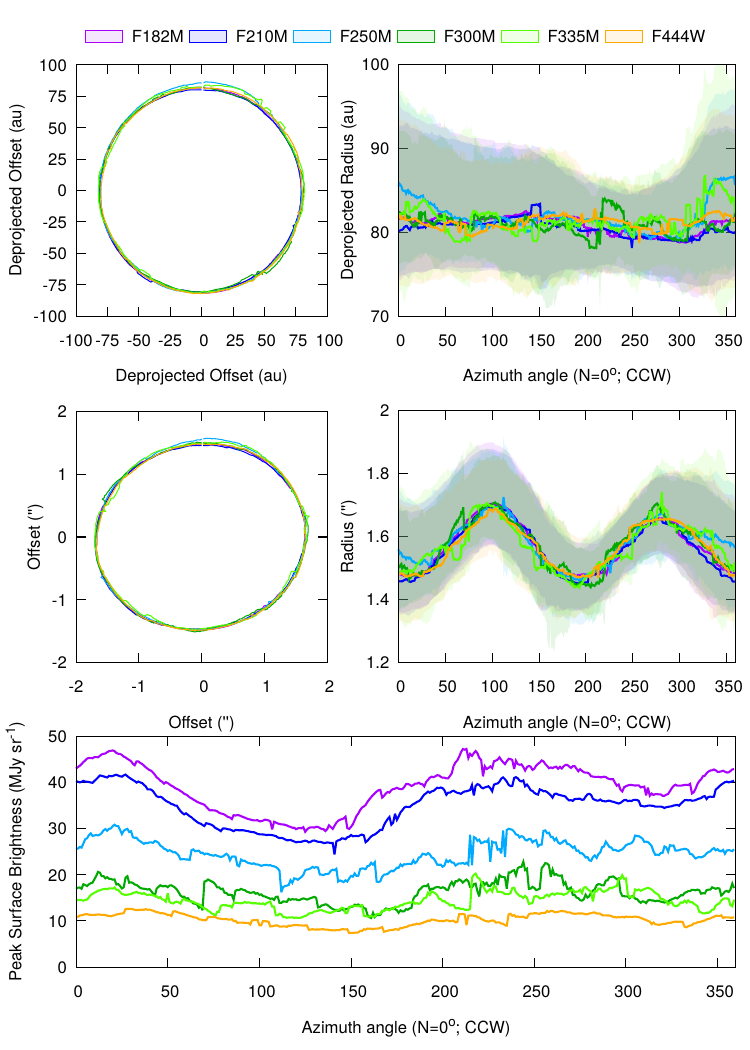}
    \caption{The peak brightness of the HD~181327 ring, located in 1 degree steps,
    using 30 degree bins at each step. The {\it top row} displays the values for the 
    deconvolved and the {\it middle row} for the observed images. The shaded areas
    represent the locations that are within 0.5 $\times$ the peak brightness, which is
    displayed in the {\it bottom panel}. The disk presents a small offset near 0$^{\circ}$
    azimuth at 2.5 $\micron$. The brightness is asymmetric at the shorter wavelengths, with
    peaks near 20$^{\circ}$ and 230$^{\circ}$.}
    \label{fig:hd181327rings}
\end{figure*}

\begin{figure*}
    \centering
    \includegraphics[width=0.99\linewidth]{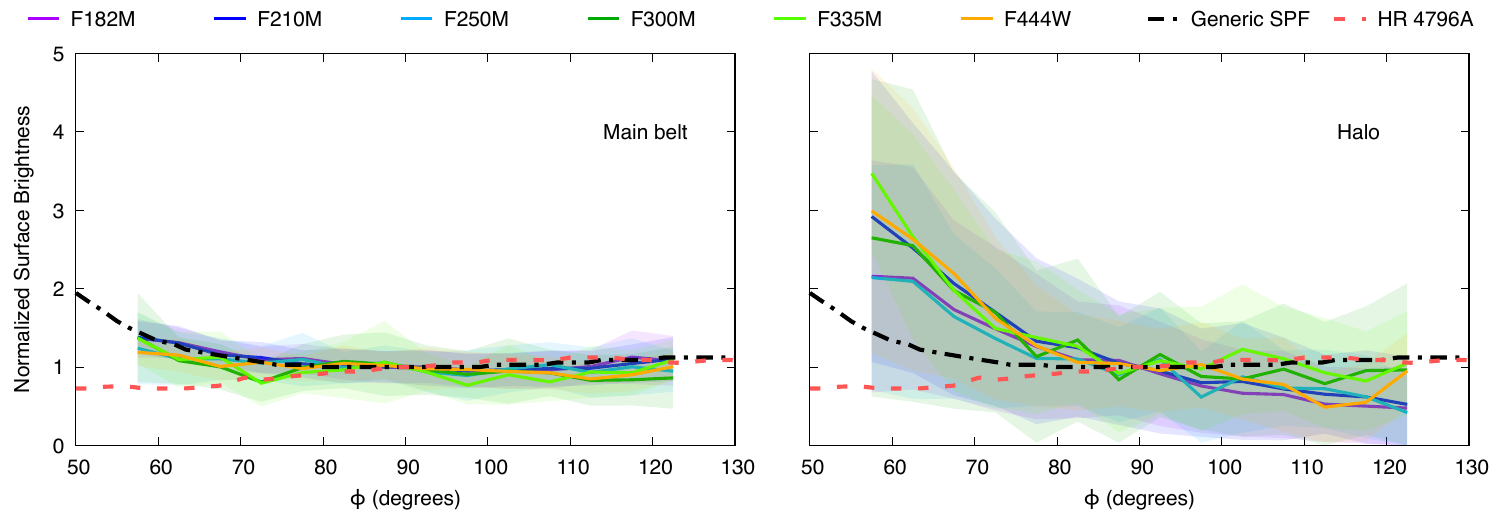}
    \caption{The normalized (at $90^{\circ}$) scattering phase functions of the HD~181327 system at each observed wavelength,
    with a generic SPF compiled by \cite{2024MNRAS.528.6959H} and the observed SPF of HR~4796A \citep{2017A&A...599A.108M} 
    overplotted for comparison. In the {\it left panel:} we show the SPF for the main belt, while in the {\it right panel:} 
    we show it for the disk halo. The halo region is significantly more forward scattering than the main belt and we 
    do not see significant  variations as a function of wavelength.}
    \label{fig:hd181327_SPF}
\end{figure*}

At longer wavelengths, \citet{2008ApJ...689..539C} report Gemini South T-ReCS Q$_{a}$-band (18.3 $\mu$m) images that, 
within the limited signal to noise, show that the emission is centered near the 90 au ring. \citet{marino2016} 
present an ALMA image. It shows a uniform ring with no notable asymmetries, with a radius of 86.0 $\pm$ 0.4 au. 
There is a low surface brightness halo (or second broad ring)  extending to $\sim$ 200 au. 
The ring radii reported by \cite{2014ApJ...789...58S} and \citet{marino2016} assume a distance of 51.8 pc; 
correcting to the modern value of 47.78 pc \citep{gaia22} reduces the radii by a factor of 0.92.

\begin{figure}
    \centering
    \includegraphics[width=0.99\linewidth]{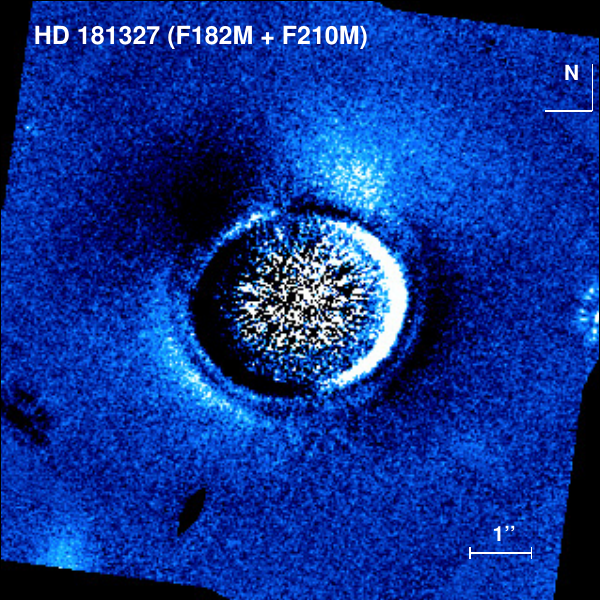}
    \caption{This figure shows the gradient in the halo brightness, by subtracting
    an image of the disk from itself, which was mirrored along the disk minor axis at ${\rm PA} = 10\fdg39$. We used
    a combined F182M and F210 image (non deconvolved), to increase the SNR and to avoid
    artifacts from deconvolution. The N-W lobe of the halo shows a small level of brightness enhancement relative to the N-E side.}
    \label{fig:HD181327flip}
\end{figure}

\begin{figure}
    \centering
    \includegraphics[width=0.99\linewidth]{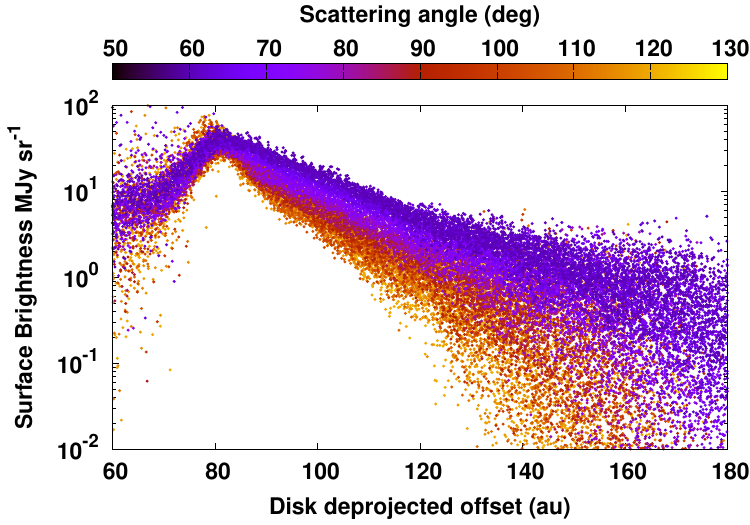}
    \caption{The disk surface brightness as a function of deprojected 
    distance and scattering angle with the F182M filter for each pixel. 
    The forward scattering parts of the halo are significantly brighter than
    for the main ring, due to the blowout of smaller grains in the disk.}
    \label{fig:hd181327_RFSPF}
\end{figure}

\begin{figure*}
    \centering
    \includegraphics[width=0.49\linewidth]{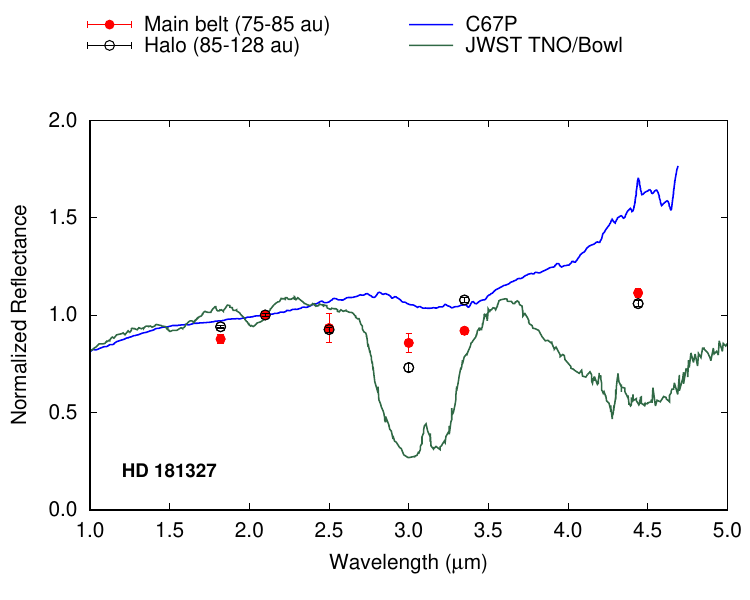}
    \includegraphics[width=0.49\linewidth]{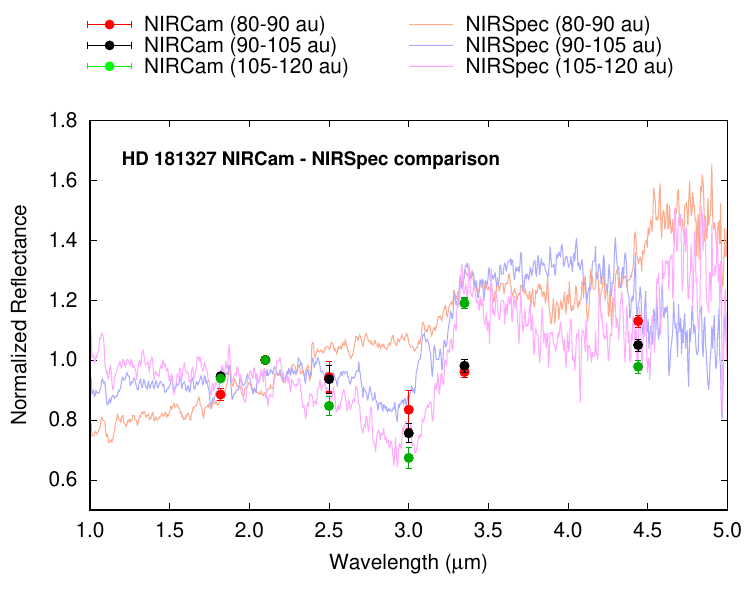}
    \caption{{\it Left panel:} The normalized (at $2.1~\micron$) reflectance spectra of the
    HD~181327 system, using the deconvolved images and scaled by deprojected $R^2$, compared to
    the observed reflectance of Solar System objects. The measurements are comparable to cometary
    objects, with a small but measurable dip at 3 $\micron$. {\it Right panel:} The observed reflectance
    spectra compared to the NIRSpec measurements presented in \cite{xie2025}. The agreement between
    the NIRSpec spectra and NIRCam broadband photometry is excellent for the outer regions (90-105 and 105-150 au). 
    However, we see a deviation in
    the main belt measurements (80-90 au), where the NIRCam data shows
    broadband photometry similar to the halo region, while the NIRSpec spectrum does not reveal the water
    absorption feature at 3 $\micron$. 
    }
    \label{fig:hd181327_spectra}
\end{figure*}

A number of preliminary conclusions have been drawn from these images. It has been suggested that the larger size 
of the scattered light ring (HST) than that seen with ALMA is consistent with the radiation pressure blowout of 
small grains from the parent-body ring seen by ALMA \citep{marino2016}. The lack of asymmetry in the ALMA ring 
would appear to rule out a recent massive collision as the source of the structure seen with HST \citep{marino2016}. 
It is argued to be more likely that the disk is warped by interaction with the interstellar medium \citep{marino2016}, 
although the asymmetry could arise from very small grains being eroded and released from an orbiting object as 
suggested in another context by \citet{rappaport2014}.

While detections of gas in a debris disks of this age are rare (and restricted to A-type stars), 
\citet{marino2016} report $^{12}{\rm CO}(2-1)$ emission in the HD~181327 disk detected with ALMA. The gas is 
co-located with the dust in continuum emission, suggesting an exo-cometary origin. Using the Spitzer MIPS 
spectral energy distribution (SED) mode, \citet{2008ApJ...689..539C} made a tentative identification of water 
ice in the disk; this has been confirmed by \citet{xie2025}, who detected the 3 $\mu$m ice feature in the 
90 au ring and just outside it.  The luminosity of a F6V star is about 3 L$_\odot$, showing that the region 
with icy grains is analogous thermally to our Kuiper Belt, which is populated with icy planetesimals. The 
bodies releasing the icy grains in HD~181327 are probably analogous. 

\citet{2012AA...539A..17L} report measurements of the system with Herschel and at 3.2 mm with the 
Australia Telescope Compact Array (ATCA). Combining with data from the literature, the SED is 
constrained by 38 measurements.  Their modeling of this detailed  SED indicates that the dust grains 
are dominated by water ice ($\sim$ 65\%) with porous amorphous silicates and carbonaceous material. 
Their model indicates a minimum grain radius of $0.81 \pm 0.31~\mu$m, and a power law  size 
distribution with exponent $ 3.41 \pm 0.09$. 

Recently, \citet{milli2024} presented both total and polarized intensity observations of 
HD~181327 with SPHERE/IRDIS, combined with archival HST/NICMOS observations reprocessed as part of the 
ALICE program \citep[][]{2014SPIE.9143E..57C, 2018AJ....155..179H}. The polarized fraction ranges 
from 23.5\% to 18\% for phase angles from 60$^\circ$  to 120$^\circ$.  Their models indicate a minimum 
grain size of $\sim$~0.1-0.2~$\mu$m, well below the blowout size around this star. This result may 
suggest a very high collision rate within the debris ring, with an accompanying high rate of production 
of tiny grains \citep[e.g.,][]{thebault2019}. The polarimetric data also require presence in the 
grains of a highly refractive material ($n \ge 3$), possibly an iron compound \citep{milli2024}.

No planetary companions have been detected around HD~181327 
\citep[e.g.][]{2022A&A...667A..63Z,2021A&A...651A..71L}. 
Sparse Aperture Masking with the VLT provides detection limits of $M >10~{\rm M}_{\jupiter}$ from 5 - 20 au 
\citep{2016A&A...595A..31G}. 

\subsubsection{JWST Results}

We present the six filter images of the HD~181327 system in Figure \ref{fig:hd181327_gallery}, showcasing both 
the observed and deconvolved versions. The JWST PSF leaves an imprint on the observed long wavelength images of the disk,
yielding a hexagon shaped ring; this imaging artifact is rectified by the deconvolution step. The debris ring
is bright and spatially well resolved at all wavelengths (at high SNR) as is the azimuthally asymmetric 
halo, previously seen with HST/STIS imaging \citep{schneider14,2014ApJ...789...58S}. The images do not hint at the 
presence of additional components (rings or gaps) or associated point sources.

The narrow bright ring provides us the opportunity to study the orbital parameters of the main belt and
possible variations as a function of observing wavelength. We perform orbit fitting to the brightest part 
of the main belt using the deconvolved F182M image to determine the azimuthal symmetry and obliquity 
of the disk. We find the following orbital fit to the brightest central peak of the ring, using the latest
GAIA distance measurement to convert angular sizes to physical scales:
${\rm a} = 80.07 \pm 0.14~{\rm au}$, $e=0.016\pm0.001$, $\iota = 28\fdg{54}\pm0\fdg31$, and 
${\rm PA}=100\fdg39\pm0\fdg63$. In Figure \ref{fig:hd181327orbfit}, we show the resulting cornerplot of the 
orbital fits, visualizing the covariance matrix of the fitted orbital variables. The disks imaged with the
NIRCam and HST observations \citep{2014ApJ...789...58S} are at similar radial distance (80.07 vs.\ 83.47 au, correcting for the modern
distance value) and at the same inclination (both at $28\fdg5$). Figure \ref{fig:hd181327rings} presents the 
location of the central peak in the deprojected (top row) and observed frames (middle row) and the azimuthal
variation of the disk widths as a function of observed wavelength. The architecture of the disk is mostly 
independent of wavelength and becomes wider towards the N direction (azimuth angle $\sim~350^{\circ}$). The 
disk profile varies as a function of wavelength towards the N and is widest at 3.35 $\micron$. The halo extends towards the N direction at a greater extent than
the S. We also notice a small asymmetry in the halo towards the N. To highlight this asymmetry, we subtract a mirrored image of the disk from itself, with the mirroring performed along the minor axis towards ${\rm PA} = 10\fdg39$. We show 
this image in Figure \ref{fig:HD181327flip}.

We analyze the scattering properties of the dust in the system in Figures
\ref{fig:hd181327_SPF} and \ref{fig:hd181327_RFSPF}. Figure \ref{fig:hd181327_SPF}
shows the scattering phase function of the system, analyzed using the deconvolved 
images and scaled by $R^2$, separated for the main belt (75-85 au) and the 
halo, located further out (85-128 au). The SPF of the main belt is very similar to the Generic SPF profile compiled
by \citet{2024MNRAS.528.6959H}, and more forward scattering than that observed
for HD~4796A \citep{2017A&A...599A.108M}. The SPF of the halo, shown in the {\it right}
panel of the figure, reveals an even stronger forward scattering component 
(at all wavelengths), indicating the presence of submicron sized particles, 
likely being blown out by radiation pressure. Figure \ref{fig:hd181327_RFSPF} shows
brightening of the dust particles towards the forward direction using the F182M
filter image, as a function of deprojected radial offset. The main belt and regions
inwards show an even brightness distribution as a function of scattering angle, while
the domains further out are clearly separated. This can be explained by a variation
in the dust size populations between the regions, with the halo dominated by
sub-micron size particles.

As for the other sources in the program, we analyze the reflectance photo-spectra 
of the system using the NIRCam six band observations in Figure 
\ref{fig:hd181327_spectra}. The {\it left} panel of the figure shows the normalized 
(at 2.1 $\micron$) reflectance spectra, multiplied by $R^2$, determined using 
the deconvolved images. Errors are calculated using the methods described in Appendix \ref{app:A_pt_3}. The system exhibits a shallow
$3~\micron$ absorption dip, with the halo and the main disk having similar profiles.
In the {\it right} panel of the same figure, we compare our observed 
photo-spectra to the NIRSpec observations presented by \cite{xie2025}. We determined
the main belt to peak at 80 au, therefore in the solar system comparison ({\it left}
panel), we set the main belt to be between 75 and 85 au. For a proper comparison
with the \cite{xie2025} results, we re-evaluate the photometry within the distances they
used. The NIRSpec spectra were fitted to the corresponding NIRCam observations by multiplying and integrating them with the appropriate filter throughput curves and performing $\chi^2$ minimization.
While the NIRCam photometry agrees with the NIRSpec spectra determined outside of 90 au,
there is a considerable discrepancy in the 80-90 au bin, where the NIRSpec observations
are flat over the water-ice absorption feature at 3 $\micron$, while the NIRCam photo-spectra show the
same dip for this region as they did for the outer ones. We do not know the source of this discrepancy,
but given the significantly larger sample area of the NIRCam observations relative to the
narrow wedge sampled by NIRSpec and the difficulties presented in the post-processing
of the NIRSpec IFU data, it is likely that water-ice is also present in the main-belt
of the system. We summarize the photometry values in Table \ref{tab:hd181327phot}.

\begin{table}[!t]
\setlength{\tabcolsep}{2pt}
\begin{center}
\caption{HD~181327 Photometry\label{tab:hd181327phot}}
\begin{tabular}{lcccccc}
\hline\hline
                   & F182M  & F210M & F250M & F300M & F335M & F444W \\
\hline
F$_{\nu,\star}$ (Jy\tablenotemark{a}) & 3.63  & 2.93 & 2.18 & 1.61 & 1.32 & 0.79 \\
\hline
\multicolumn{7}{c}{Main belt ($75 - 85$ au)}\\
\hline
F$_{\rm sc}$ ($\mu$Jy)      & 1473 & 1357 & 942 & 640 & 563  & 406 \\
$\sigma_{\rm sc}$ ($\mu$Jy) & 40   & 39   & 73  & 35  & 11 & 8 \\
\hline
F ($\mu$Jy)                 & 1416 & 1306 & 905 & 615 & 541 & 390 \\
$\sigma$ ($\mu$Jy)          & 39   & 38   & 73  & 33  & 11  & 8 
\\
\hline
\multicolumn{7}{c}{Halo ($85 - 128$ au)}\\
\hline
F$_{\rm sc}$ ($\mu$Jy)      & 2686 & 2307 & 1587 & 927 & 1122 & 656 \\
$\sigma_{\rm sc}$ ($\mu$Jy) & 19   & 15   & 22   & 24  & 15   & 9   \\
\hline
F ($\mu$Jy)                 & 1702 & 1470 & 1019 & 600 & 704  & 421 \\
$\sigma$ ($\mu$Jy)          & 13   & 12   & 15   &  21 & 11   & 6 \\
\hline
\end{tabular}
\end{center}
\tablecomments{Photometry of the HD~181327 disk rings. 
The ``{\it sc}'' scatter values are scaled by $(R/R_0)^2$, where $R_0$ 
is set at 80 au.}
\tablenotetext{a}{Stellar photosphere values.}
\end{table}

\begin{figure}[!t]
    \centering
    \includegraphics[width=0.99\linewidth]{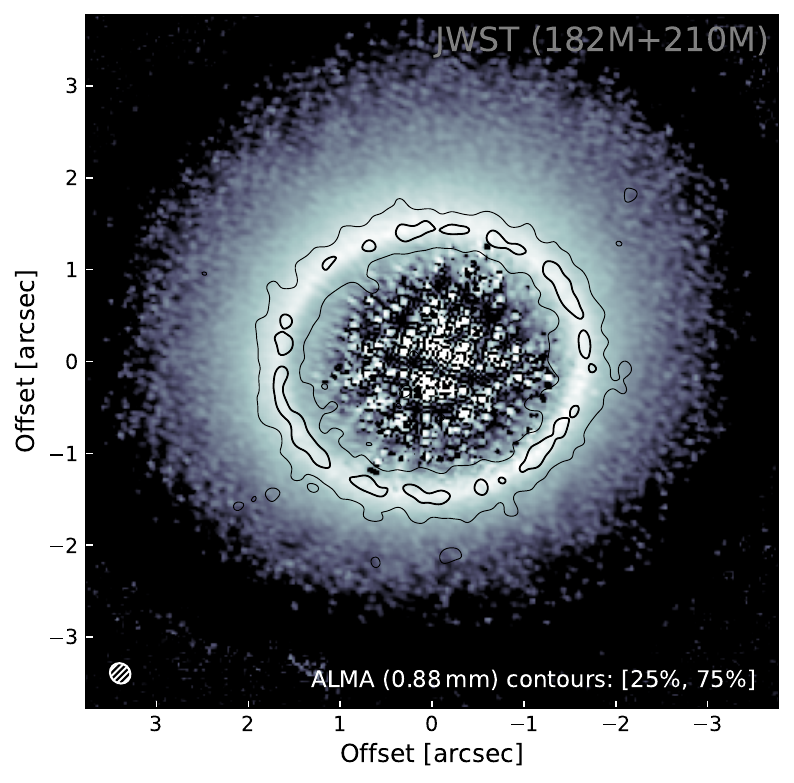}
    \caption{Contours from the ALMA observations obtained within the ARKS program \citep{2026A&A...705A.195M}, overlaid
    on the combined F182M and F210M JWST/NIRCam observations for HD~181327.}
    \label{fig:hd181327_alma}
\end{figure}

Finally, we show the F182M+F210M combined image, overlaid with the 0.88\,mm 
brightness contours observed with the REASONS survey \citep{matra2025}
in Figure \ref{fig:hd181327_alma}. The scattered light disk, observed with NIRCam,
is co-located with the planetesimal belt, as observed with ALMA, with similar narrow
profiles, hinting at the narrow ring being the site of active dust production and
continuous removal via radiative forces. Unlike the NIRCam data, the ALMA observations
do not show azimuthal asymmetry. 

\section{Survey results}
\label{sec:compare}

\subsection{Comparative study of reflectance brightness}

The primary goal of the NIRCam Scattered Light Disks Survey was to probe the
prevalence of water-ice in the cold planetesimal belts of mature planetary systems.
At low temperatures and pressures, crystalline ice forms
on the small dust particles within the planetesimal belts, which is an effective absorber
of 3.12 $\micron$ (3200 cm$^{-1}$) infrared emission, due to an O-H stretch mode \citep{2001JGR...10633333D}.
Subbands of this stretch mode, resulting from strong intermolecular coupling, are 
present from 2.94 $\micron$  (3400 cm$^{-1}$) to 3.2 $\micron$ (3100 cm$^{-1}$) and 
produce a wide near-infrared absorption band. This water-ice absorption band has
been observed for a number of solar system objects over the years, such as Sedna 
\citep{emery24}, Comet C67 \citep{quirico16}, and various categories of TNO objects 
recently compiled by \citet{pinilla25}. While comet C67 only has a weak absorption
band, objects in the Kuiper belt reveal a much higher water-ice content.

\begin{figure*}
    \centering
    \includegraphics[width=0.99\linewidth]{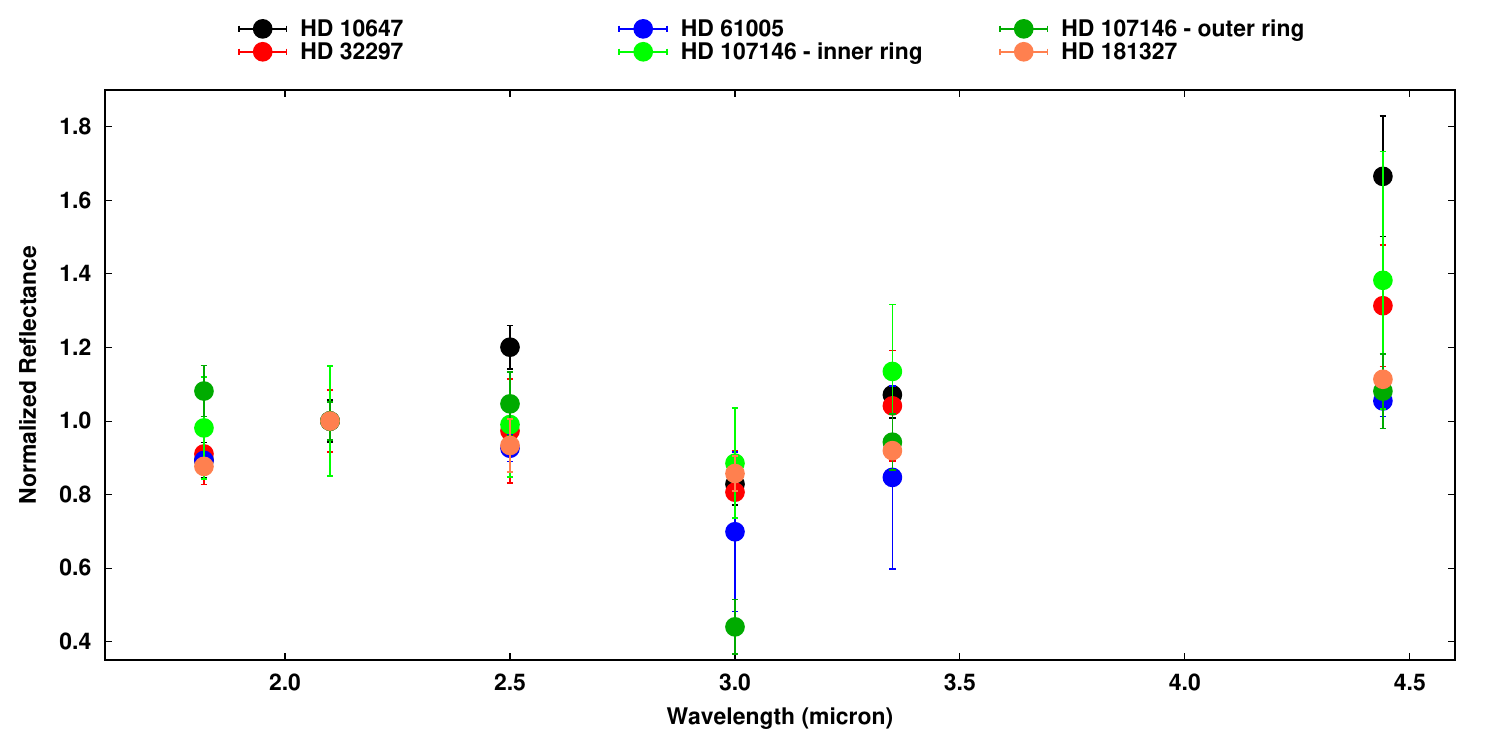}
    \caption{Comparison of the reflectance spectra of all systems in
    our survey sample, normalized at 2.1 $\micron$. For HD~10647, we chose the 20$^{\circ}$-30$^{\circ}$ 
    scattering angle range, as it provides the highest SNR while avoiding the bulk of
    the speckle dominated regions. For HD~32297, HD~61005, and HD~181327, the plotted photometry was extracted from the main belt regions, whereas for HD~107146 we show measurements from both the inner and outer rings. The normalized reflectance spectra of the systems are quite similar, with all of them showcasing various degrees of water-ice absorption at 3 $\micron$. The deepest water-ice detection is around the outer
    ring of HD~107146. The 2.5 $\micron$ peak of HD~10647 stands out as an anomaly relative to the other measurements.}
    \label{fig:alldisks_spectra}
\end{figure*}

The six-filter NIRCam coronagraphic observations -- spanning from 1.82 to 4.44 $\micron$, 
with data at 3.0 and 3.35 $\micron$ -- allowed us to probe this water-ice absorption/reflectance 
band in systems that are challenging/impossible to observe with any other instrument or 
observatory due to their low surface brightness and proximity to their bright host stars. 
In Figure \ref{fig:alldisks_spectra}, we 
compile the reflectance spectra of all of the sources in the survey, scaled by $R^2$ and 
normalized at 2.1 $\micron$. The systems showcase a generic profile and reveal a uniform
water-ice absorption band, present in all of the observed photo-spectra. The deepest
absorption is measured for the outer ring of the HD~107146 system and the main 
belt of the HD~61005 system, while the inner ring of
HD~107146, the well defined belt HD~181327, and the main belt of HD~10647 reveal shallower features. 
The NIRSpec observations of the HD~181327 system \citep{xie2025} confirm the presence of 
water-ice through the presence of the Fresnel peak at 3.1 $\micron$, therefore even the shallow
absorptions detected by the NIRCam data are of high confidence. The 2.5 $\micron$ peak
observed in the HD~10647 system is an anomaly that stands out relative to the other 
observations. The 4.44 $\micron$ data for HD~61005, 181327 and the outer ring of HD~107146 show the characteristic turnover also produced due to the presence of water-ice.

\subsection{Comparative study of Scattering Phase Functions}

Only two of the sources in the sample, HD~10647 and HD~181327, are oriented in such a way
that their scattering phase functions can be studied. The former provides a large
range of angles, from 13 to 167$^{\circ}$, while the latter only a narrow one, from
60 to 120$^{\circ}$, within which the scattering properties of the dust can be analyzed. 
The SPFs of the main belts in both systems are wavelength-independent and show
remarkably good agreement with the Generic solar system SPF compiled by \cite{2024MNRAS.528.6959H}.
In contrast, the halo region of HD~181327 showcases a stronger forward scattering than the
generic SPF,
likely because it is dominated by submicron-sized dust grains.

\section{Summary}
\label{sec:summary}

We executed a 57.5 hour NIRCam coronagraphic survey (GO 2780, PI A.\ G\'asp\'ar, Co-PI J.\ Leisenring) 
in Cycle 2, within the NIRCam GTO program (PI M.\ Rieke), to observe five bright circumstellar debris 
disks using five medium and one wide band filters. Two of the sources were also observed with
MIRI, using the F2100W filter. All five of the disks were detected at all six filters, allowing us
to characterize the architectures of the systems at near-IR wavelengths as well as their water-ice 
content. All five of the systems reveal the presence of water-ice at various levels. 
The two systems where scattering phase function studies could be performed revealed scattering properties very similar to what we observe for solar system objects. All observations were
checked for co-moving point sources; unfortunately none of the images revealed any exoplanet candidates. 
Below, we provide a short summary of the most important results for each system.

\begin{itemize}
\item {\bf HD~10647} is the oldest system in the sample, with a solar-type star (F9V) as its host.
The disk is spatially extended and has a low surface brightness, therefore we did not analyze
deconvolved images for this source. The SPFs show strong forward scattering and brighter levels
for the three shortest wavelength filters, while the profiles -- when normalized -- agree with
a generic Solar System measurement \citep{2024MNRAS.528.6959H}. The reflectance spectrum, while rather noisy for scattering
angles where the coronagraphic mask interferes, shows that the system contains water-ice in the
outer regions. Mysteriously, the system shows an elevated 2.5 $\micron$ flux. The MIRI observations
reveal a central spatially unresolved component, similar to the many other debris disks observed with
MIRI, likely produced by dragged in particles. 

\item {\bf HD~32297} is a young (15-30 Myr) debris disk system, viewed edge-on, 
hosted by an early-type star (A6V), making it a distant (129 pc) analog of the
famous $\beta$ Pictoris disk. Given the compact nature and edge-on orientation of the
disk, analysis was carried out on deconvolved images. The disk is flared towards the
N direction at all NIRCam bands, much like it was seen with HST images 
\citep{kalas05,debes09}. This feature likely arises from a mix of intense dust production
via giant impacts within the system as well as interactions with the local ISM 
\citep{jones2023}. Water-ice is detected in the system at levels similar to the other
targets within the survey.

\item {\bf HD~61005} is a young ($\sim 100~{\rm Myr}$) solar-type (G8V) star with
an extended flared debris disk, giving rise to its nickname ``The Moth''. This structure
is likely produced by interactions with a dense local ISM cloud \citep[e.g.][]
{pastor2017,jones2023}.  Water-ice is detected both in the disk and the swept wing
halo, with the halo also exhibiting a downturn in the 4.44 $\micron$ reflectance flux
in the deconvolved images. The halo is more extended in the shorter wavelength images,
indicative of it consisting of smaller fragments.

\item {\bf HD~107146} is an adolescent (100-300 Myr) solar-analog (G2V) star, with
a double ring debris disk system, located at $\sim 50$ and $\sim 120~{\rm au}$ from
the central star \citep{marino2018}. Given its face-on orientation and large spatial 
extent, the disk has a very low near-IR reflected surface brightness, only detected 
due to its otherwise relatively large mass and fractional infrared luminosity 
($L_{\rm disk}/L_{\ast} \sim 10^{-3}$). The JWST/NIRCam images place the inner ring
slightly closer in at 35-55 au and the outer ring at 90-130 au. The MIRI images also
resolve these components as well as an unresolved central component, possibly 
the same source ALMA observations placed at $\leq 20~{\rm au}$. The reflectance 
spectra of the two rings observed with NIRCam both show the presence of water-ice,
with the outer ring displaying a deeper absorption feature. The presence of the three ring 
components, along with the spectral-type of the central star and the significant presence
of water-ice in the outer regions, further deepens the similarities between HD~107146 
and the solar system.

\item {\bf HD~181327} is a $\beta$ Pictoris group member ($\sim$ 18 Myr) F5/F6V-type star,
with a well characterized narrow ($\sim 10~{\rm au}$ wide) debris ring, 
located at $1\farcs69$ (80 au, using the latest GAIA distance measurement) from its
central star. Its estimated distance
from the Sun varies in the literature, resulting in discrepancies in the absolute 
scaling of the system. The disk has a halo extending outwards from the main ring,
detected with increased surface brightness towards the N direction. Confirming previous
NIRSpec observations, the NIRCam data reveal the presence of water-ice in the system,
even within the main-belt, which was missing from the NIRSpec spectra. While this system
shows lower water-ice absorption at 3.35 $\micron$ then some of the other sources in
the survey, the independent confirmation with the NIRSpec spectra provides confidence
to our conclusions.
\end{itemize}

Overall, the survey reveals a diverse set of debris disk architectures spanning a wide 
range of stellar ages, spectral types, and viewing geometries. Despite these differences, 
all five systems exhibit evidence for water ice in their scattered-light reflectance 
spectra, suggesting that icy grains are a common component of bright debris disks. 
These observations provide a valuable dataset for future modeling efforts aimed at 
constraining the composition, size distribution, and the processes shaping the spatial 
distribution of dust grains in circumstellar environments.

\begin{acknowledgments}
This paper is dedicated to the memory of Glenn H Schneider, whose extensive 
work on scattered light imaging of debris disks -- including the targets 
presented in this paper -- has been foundational to this program and who
participated in the planning of these GTO observations prior to the launch of 
JWST.

A.G., J.M.L, S.G.W., G.H.R, and M.J.R., acknowledge support from the
NIRCam Science Team contract to the University of Arizona, NAS 5-2105.

A.A.S.~is supported by the Heising-Simons Foundation through a 51 Pegasi b Fellowship.

J.B.L.~acknowledges the Smithsonian Institute for funding via a CfA Prize Fellowship (J.C. Ryan). 

The reduced and post-processed data, as presented in this paper, can be downloaded
from \dataset[10.5281/zenodo.21683353]{https://doi.org/10.5281/zenodo.21683353}.
Some/all of the data presented in this paper were obtained from the Mikulski Archive for Space Telescopes (MAST) at the Space Telescope Science Institute. The specific observations analyzed can be accessed via \dataset[https://doi.org/10.17909/6a5r-qf97]{https://doi.org/10.17909/6a5r-qf97}. STScI is operated by the Association of Universities for Research in Astronomy, Inc., under NASA contract NAS5–26555. Support to MAST for these data is provided by the NASA Office of Space Science via grant NAG5–7584 and by other grants and contracts.

We acknowledge a humble dog named Winnie, whose companionship contributed 
significantly to the creation of the software called \texttt{Winnie}, 
which is utilized herein.
\end{acknowledgments}






%
\facilities{JWST(NIRCam), JWST(MIRI)}

\software{AstroPy \citep{astropy:2013,astropy:2018,astropy:2022}, CuPy \citep{cupy_learningsys2017}, JWST Pipeline \citep{bushouse}, LMFIT \citep{2025zndo..15014437N}, Matplotlib \citep{Hunter:2007}, spaceKLIP \citep{2022SPIE12180E..3NK}}

\appendix

\begin{figure*}[!t]
    \centering
    \includegraphics[width=0.99\linewidth]{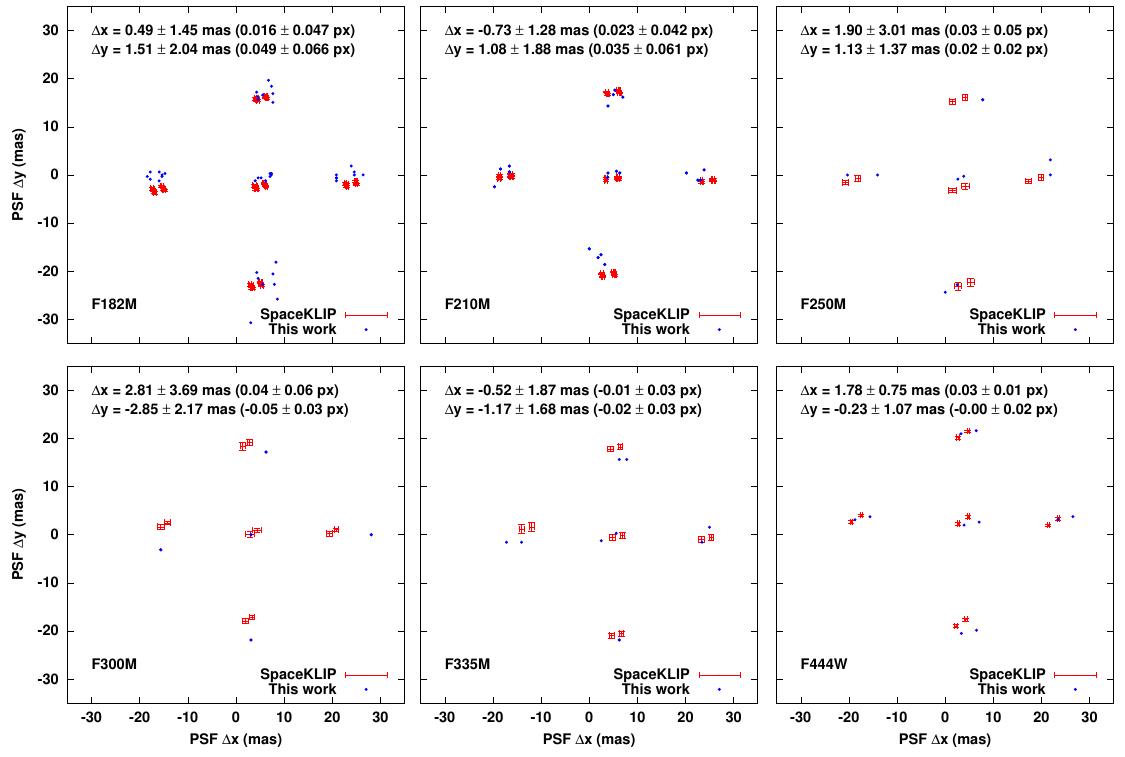}
    \caption{The offsets of the PSFs relative to the target observations for the HD~61005 dataset
    at each observed filter, as determined by SpaceKLIP and our custom by-eye method. Minor systematic
    offsets between the two methods are present; for certain pairings up to 0.2 px difference was observed.}
    \label{fig:offsets}
\end{figure*}

\section{Detailed description of the hybrid classical/LOCI RDI post-processing}
\label{sec:postproc}

We executed a custom hybrid post-processing sequence, based on the classical and locally optimized
images \citep[LOCI;][]{lafreniere07} reference differential imaging (RDI) techniques, as 
introduced in Section \ref{sec:cRDI}. The hybrid method provides results that encapsulate the benefits
of each of the methods: (i) high signal-to-noise and smooth backgrounds (from the classical method), which
are critically important for identifying faint sources, and (ii) clear subtractions in the image core (from the LOCI
method). Additionally, with proper constraints, the method mitigates oversubtraction -- a common
issue in automated post-processing techniques when extended sources are present in the target image.

Based on previous experiences with high-contrast image processing on Hubble Space Telescope 
coronagraphic observations \citep[e.g.][]{schneider14,gaspar20,wolff23} and JWST MIRI direct observations 
\citep[e.g.][]{gaspar23,su24,wolff25}, we know that classical RDI post-processing methods can produce excellent 
results where the PSFs are stable and not strongly affected by their positioning behind a mask or within
the focal plane. The NIRCam coronagraphic PSF is strongly dependent on its position behind the mask, therefore
the LOCI method provides superior results to classical RDI in the core of the PSF. However, LOCI generally
weighs a small percentage of the reference PSFs in the construction of the locally optimized best fitting 
solution, which can result in a noisier reference PSF where it enters the photon dominated region, outside
of the contrast limited core. When resolving faint sources, this is not ideal. The classical RDI
weighs all references roughly equally, thereby reducing the photon noise outside of the PSF core -- and on the flip-side
increasing the contributions of speckles in the cores. Below, we detail the steps we took with the 
hybrid method to merge the strengths of both methods.

Data reduction was executed using the SpaceKLIP/JWST pipeline up to stage 2, as discussed in 
Section \ref{sec:reds}. Additional image operations were executed with astropy, IRAF, and IDL.
Individual integrations within each sequence (action) were median combined with 3$\sigma$ clipping 
around the median for all datasets, as the observatory pointing is stable within a sequence and the 
final products are smoother with higher SNRs. The sky background levels were determined by hand for all 
target and PSF observations in clear and far locations from the central sources and subtracted from the 
individual combined images. De-striping corrections, performed later, further adjusted the sky backgrounds.

\subsection{Image Registration/PSF scaling}

Precise image registration and centering is one of the most crucial post-processing steps for 
coronagraphic imaging, consisting of finding the absolute center of at least one of the target images
to which the remaining target images can be aligned, and finding the relative offsets between target-PSF
image pairs. The former is important, as all de-rotations are performed around the target center and
any offsets are amplified during the final combination steps. Searches for point-sources are especially
sensitive to accurate center registration of all target observations, as misalignment will lead to a reduction of total signal during stacking/averaging. Accurate relative offsets between target-PSF
pairs reduce residual speckle patterns and their determination for targets with extended emissions can be 
skewed by automated processes. Therefore, following along our experiences with HST and JWST/MIRI data
processing, we attempted to find the image centers and target-PSF relative offsets by hand for the NIRCam
dataset. All offsets and PSF scaling values were determined using the IDL software 
IDP3\footnote{https://archive.stsci.edu/prepds/laplace/idp3.html}.

\begin{figure*}
    \centering
    \includegraphics[width=0.99\linewidth]{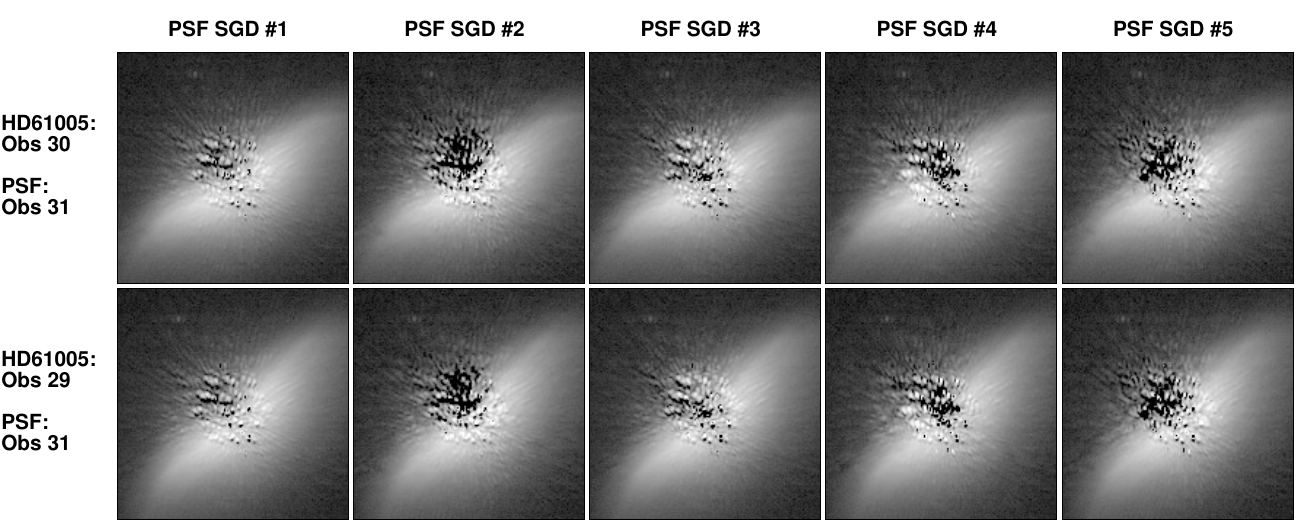}
    \caption{All possible ``classic" PSF subtraction combos for the exposures in the HD~61005 F182M dataset, shown for the
    inner 160x160 px area. 
    The displayed surface brightness is logarithmically scaled
    between -1 and 170 MJy sr$^{-1}$.
    Each column displays a different PSF small grid dither pattern position for the 
    two target rotations observed.}
    \label{fig:allsubs}
\end{figure*}

To find the image center of the first target image in a sequence, we turn to models. Using STPSF, we 
calculate the theoretical PSF at the observation date of the data sequence, given the regularly recorded
optical path difference (OPD) data of the observatory. We execute the STPSF simulations including
both the image mask (MASK335R) and pupil mask (MASKRND) as well as without the image mask (only using the pupil
mask), as the full coronagraphic simulation includes an offset from the true position, which can be determined
using the model that only includes the pupil mask (STScI, priv.\ communication). We determine the actual image 
center in the later simulation by finding the central peak and relative offsets between the former PSF model 
and the first target image. 
The image centers of all other target images are determined relative to the first target image. Relative 
offsets between all combinations of target-PSF pairs were then determined also by eye. In Figure \ref{fig:offsets},
we show these relative target-PSF offsets for the HD~61005 observations for all filters and compare them to the
offsets determined by the SpaceKLIP algorithm. The two methods are in good agreement, with only up to 2-3 
mas offsets on average, although with a few notable larger offsets. We tested a complete post-processing algorithm -- using our hybrid method -- with 
both set of offsets (spaceKLIP and our by-eye estimates) and there were no discernible difference between 
the final products, therefore we conclude that the centering algorithm employed by SpaceKLIP is as accurate as our by-eye method.

The 5 point small grid dither (SGD) employed by the coronagraphic observing sequence for the PSF observations, 
meant to provide a well sampled variety of PSFs, can be well seen in Figure \ref{fig:offsets}. A post-processing algorithm 
like KLIP can decompose the library of PSFs and assemble an ideally matching one using principle component 
analysis, while LOCI calculates an ideal linear combination of these PSFs to provide a locally optimized 
PSF subtraction. In Figure \ref{fig:allsubs}, we present all 10 target-PSF pairings for the F182M HD~61005 
observations. The target was observed at two rotations (Observations 29 and 30), providing two target images, while 
the PSF was observed at a single rotation (Observation 30), using a 5 point small grid dither pattern. The first 
column of Figure \ref{fig:allsubs} shows the subtractions using the first position SGD pattern PSFs, 
which turn out to be the best subtractions. Not surprisingly, these PSFs have the smallest positional 
offsets from the targets. Position \#2 yielded the worse subtraction -- even oversubtracting with an 
otherwise ideal scaling. In Figure \ref{fig:allsubs}, each PSF is scaled by a factor of 0.22 for the subtraction. 
Note that the hybrid method we are introducing does not use this PSF scaling to produce the final product; 
here we are just demonstrating the quality of each target-PSF pairing. 

While determining the spatial offsets between target-PSF pairs, as noted above, we also examined the 
ideal average PSF scaling necessary to achieve a clean PSF subtraction. This has proven to be difficult 
for the automated algorithms to perform when an extended component is present, as they convolve with the 
stellar PSF, while balancing over/under-subtractions by eye is relatively straight forward. 
The various flavors of automated post-processing steps shown in 
panels (a)-(d) Figure \ref{fig:allims} highlight this issue. Importantly, neither the KLIP or LOCI 
algorithms allow the coefficients themselves to be constrained, therefore the contributions of extended 
features can only be avoided by either masking them, filtering them, or removing them by modeling them. 
The hybrid fitting algorithm we employ, however, does allow us to limit oversubtration by constraining the 
sum of the coefficients -- basically setting the PSF scaling. The average scaling we determined while 
estimating the offsets yields the constraint used for the sum of the coefficients.

\subsection{The hybrid classic-LOCI post-processing method}

\begin{figure*}
    \centering
    \includegraphics[width=0.49\linewidth]{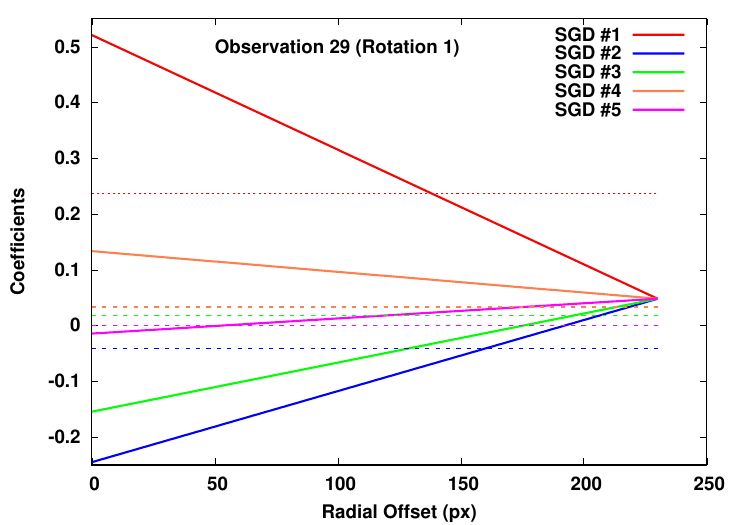}
    \includegraphics[width=0.49\linewidth]{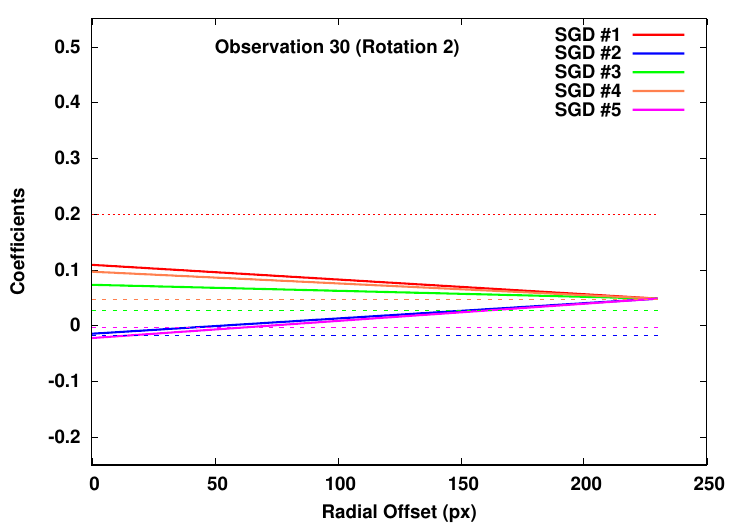}
    \caption{The PSF coefficients for the two rotations of the F182M HD~61005 observations. The solid lines 
    show the radially dependent coefficients the fitting algorithm determined for our hybrid post-processing
    method, while the dashed lines give the initial guesses, based on the LOCI algorithm. The sums of the fitted
    coefficients are 0.2417 and 0.2407 for the two rotations, respectively, at all radial offsets, while they
    are 0.2470 and 0.2524 for the LOCI solutions. Note that our by-eye best scaling was much lower at 0.22.}
    \label{fig:coeffs}
\end{figure*}

\begin{figure*}
    \centering
    \includegraphics[width=0.9\linewidth]{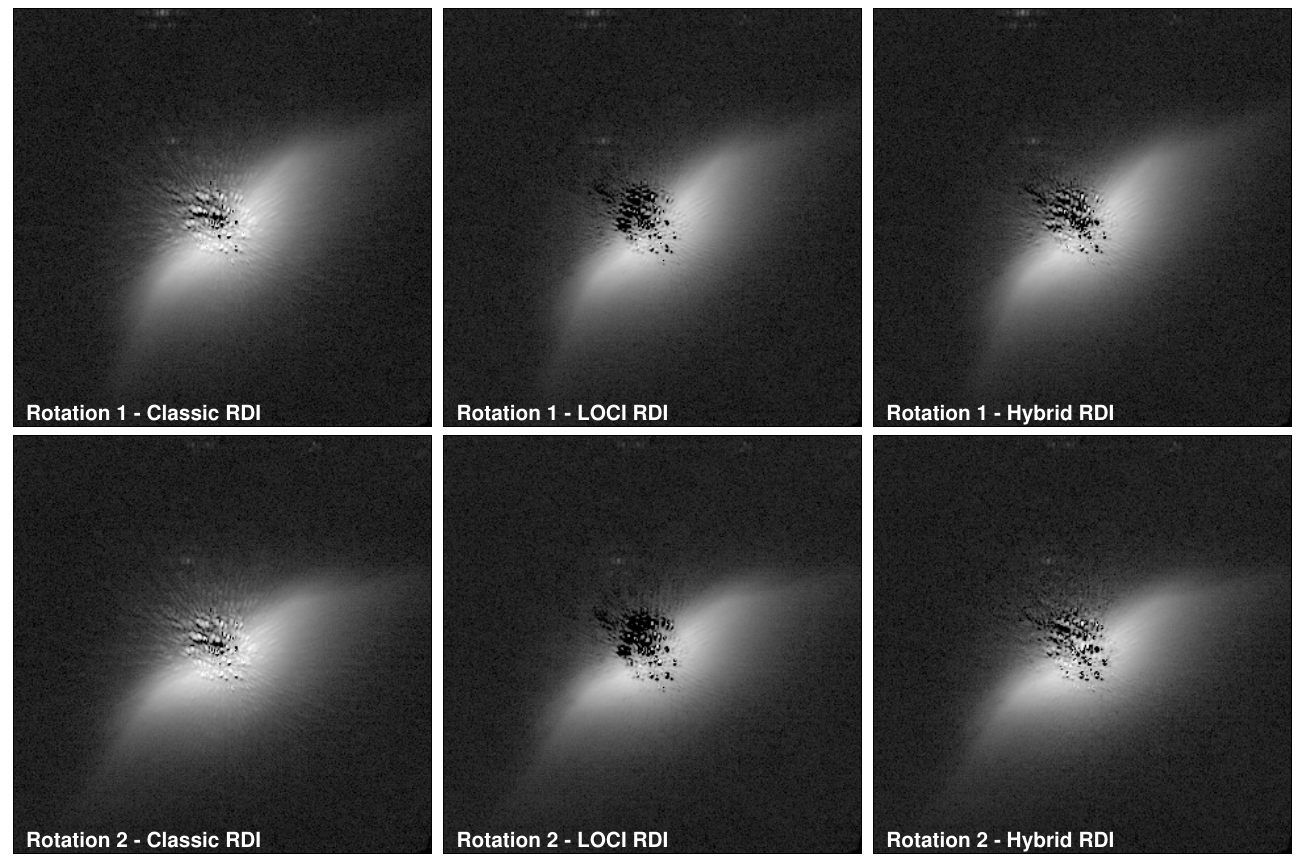}

    \includegraphics[width=0.9\linewidth]{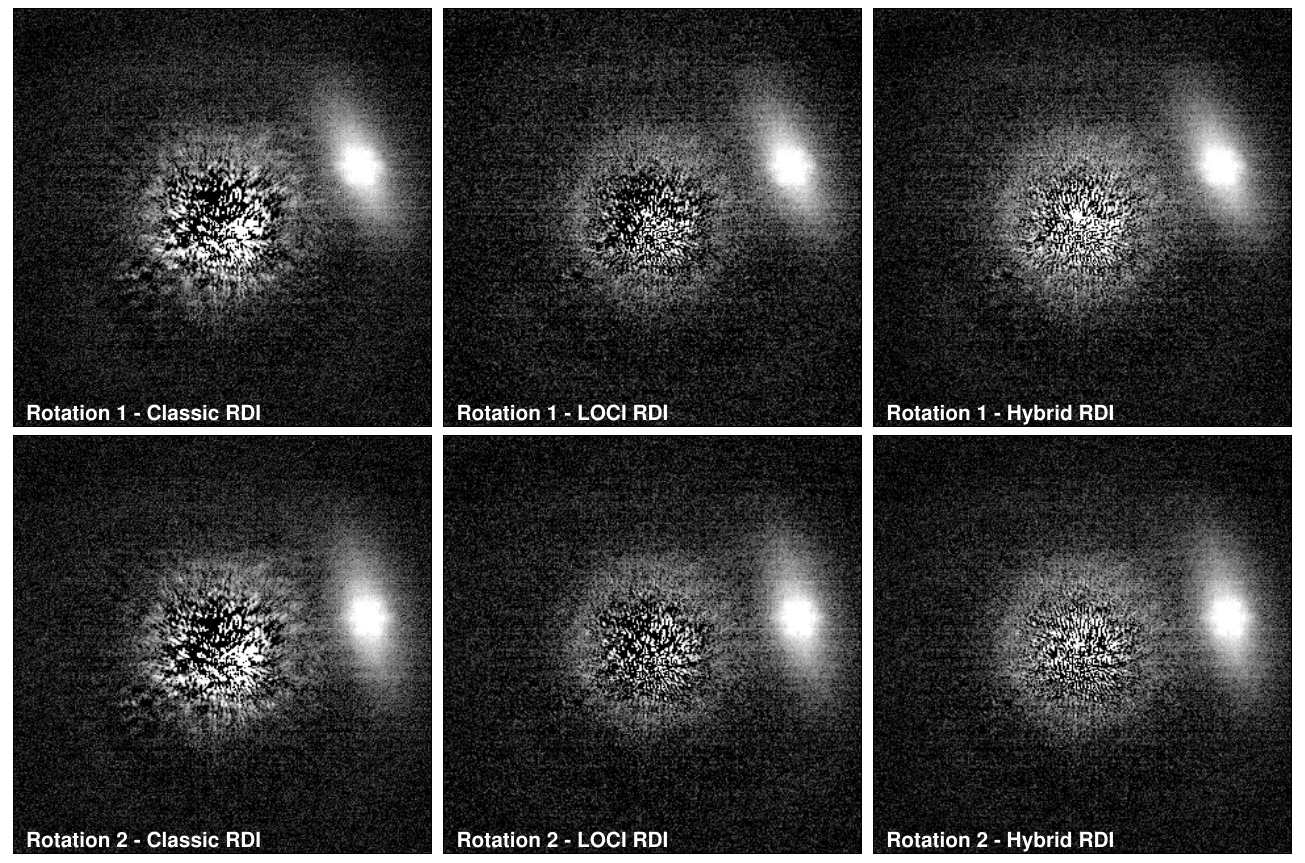}
    \caption{Comparison of the results obtained using the Classic, LOCI, and Hybrid RDI methods for both rotations
    of the F182M datasets of HD~610005 (top two panels) and HD~107146 (bottom two panels). We show the results for the HD~61005 system for consistency in showcasing the
    reduction steps for this system, while the HD~107146 results are presented to highlight the improvements we
    are able to achieve with the Hybrid RDI method, which is more apparent for fainter sources.
    The Classic RDI method results in striations, which the LOCI and Hybrid methods
    are able to remove. Due to the constraints that the Hybrid method is able to include, there is no oversubtraction
    in the core with the Hybrid method, and the SNR is higher in the outer regions.}
    \label{fig:methods_comp}
\end{figure*}

The Locally Optimized Combination of Images (LOCI) algorithm computes the linear combination
of references that minimizes residuals in an optimization zone
via a matrix inversion step \citep{lafreniere07} and then subtracts the result within a
subtraction zone. These zones are typically defined in polar coordinates, with each zone
subtending some range of azimuthal angles and extending between two radii. For detection of
point sources, LOCI generally uses a large number of zones, with each subtraction zone
spanning as little as 1.5$\lambda/D$ in radius. The RDI procedure in \texttt{Winnie} adopts a
version of the LOCI algorithm, but with settings tuned to better preserve disk signal against
oversubtraction by adopting a single optimization / subtraction zone by default. For
simplicity, we refer to this implementation as LOCI throughout, but acknowledge that the 
resulting procedure is no longer truly ``locally optimized".

\begin{figure*}
    \centering
    \includegraphics[width=0.99\linewidth]{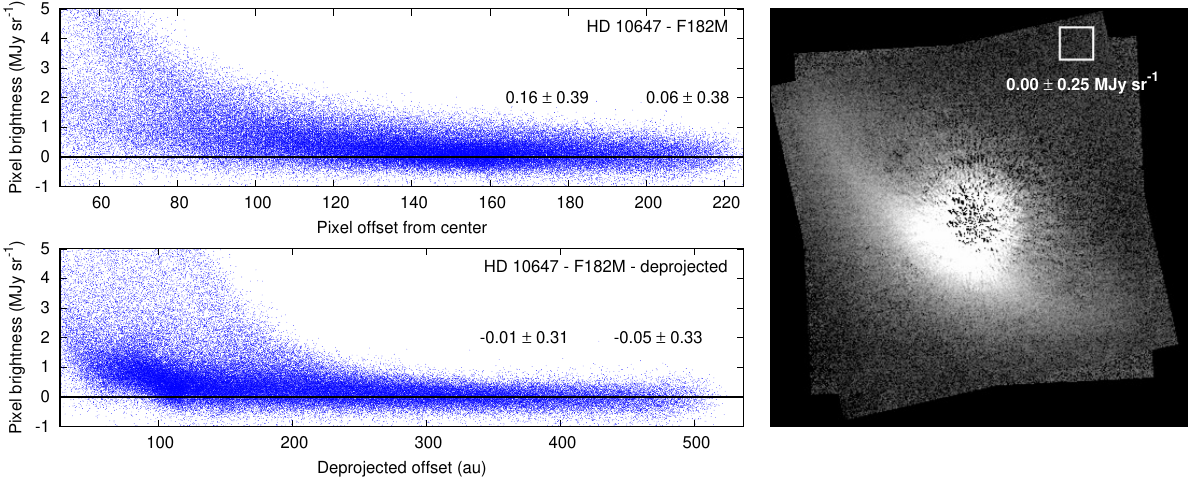}
    \caption{Estimating the sky background value and its error for the F182M observations of HD~10647.
    Left panels: The pixel surface brightness values as a function of offset from the star
    in image coordinate ({\it top} panel) and deprojected distance ({\it bottom} panel). The inset values show
    the sky values at various outer locations. The background value levels off to approximately zero in the 
    deprojected coordinates. {\it Right} panel: Determined in a clear region of
    the image far from the central star, the background is also approximately zero with
    an even lower scatter.}
    \label{fig:errorsky}
\end{figure*}

For JWST/NIRCam, the diversity in the stellar PSF among observations with dedicated (color-matched) reference observations will generally be dominated by the position of the star behind the coronagraph. As such, it is often the case that a small number of reference dithers will dominate the stellar PSF model — acquiring much larger coefficients in the least-squares calculation. For example, if the target acquisition is accurately repeated between the science and reference observations, then the central position of the reference dithers will be significantly over-represented in the final model. While this produces a strong match for the stellar diffraction pattern, the resulting PSF model is much noisier than if a more equitable weighting had been used. Thus, the resulting PSF model permits a good result at small separations — in the contrast dominated region — but compromises SNR at wider separations (assuming reference PSFs of comparable SNR to the science PSF).
Therefore, we employ 
an algorithm where the coefficients are not a single constant across a reference PSF, but vary radially, 
approaching a uniform value at the edge of the field. The coefficients are set to vary linearly as a 
function of the radial offset from the location of the central target, as
\begin{equation}
    c_i (r) = c_{i,0}+\frac{r\left(C - c_{i,0}\right)}{R},\
\end{equation}
where $c_{i,0}$ is the peak value of the coefficient for reference image $i$ in the center of the image,
$C=\sum_N c_{i,0}/N$ is the average of the central values, and $R$ is the edge of the field in the same units 
as $r$, which is the radial offset of the pixel. We chose a linear function as it was simple to employ, 
provided good results, and with the above 
formulation will ensure that the sum of the coefficients remains a constant in the entire field. Given that number of unknowns (the coefficients and their sum) are now larger than the number of equations, we cannot solve for their values via a simple matrix inversion,
and must therefore search numerically. We can, however, use this to our advantage and further improve on
the subtractions, by placing various constraints/limits on the fitting procedure. Based on the by-eye
position and scaling fitting we conducted, the total scaling of the ideal PSF is known to within a certain
degree. We constrain the coefficient fitting to give a sum of the central coefficient values within 10\%
of the best scaling solution we determined by eye. For the F182M dataset of the HD~61005 observations,
we determined a PSF scaling factor of 0.22 by eye, thereby restricting the sum of the central coefficient
values between 0.198 and 0.242. The LOCI algorithm yields a sum for the coefficients of 0.247 and 0.2524 for
the two rotations, respectively, slightly higher than the sum of the central values of our fits of 0.2417 
and 0.2407. Additionally, we placed constraints on the allowed median values within 3 pixel
radial bins, requiring them to be larger than $-1\times$ the median absolute deviation (MAD) of the pixels 
in the bins. {\it The best fitting central coefficient ($c_{i,0}$) values were found by minimizing the
integrated radial median absolute deviation curves in 2D.} We used the basinhopping algorithm within the 
scipy.optimize package through a custom python code to search for the coefficient solutions, with the solutions
from the standard LOCI algorithm providing the initial guesses for the values. The fits converged typically
within a few hundred iterations, using the L-BFGS-B minimization function, with an initial stepsize equal to
the 10\% of the initial PSF scaling value.

In Figure \ref{fig:coeffs}, we show the coefficients fitted for the F182M HD~61005 observations, showing
both the traditional LOCI solutions with dashed lines and our radially dependent hybrid solutions with solid
lines for the five SGD positions. While in the image center SGD position \#1 provides the majority of the 
contribution, thereby reducing the speckle noise, at the edge of the field all 5 PSFs contribute equally,
thereby reducing the photon noise. In Figure \ref{fig:methods_comp}, we compare the post-processing results
for the two rotations of this dataset, using the Classic, LOCI, and Hybrid methods. For the LOCI and Hybrid 
methods, we masked areas with bright extended background features when determining the coefficients, when possible.
For completely face-on systems, like HD~107146 and HD~181327, we did not mask the extended features as they
affected radial bins uniformly. The HD~61005 disk is one of the brightest in our survey, therefore the reduction of
background noise with the hybrid method -- relative to LOCI -- is not as apparent as it is for the fainter targets.

\begin{figure*}
    \centering
    \includegraphics[width=0.99\linewidth]{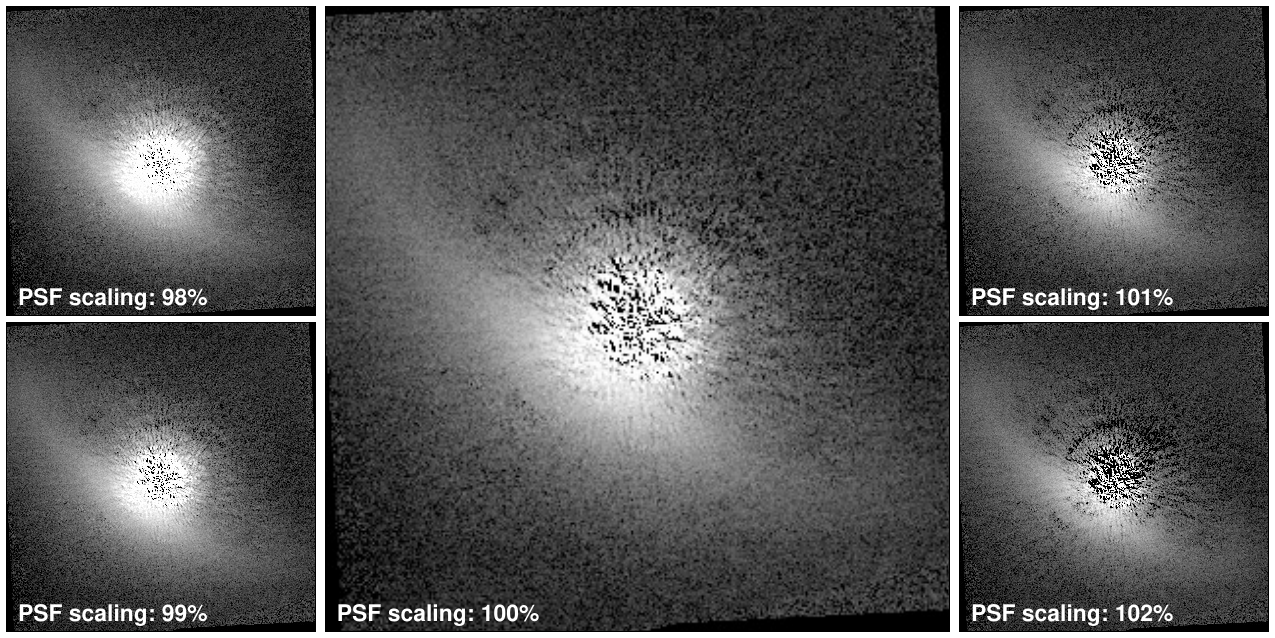}
    \caption{Visual representation of the post-processed HD~10647 F182M dataset using a PSF
    scaling offset by $\pm1$ and $\pm2\%$ from the nominal solution. Following PSF
    subtraction, the backgrounds were adjusted to zero. The demonstration shows that a
    PSF scaling over 1\% from the nominal case (middle panel) results in visible over- and under-subtraction
    residuals in this dataset. We used this visual method to estimate the statistical 
    error resulting from the uncertainty in the PSF scaling.}
    \label{fig:PSFscaling}
\end{figure*}

The resulting final image (shown in the large right panel of Figure \ref{fig:allims} for the F182M observations of 
HD~61005) was produced by averaging the two rotations, using masks to block bright background sources in
the PSF data. As a last step, to reduce the effects of unrealistic negative pixels resulting from
lingering oversubtraction and/or speckle noise in the core, if a pixel's value was less than -3$\sigma_{\rm bckg}$ (where 
$\sigma_{\rm bckg}$ is the standard deviation of background pixels determined at a clean location in 
the image), it was not included in the average, unless all (both) pixels averaged were below it.

\subsection{Photometric error estimates} \label{app:A_pt_3}

Estimating the photometric error of faint extended sources surrounding bright targets imaged via coronagraphy
is a complex task. Contributions to the total error include terms from uncertainty in the mean sky background level
(not the same as background noise level), photon noise from the bright central source, uncertainty in the 
post-processing algorithm (such as the values of the scaling coefficients), and correlated noise (such as speckles). 
Below, we use the F182M observations of HD~10647 to understand the contribution of each term to the total error and 
establish a process for estimating the photometric errors of the survey. In this example, we will calculate the 
photometric error of the main belt of HD~10647, located between deprojected distances of 70-100 au. The belt is 
imaged by $N_{\rm px;target} = 13016$ pixels with the NIRCam SW filter with a total flux of 0.604 mJy, assuming a pixel 
area of $2.22\times10^{-14}~{\rm sr}$.

\begin{figure*}
    \centering
    \includegraphics[width=0.99\linewidth]{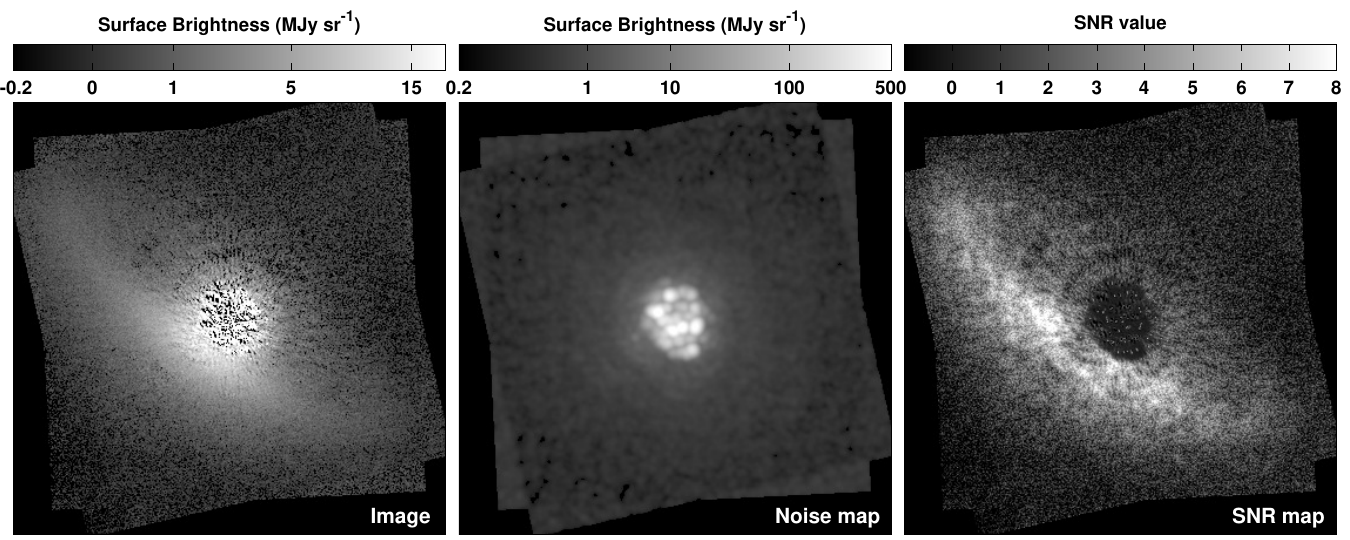}
    \caption{The {\it left} panel shows the observed F182M image of the HD~10647
    system, the {\it middle} panel shows the noise map characterizing the correlated errors, while the {\it right} panel shows the SNR map, i.e.\ the ratio of the first two images. The disk is detected at a high SNR.}
    \label{fig:SNR}
\end{figure*}

The first error term we investigate results from uncertainty in the mean sky background level, providing a systematic
error to the photometric measurement. In the {\it left} panels of Figure \ref{fig:errorsky},
we analyze the per-pixel surface brightness levels for the test case observation. The simplest way to characterize it
would be to plot the levels as a function of offsets from the stellar center location ({\it top left panel}), which yields a value
of $\approx 0.06 \pm 0.38~{\rm MJy~sr}^{-1}$ at the edge of the image. Since the target fills most of the FOV,
a better way is to plot the values as a function of the deprojected offset from the central star ({\it bottom left panel}). 
This estimate
narrows the distribution, by removing the brighter disk regions that are present at the edge of the image across 
the projected major axis. The surface brightness drops to $\approx -0.01 \pm 0.31~{\rm MJy~sr}^{-1}$ at 350 au
and $\approx -0.05 \pm 0.33~{\rm MJy~sr}^{-1}$ at 450 au. Choosing an area -- by-eye -- towards the 
back-scattering regions, where we anticipate lower disk flux, the sky background reaches the lowest per-pixel noise
level at $\pm 25~{\rm MJy~sr}^{-1}$. Once we understand the per-pixel variance, we can determine the error on estimating
a mean sky background value. Since we have a limited clean sky background area, we simulated the average 
background measurement we would measure with 1000 observations, each executed within a 20 px $\times$ 20 px square area, 
with values generated with a $\sigma_{\rm px}=0.31~{\rm MJy~sr}^{-1}$ standard deviation. The variance, 
around zero, was $\sigma_{\rm sky} = 0.016~{\rm MJy~sr}^{-1}$, which is smaller than the variation we see in the 
mean background value in the deprojected image from 350 au to 450 au, showcasing that there is also natural 
fluctuation in the background and that the NIRCam FOV is too small -- at short wavelengths -- for this extended 
object to provide a clear sky area. Therefore, as a conservative estimate, we give 
3$\sigma_{\rm sky}=0.048~{\rm MJy~sr}^{-1}$ as the $1\sigma$ error in estimating the baseline sky correction value. 
This value is also in agreement with the mean variation we measured in random background apertures.
The contribution to the total error from the uncertainty in the sky mean value will be 
$N_{\rm px;target}\sigma_{\rm sky} = 0.014~{\rm mJy}$, or 2.3\% of the total flux.

Next, we investigate the error resulting from the scaling of the reference PSF. The hybrid classical-LOCI 
post-processing technique we introduced allowed us to place constraints on the sum of the coefficients as well
as the on the subtraction residuals and their errors. The technique results in clean and flat subtractions in the
stellar core and high accuracy PSF construction and scaling factor. Regardless, the bulk of the total error in the
photometry of the extended disk structure will be from the inaccuracies of the removal of the stellar contribution. 
The sum of the PSF coefficients of the hybrid classical-LOCI method in theory equals the ratio of the fluxes of the
target vs.\ reference sources. To estimate the error from the PSF scaling, we adjusted the coefficients given by
our pipeline by various factors and analyzed the results. In Figure \ref{fig:PSFscaling}, we show the outcome
of varying the coefficients -- and therefore the PSF scaling -- by $\pm 1$ and $\pm 2\%$ from the nominal solution. Note,
the sky background values were determined and subtracted from each solution to ensure they don't introduce additional errors. 
The $\pm 2\%$ solutions are clearly under- and oversubtracting the reference PSF, producing unacceptable results.
The $99\%$ scaled PSF subtraction is undersubtracting, producing a brighter core, but at a level that is acceptable
and that could be explained by an unresolved inner component or chromatic mismatch. Similarly, the $101\%$ scaling
is also acceptable, with lower levels of oversubtraction that could be explained by a chromatic mismatch or
a not ideal small gird dither pattern of the PSF observation. Therefore, we place a 1\% uncertainty on the PSF scaling
of this particular target-reference pair. We measure the total disk flux at 0.527 and 0.681 mJy for the 101 and 99\%
scaled PSF solutions, i.e.\ at $\pm 0.077~{\rm mJy}$ values, respectively, corresponding to an error of 12.7\%. The error
from this term is 5.5$\times$ larger than from the error in the mean sky value. Admittedly, the size of this
error is greatly dependent on what we determine to be ``acceptable'' after a visual inspection, therefore
we aimed to be conservative with the $\pm1\%$ scaling. The main belt does intersect the core of the PSF for this example
case, therefore we anticipate larger errors for the test case than for some of the other systems in the survey.

Finally, we investigate the error from photon and correlated noise factors, such as speckles. This is done by 
estimating an error for all pixels by calculating the standard deviation of all pixels within a distance of 
4$\times$ the FWHM of the stellar PSF ($=4\times2~{\rm px}$ at F182M), weighted by their offset using a gaussian 
kernel. This algorithm incorporates a rough estimate of both error components. In Figure \ref{fig:SNR}, we show
the SNR errormap produced by this method, showcasing the high confidence in the detection of the disk around
HD~10647 with NIRCam at 1.82 $\micron$. The total error from this term is equal to the root-sum-square of the individual
errors over the 13016 pixels within the main belt, yielding 2.5 mJy or 0.4\%, which is negligible next to the 
error resulting from the uncertainty in the mean sky background value (2.3\%) or the statistical errors (12.7\%).
The final total error is an RSS of the individual errors or 12.9\%, driven by the uncertainties in the PSF scaling.
We will finally note that for the reflectance spectra calculations, the photometry is appropriate scaled by the
deprojected $R^2$ pixel values.

\begin{figure*}
    \centering
    \includegraphics[width=0.99\linewidth]{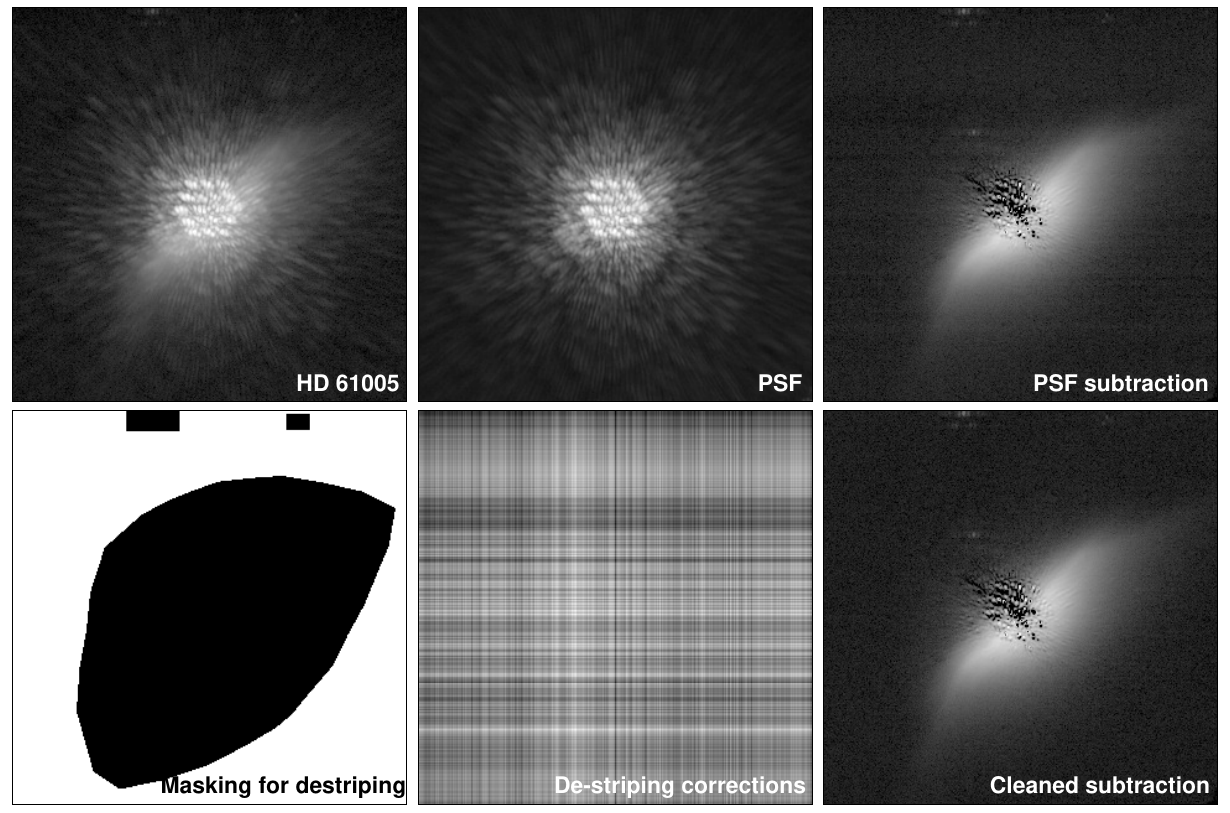}
    \caption{The figure demonstrates the steps taken to de-stripe (remove the row and column artifacts) the observations taken at the first rotation of the HD~61005 F182M dataset. {\it Top-Left:} Target image in log scale; {\it Top-Middle:} 
    Scaled PSF image in log scale, {\it Top-Right:} Initial
    PSF subtracted image, {\it Bottom-Left:} Custom mask used for this frame to separate areas that determined the
    corrections; {\it Bottom-Middle:} The de-striping correction (linear scale); {\it Bottom-Right:} Cleaned subtraction.}
    \label{fig:destriping}
\end{figure*}

\subsection{Row/Column corrections (de-striping)}

The NIRCam H2RG detectors present a low level row and column offset noise. The rows' offsets, known as the
$1/f$ noise \citep{schlawin20}, is a read-out pattern that originates from the read out circuits (ROICs) and
the direct current (DC) biases in the SIDECAR ASIC electronics \citep{raucher11}. Column offsets also appear,
due to minor bias voltage variations. While these row and column offsets are reduced by the pipeline using the 
detector reference pixels, which are read out by the same electronics at the same time, residual noise is still 
apparent in the stage 2 data, once the PSFs are subtracted to study low surface brightness features.
In Figure \ref{fig:destriping}, we present the steps we further take to reduce the latent row/column noise
in the datasets. The first two panels show the target and the hybrid combined PSF images at one of the rotations, 
while the third panel in the top row displays their difference image with the residual noise. This latent pattern depends 
on not only the inherent patterns present in the individual images, but also the two-directional PSF offset and
scaling applied, therefore striping is removed in the detector frame, without rotations. We determine the striping 
residual by constructing a custom mask for each individual target-PSF observation pair and calculating the median row 
and column values. The custom masks block the extended features, background point sources present in the target and 
PSF images, and any other detector patterns from affecting the corrections. We verify the correction in the last 
panel of Figure \ref{fig:destriping}, and apply the correction to the original image (first panel), as we translate 
and rotate all images to their final locations prior to PSF subtractions, as smooth patterns behave better under 
numerical interpolations.

\end{document}